%% file: KiDS-photo-z-ANNs_accepted.tex
\documentclass[longauth]{aa} % for the long lists of affiliations 

\usepackage{graphicx}
\usepackage{txfonts}
\usepackage{arydshln}
\usepackage[colorlinks=true,citecolor=blue,urlcolor=blue]{hyperref}
\usepackage{url}
\usepackage[usenames,dvipsnames]{xcolor}

\newcommand{\phz}{photo-$z$}
\newcommand{\phzs}{photo-$z$s}
\newcommand{\spz}{spec-$z$}
\newcommand{\spzs}{spec-$z$s}
\newcommand{\zsp}{z_\mathrm{spec}}
\newcommand{\zph}{z_\mathrm{phot}}
\newcommand{\ttt}{\texttt}
\newcommand{\mtt}{\mathtt}
\newcommand{\mrm}{\mathrm}
\newcommand{\meanz}{\langle z \rangle}

\begin{document} 

\title{Photometric redshifts for the Kilo-Degree Survey}
\subtitle{Machine-learning analysis with artificial neural networks}
\titlerunning{Kilo-Degree Survey photometric redshifts with artificial neural networks}

\author{
M.~Bilicki\inst{\ref{Leiden},\ref{NCBJ},\ref{ZielonaGora}}
\and H.~Hoekstra\inst{\ref{Leiden}}
\and M.~J.~I.~Brown\inst{\ref{Monash}} 
\and V.~Amaro\inst{\ref{Pancini}}
\and C.~Blake\inst{\ref{Swinburne}}
\and S.~Cavuoti\inst{\ref{Pancini},\ref{Capodimonte},\ref{IFNF}}
\and J.~T.~A.~de~Jong\inst{\ref{Kapteyn},\ref{Leiden}}
\and C.~Georgiou\inst{\ref{Leiden}}
\and H.~Hildebrandt\inst{\ref{Bonn}}
\and C.~Wolf\inst{\ref{ANU}}
\and A.~Amon\inst{\ref{Edinburgh}}
\and M.~Brescia\inst{\ref{Capodimonte}}
\and S.~Brough\inst{\ref{NSW}}
\and M.~V.~Costa-Duarte\inst{\ref{SaoPaulo},\ref{Leiden}}
\and T.~Erben\inst{\ref{Bonn}}
\and K.~Glazebrook\inst{\ref{Swinburne}}
\and A.~Grado\inst{\ref{Capodimonte}}
\and C.~Heymans\inst{\ref{Edinburgh}}
\and T.~Jarrett\inst{\ref{UCT}}
\and S.~Joudaki\inst{\ref{Oxford}}
\and K.~Kuijken\inst{\ref{Leiden}}
\and G.~Longo\inst{\ref{Pancini}}
\and N.~Napolitano\inst{\ref{Capodimonte}}
\and D.~Parkinson\inst{\ref{Queensland},\ref{KASSI}}
\and C.~Vellucci\inst{\ref{Pancini}}
\and G.~A.~Verdoes Kleijn\inst{\ref{Kapteyn}}
\and L.~Wang\inst{\ref{SRON},\ref{Kapteyn}}
}

\institute{
Leiden Observatory, Leiden University, P.O. Box 9513, NL-2300 RA Leiden, The Netherlands \label{Leiden} %1
\and
National Centre for Nuclear Research, Astrophysics Division, P.O. Box 447, PL-90-950 {\L}\'{o}d\'{z}, Poland \label{NCBJ} %2
\and
Janusz Gil Institute of Astronomy, University of Zielona G\'ora, ul. Szafrana 2, 65-516 Zielona G\'{o}ra, Poland \label{ZielonaGora} %3
\and
School of Physics and Astronomy, Monash University, Clayton, Victoria 3800, Australia \label{Monash} %4
\and
Department of Physics ``E.Pancini'', University Federico II, via Cinthia 6, I-80126 Napoli, Italy \label{Pancini} %5
\and
Centre for Astrophysics \& Supercomputing, Swinburne University of Technology, P.O. Box 218, Hawthorn, VIC 3122, Australia \label{Swinburne} %6
\and
INAF - Astronomical Observatory of Capodimonte, Via Moiariello 16, I-80131 Napoli, Italy \label{Capodimonte} %7
\and
INFN section of Naples, via Cinthia 6, I-80126, Napoli, Italy \label{IFNF} %8
\and
Kapteyn Astronomical Institute, University of Groningen, Postbus 800, 9700 AV, Groningen, The Netherlands \label{Kapteyn} %9
\and
Argelander-Institut f\"{u}r Astronomie, Auf dem H\"{u}gel 71, 53121 Bonn, Germany \label{Bonn} %10
\and
Research School of Astronomy and Astrophysics, Australian National University, Canberra, ACT 2611, Australia \label{ANU} %11
\and
School of Physics, University of New South Wales, NSW 2052, Australia \label{NSW} %12
\and
Instituto de Astronomia, Geof\'isica e Ci\^encias Atmosf\'ericas, Universidade de S\~ao Paulo, R. do Mat\~ao 1226, 05508-090, S\~ao Paulo, Brazil \label{SaoPaulo} %13
\and
Scottish Universities Physics Alliance, Institute for Astronomy, University of Edinburgh, Royal Observatory, Blackford Hill, Edinburgh EH9 3HJ, UK \label{Edinburgh} %14
\and
Department of Astronomy, University of Cape Town, Private Bag X3, Rondebosch, 7701, South Africa \label{UCT} %15
\and
Department of Physics, University of Oxford, Denys Wilkinson Building, Keble Road, Oxford OX1 3RH, U.K. \label{Oxford} %16
\and
School of Mathematics and Physics, University of Queensland, Brisbane, QLD 4072, Australia \label{Queensland} %17
\and
Korea Astronomy and Space Science Institute, Daejeon 34055, Korea \label{KASSI} %18
\and
SRON Netherlands Institute for Space Research, Landleven 12, 9747 AD, Groningen, The Netherlands \label{SRON} %19
}

\authorrunning{M.~Bilicki et al.}

\offprints{M.~Bilicki,\\ \email{\url{bilicki@strw.leidenuniv.nl}}}

\date{ Accepted for publication in A\&A on 30 April 2018}
\abstract{
We present a machine-learning photometric redshift (ML \phz) analysis of the Kilo-Degree Survey Data Release 3 (KiDS DR3), using two neural-network based techniques: ANNz2 and MLPQNA. Despite limited coverage of spectroscopic training sets, these ML codes provide \phzs\ of quality comparable to, if not better than, those from the Bayesian Photometric Redshift (BPZ) code, at least up to $\zph\lesssim0.9$ and $r\lesssim23.5$. At the bright end of $r\lesssim20$, where very complete spectroscopic data overlapping with KiDS are available, the performance of the ML \phzs\ clearly surpasses that of BPZ, currently the primary \phz\ method for KiDS.

Using the Galaxy And Mass Assembly (GAMA) spectroscopic survey as calibration, we furthermore study how \phzs\ improve for bright sources when photometric parameters additional to magnitudes are included in the \phz\ derivation, as well as when VIKING and WISE infrared (IR) bands are added. While the fiducial four-band $ugri$ setup gives a \phz\ bias $\langle \delta z / (1+z) \rangle = -2\times 10^{-4}$ and scatter $\sigma_{\delta z / (1+z)} < 0.022$ at mean $\meanz = 0.23$, combining magnitudes, colours, and galaxy sizes reduces the scatter by $\sim7\%$ and the bias by an order of magnitude. Once the $ugri$ and IR magnitudes are joined into 12-band photometry spanning up to 12 $\mu$m, the scatter decreases by more than $10\%$ over the fiducial case. Finally, using the 12 bands together with optical colours and linear sizes gives $\langle \delta z / (1+z) \rangle < 4\times 10^{-5}$ and $\sigma_{\delta z / (1+z)} < 0.019$.

This paper also serves as a reference for two public \phz\ catalogues accompanying KiDS DR3, both obtained using the ANNz2 code. The first one, of general purpose, includes all the 39 million KiDS sources with four-band $ugri$ measurements in DR3.
The second dataset, optimized for low-redshift studies such as galaxy-galaxy lensing, is limited to $r\lesssim20$, and provides \phzs\ of much better quality than in the full-depth case thanks to incorporating optical magnitudes, colours, and sizes in the GAMA-calibrated \phz\ derivation.
}


\maketitle 

%%%%%%%%%%%%%%%%%%%%%%%%%%%%%%%%%%%%%%%%%%%%%%%%%%

\section{Introduction}

The distance to an astronomical object is arguably one of the most important quantities that we want to measure. In extragalactic studies, except for sparse and mostly local samples of redshift-independent `distance indicators', the best way of estimating source distance is via its redshift. Redshifts can be measured precisely only from spectroscopy, and massive dedicated spectroscopic surveys have been very successful in obtaining them for millions of galaxies. But even the most advanced techniques, such as multi-fibre spectroscopy, have their limitations: obtaining spectroscopic redshifts (\spzs) is expensive and time-consuming.  
Today's largest {imaging} surveys already include hundreds of millions galaxies, and this number is expected to grow by at least an order of magnitude in the coming years. It is already now infeasible to obtain spectra for even a significant fraction of catalogued galaxies.

Fortunately, {many applications do not require the redshift precision available from spectroscopy.}
Various approaches {can be employed} instead to \textit{estimate} redshifts, both on an individual basis, as well as for redshift distributions of particular samples. 
As far as the individual redshifts are concerned, broad-band photometry can be used to derive \textit{photometric redshifts} (\phzs; \citealt{Baum57,Koo85,LS86}), using two main approaches, sometimes in concert {\citep{Brodwin06,PHAT}}: \textit{i)} empirical, usually machine-learning (ML); and \textit{ii)} source energy distribution (SED), or template, fitting. 

In the ML domain, techniques such as artificial neural networks \citep[ANNs,][]{Tagliaferri03,Firth03}, boosted decision/regression trees \citep[BDTs,][]{Gerdes10}, Gaussian processes \citep{Way09}, or genetic algorithms \citep{GAz}, to list just a few, are calibrated (\textit{trained}) on \spz\ samples, which have the relevant set of passbands measured, to derive the mapping from photometry to \spzs, and the best-fit solution is then propagated to the target data with photometry only. These methods are usually agnostic to any physics, and thus need well-controlled and representative training sets to work properly. If the latter are available, the ML \phz\ approaches usually provide both very accurate (minimal bias) and precise (low scatter) estimates. In addition to magnitudes, they can also {directly} use other galaxy observed properties as inputs, such as colours, sizes, half-light radii, and so on \citep[e.g.][]{ANNz,Wadadekar05,Wray08}. {A recently proposed extension of ML \phz\ estimation is by working directly on imaging data instead of using post-processed source catalogues; this is possible thanks to `deep learning' \citep[e.g.][]{Hoyle16,DIsanto17}.}

Among the advantages of ML methods (MLMs) is their ability to automatically handle some systematics in the data, such as varying aperture bias as a function of wavelength, which can produce errors in SED fitting if not dealt with correctly. Last but not least, the empirical methods are able to `learn from data' -- their performance gets increasingly better as the training data improve in quantity and quality. The major drawback of MLMs is their poor performance in extrapolation, that is, ML \phzs\ are usually not reliable beyond the range of magnitudes, colours, etc., spanned by the training sets.

SED-fitting, on the other hand, uses a more direct and physically motivated approach of matching the measured multi-band magnitudes, or fluxes, to the best-fit redshifted spectrum, the latter coming from libraries of either real galaxy spectra and/or artificial ones \citep[e.g.][]{BPZ,hyperz,EAZY}. The main advantage of these methods is that they are largely independent of spectroscopic calibration, although they might require priors to avoid assigning unrealistically high redshifts to galaxies of bright observed magnitudes \citep{Kodama99,EAZY}. The two main drawbacks of SED-fitting \phzs\  are: \textit{i)} template model dependence, which requires knowledge of realistic galaxy SEDs at various redshifts; \textit{ii)} their general inability to use parameters other than magnitudes/fluxes (such as galaxy sizes or shapes).

The empirical methods for deriving  individual \phzs\ {always require} spectroscopic calibration data, even if the requested properties of these data differ for various techniques. Overlapping \spzs\ are also needed to judge the performance of the methods, and this includes the SED-fitting ones as well. Generally, it can be stated that every approach for redshift estimation requires \spz\ samples at some stage of its application or performance testing. 

In this paper we present a machine-learning \phz\ analysis for the Kilo-Degree Survey \citep[KiDS,][]{KiDS}. KiDS is one of the major wide-angle photometric surveys currently undertaken, along with the Dark Energy Survey \citep{DES} and the Hyper Suprime-Cam Subaru Strategic Program \citep{HSC}, and all three are precursors for even more ambitious efforts such as the Large Synoptic Survey Telescope \citep{LSST} and Euclid \citep{Euclid}. These surveys face a common challenge of the necessity of using \phzs\ for scientific analyses, as \spzs\ are and will be available only for a very small fraction of detected sources.

{The KiDS pipeline \phz\ solution, used in most of the scientific analyses so far,} comes from the Bayesian Photometric Redshift \citep[BPZ,][]{BPZ} SED-fitting code. 
However, two ML approaches are also used for deriving alternative \phzs\ {in KiDS}: MLPQNA \citep{Cavuoti12}, and ANNz2 \citep{ANNz2}. 
This paper {aims at quantifying the performance of these MLMs in the most recent Data Release 3 (DR3) of KiDS.} 
{This has already been briefly presented in the DR3 publication \citep{KiDS-DR3} and here we provide a more detailed discussion. The paper is accompanied by two ANNz2-based KiDS \phz\ catalogues and serves as a reference for their end-users.}

{The overall structure of this paper is the following. First, in \S\ref{Sec:codes_used} we present the \phz\ codes used in this work: ANNz2 (\S\ref{Sec:ANNz2}), and MLPQNA (\S\ref{Sec:MLPQNA}). Next, in \S\ref{Sec:Input_data} we describe the data employed in our studies: photometric from KiDS (\S\ref{Sec:KiDS-photometric}), VIKING (\S\ref{Sec:VIKING}), and WISE (\S\ref{Sec:WISE}), as well as spectroscopic coming from various samples overlapping with KiDS (\S\ref{Sec:KiDS spectroscopic}). A summary of the joint photo-spectro sample is provided in \S\ref{Sec:KiDS_photo-spectro}.}

{We then explore ML \phzs\ in two different regimes and setups of the KiDS data. First, in \S\ref{Sec:KiDS-DR3-photozs} we study the performance of the two ANN-based algorithms at almost the full depth of KiDS, using various overlapping \spz\ datasets as training and test samples; we also compare the results with those from the fiducial KiDS \phz\ solution from BPZ (\S\ref{Sec:Random_DR3_subsample}-\ref{Sec:COSMOS-CDFS-tests}). We conclude that Section by describing in \S\ref{Sec:DR3-release} the publicly released KiDS DR3 full-depth \phz\ catalogue obtained by applying the ANNz2 algorithm. An earlier version of that dataset was already made available with the DR3 release\footnote{\url{http://kids.strw.leidenuniv.nl/DR3/ml-photoz.php\#annz2}} \citep{KiDS-DR3} and is now updated with this paper.} 

{In the second set of experiments, described in \S\ref{Sec:GAMA-depth}, we use ANNz2 for the bright end of KiDS, for which there is very complete spectroscopic training data from the Galaxy And Mass Assembly \citep[GAMA,][]{GAMA} survey.} 
{We study how the basic KiDS $ugri$ parameter space can be extended to improve \phzs\ at the GAMA depth, by adding further imaging information, such as galaxy morphology.} 
{We also examine what can be gained in terms of \phz\ quality if the wavelength range is extended by adding VIKING near-infrared (IR) and WISE mid-IR information. This is of particular importance because dedicated reductions of the relevant data are either ongoing (KiDS-VIKING) or planned (KiDS-WISE). The results of these tests are detailed in \S\ref{Sec:GAMA magnitude type}-\S\ref{Sec:GAMA-depth_mags+more_sets}.} 
{The GAMA-based analysis is also accompanied by a public catalogue release, in this case  limited to $r \lesssim 20$ mag, with much more accurate and precise \phzs\ than in the global solution; see \S\ref{Sec:GAMA catalogue release}. Such a sample with precise and accurate \phzs\ is of particular interest for studies such as galaxy-galaxy lensing, which require foreground data with well-constrained redshift estimates.} 

{In \S\ref{Sec:conclusions_and_prospects} we conclude and mention future prospects regarding KiDS \phzs.}

%%%
\section{Photometric redshift algorithms used}
\label{Sec:codes_used}

{In this Section we provide details of the two approaches used to obtain KiDS ML \phzs, ANNz2 and MLPQNA. The results from these two codes will be compared to the KiDS pipeline solution derived with the Bayesian Photometric Redshift algorithm \citep[BPZ,][]{BPZ}, and made publicly available together with the DR3 photometric data \citep{KiDS-DR3}. For the details of how BPZ was implemented in the KiDS pipeline, please see the relevant papers: \cite{KiDS-GL} and \cite{KiDS-DR3}.}

\subsection{ANNz2}
\label{Sec:ANNz2}

Most of the analysis of this paper, as well as two accompanying \phz\ catalogues, are based on the ANNz2 code \citep{ANNz2}. ANNz2 is a versatile ML package\footnote{Available from \url{https://github.com/IftachSadeh/ANNZ}; we used versions $\leq2.2.1$.}, originally designed as a successor of the ANNz software \citep{ANNz}. However, unlike its predecessor, ANNz2 is not limited to using only {artificial neural networks (ANNs)} but it also incorporates other {machine-learning methods (MLMs)}, such as {boosted decision/regression trees (BDTs)}. 
{ANNz2 is} based on the {Toolkit for Multivariate Data Analysis (TMVA)} package\footnote{\url{http://tmva.sourceforge.net/}} \citep{TMVA}, which itself is part of the ROOT C++ software\footnote{\url{https://root.cern.ch/}} \citep{ROOT}, {and therefore} allows the user to use various MLMs. In this study we have limited ourselves to exploring only the fiducial MLMs of ANNz2, i.e.\ ANNs and BDTs.
ANNz2 provides also other important improvements over ANNz. The first one is a high level of work automatisation via Python scripts, thanks to which the user does not have to define the individual MLM {properties}, allowing the software to generate {their architectures} randomly (which we applied here). By training a large ($\gtrsim 100$) number of ANNs and/or BDTs with various architectures -- in the Randomized Regression mode which we employed in our study -- the \phz\ derivation can be optimised both by using the `best' solution, as well as by folding all or part of all the solutions from each run. This allows for an overall improvement in the \phz\ quality without much user involvement in the training procedure.

The Randomized Regression mode of ANNz2 allows for deriving the probability distribution functions (PDFs) of the computed \phzs, by folding selected individual MLM results with their uncertainty estimates, the latter being derived using a k-nearest neighbours (kNN) estimator \citep{Oyaizu07}. However, these PDFs should not be treated as {actual} error distributions {with respect to the true redshift (which is unknown)} but rather as quantification of the uncertainties of {the \phz\ derivation method.}  
This will however apply to most \phz\ techniques that derive PDFs, including the fiducial KiDS method, BPZ (see the accompanying analysis by \textcolor{blue}{Amaro et al., subm.}). In general, we do not store these PDFs in the catalogues presented here, but they can be generated on request.

Last but not least, a major improvement in ANNz2 over ANNz (and several other ML \phz\ codes) is the possibility to weight the training data to mimic the target set. These weights can then be propagated throughout the training and evaluation procedure, by assigning a correction factor to the training objects depending on the input parameters. The weighting is done via the kNN method in the parameter space chosen by the user (for instance magnitudes, colours) by comparing the density of input sources to that of the target ones \citep{Lima08}. A similar approach was taken {in the KiDS cosmic shear analysis by} \cite{KiDS-450} to estimate the true redshift distributions of KiDS sources from the matched spectroscopic catalogues (the `DIR' calibration method therein).

The general framework of ANNz2 is similar to most other \phz\ MLMs. The code is fed with \textit{training} and \textit{validation} sets that have both the input (e.g. photometric) and output (e.g. redshift) parameters. If weighting of the training and validation data is requested, this is done at the beginning in the `\ttt{generate input trees}' stage of the procedure. A user-defined number and type of MLMs are first trained and then validated on the relevant data; the latter procedure is called \textit{optimisation} in ANNz2. Thus trained and validated MLMs can then be applied to `blind' data -- \textit{evaluation} sets -- either including \spzs\ for performance checks, or photometric-only for generating the final catalogues.

We followed the recommenations of \cite{ANNz2} to use at least 100 MLMs for Randomized Regression. Training BDTs is much faster than training ANNs for the same number of MLMs; on the other hand, the former requires more storage space and more memory in the optimization and evaluation process than the latter. The two types of MLMs also differ in performance: our experiments show that using BDTs generally gives worse results than ANNs, even if the number of the former is (much) larger than of the latter. In this paper we thus present results based on ANNs only; in most cases we used 250 ANNs for each experiment, with architectures always defined randomly within the code. 
{We note that a different, perhaps more optimal, setup of ANNz2 is possible if the ANNs are not generated randomly by the code but rather defined by the user, adjusted to the properties of the data (e.g. to the number of input parameters). In such a case, using fewer ANNs could give similar results to the approach we adopted here (\textcolor{blue}{John Soo, priv. comm.}). However, running ANNz2 would then require more user supervision; we thus opted for the fully randomised approach which allowed us to execute the computations in the background.}

ANNz2 provides various parameters to be set up by the user. We tested the influence of several of them on the final results and we eventually decided for the following configuration (see \citealt{ANNz2} as well as the ANNz2 online documentation for details):

\begin{itemize}

\item \ttt{optimCondReg}: a metric used to rank the performance of individual MLMs, its options are the bias, the 68th percentile scatter, or the outlier fraction;  in our experiments we found no significant difference between results for the `\ttt{sig68}' and `\ttt{bias}' options, and we used \ttt{optimCondReg = bias} everywhere;

\item \ttt{optimWithScaledBias}: used as an optimization criterion for the best MLM and the PDFs; we used \ttt{True}, i.e. the normalised bias $(\zph-\zsp)/(1+\zsp)$ was employed for optimization; 

\item \ttt{optimWithMAD}:  we used \ttt{True}, i.e. the best MLM and the PDFs were optimized using the MAD (median absolute deviation) rather than the 68th percentile of the bias distribution;

\item splitting of the training+validation data into separate training and validation sets was done randomly into two halves using the ANNz2 option \ttt{glob.annz["splitType"]  = "random"}

\item by default, ANNz2 does not use the actual errors of the training parameters but derives an error model from the data using the kNN-error method; the user can, however, propagate the actual parameter errors directly; we have tested this latter option {for our deep calibration data (zCOSMOS; \S\ref{Sec:COSMOS-CDFS-tests}), as well as for the case when low signal-to-noise WISE data were additionally used (\S\ref{Sec:GAMA-depth_mags+one_set})} and found only slight improvements in the results, {or none at all;} therefore, we used the default setup;

\item in some cases, as described in the text, we applied weighting of the training data (\ttt{useWgtkNN = True}) using a relevant reference sample; these weights were then used in the whole \phz\ estimation procedure;

\item ANNz2 outputs five types of point estimates of \phzs; the first of them, \ttt{ANNZ\_best}, comes from the single MLM which provides the best combination of performance metrics; the remaining ones are based  on \phz\ PDFs which are derived internally but do not have to be stored by the user (\ttt{glob.annz["doStorePdfBins"] = False}); the PDFs come in two options (one based on the true target as known from the training data, the other based on the results of the best MLM) and two pairs of related \phz\ point estimates are derived: \ttt{ANNZ\_PDF\_avg\_0} and \ttt{ANNZ\_PDF\_avg\_1} -- averages of the first / second PDF types (using the full weighted set of MLMs, convolved with uncertainty estimators), as well as \ttt{ANNZ\_MLM\_avg\_0} and \ttt{ANNZ\_MLM\_avg\_1} -- unweighted averages of all the MLMs which have non-zero PDF weights, i.e. of those MLMs that have good performance metrics; our experiments show that the best performance is usually achieved by \ttt{ANNZ\_MLM\_avg\_1} and we will be reporting statistics based on this point estimate; 

\item we do not use full PDFs in any other way than by employing point estimates based on them as described above; the PDFs for the published datasets can however be derived on request. 

\end{itemize}

{All the input features used in training as well as in kNN-weighting were normalised to the range $[-1;1]$ via linear rescaling; this is the default ANNz2 setup (\ttt{doWidthRescale = True}).} 

\subsection{MLPQNA}
\label{Sec:MLPQNA}

In the KiDS DR3 experiments of \S\ref{Sec:KiDS-DR3-photozs} we compare the ANNz2 results with those from another machine-learning approach used in the survey, namely MLPQNA {\citep{Cavuoti12}, which stands for the Multi Layer Perceptron feed-forward neural network (MLP; \citealt{Rosenblatt62}), trained by the Quasi Newton Algorithm (QNA; \citealt{Byrd94}) learning rule. This ML model is among the most efficient optimization methods searching for the minimum of the MLP training error function, since it makes use of a statistical approximation of the Hessian of this error, obtained by an iterative MLP network error gradient calculation.}
MLPQNA makes use of the L-BFGS algorithm (Limited-memory Broyden-Fletcher-Goldfarb-Shanno; \citealt{Byrd94}), originally designed for problems with a wide parameter space.
 
The analytical details of the MLPQNA model, as well as its performance for \phz\ estimation, have been extensively discussed elsewhere \citep{Cavuoti12,Brescia13,Cavuoti15a}, and the method has been to an earlier KiDS data release, DR2 \citep{Cavuoti15b}. Within KiDS DR3, it is embedded as a \phz\ prediction kernel into the METAPHOR (Machine-learning Estimation Tool for Accurate PHOtometric Redshifts) pipeline \citep{Cavuoti17}, able to extend the \phz\ estimation by providing also their error PDFs. The details of its application to the DR3 data are discussed in \cite{KiDS-DR3} and the resulting catalogue was released together with the overall DR3 data\footnote{\url{http://kids.strw.leidenuniv.nl/DR3/ml-photoz.php\#mlpqna}}.

MLPQNA is publicly available through the DAMEWARE (DAta Mining \& Exploration Web Application REsource; \citealt{Brescia14}) web-based infrastructure\footnote{\url{http://dame.dsf.unina.it/dameware.html}}.

%%%
\section{Input data}
\label{Sec:Input_data}

In this Section we present the data used in our studies. Most of the results described here are based on public photometric data from the KiDS DR3 \citep{KiDS-DR3}, supplemented with some additional photometry outside of the nominal KiDS footprint, as well as with public and proprietary spectroscopic datasets. Part of the analysis also uses  infrared photometry derived from VIKING and WISE surveys. Below we provide the details of the samples used in this paper.

\subsection{KiDS photometric data}
\label{Sec:KiDS-photometric}

The Kilo-Degree Survey \citep[KiDS,][]{KiDS} is a wide-angle imaging campaign being conducted with the OmegaCAM camera \citep{Kuijken11} at the VLT Survey Telescope \citep{Capaccioli12}, using four broad-band optical filters ($ugri$). The target area of the survey is $\sim1500$ deg$^2$ in two patches, one on the celestial Equator, and the other in the South Galactic Cap. The main science goal of KiDS is to map the large-scale distribution of matter, and extract related cosmological information, using weak lensing techniques \citep{KiDS-450,Joudaki17,Joudaki18,Koehlinger17,vanUitert17b}, it is however also perfectly suitable for studying galaxy evolution \citep{Tortora16}, structure of the Milky Way \citep{PilaDiez}, detecting galaxy clusters \citep{Radovich17} and high-redshift quasars \citep{Venemans15}, as well as looking for strong lenses \citep{Petrillo17}, or even Solar System objects \citep{Mahlke17}, to name just a few applications.

KiDS has had three data releases so far \citep{KiDS-DR2,KiDS-DR3} and DR3 includes about 450 deg$^2$ of photometric data, with typical $5\sigma$ depth of 24.3, 25.1, 24.9, 23.8 mag in $2\arcsec$ apertures in $ugri$, respectively. 
Accurate colours and absolute photometric calibration down to $\sim 2\%$ in $gri$ and $\sim 3\%$ in $u$ are ensured via a specific photometric homogenization scheme. In the $r$ band, which is used for galaxy shape measurements, the typical PSF size is below $0.7\arcsec$; sub-arcsecond seeing is also used for the $g$ and $i$ band observations, while in $u$ the mean PSF is $1\arcsec$. 
All this guarantees excellent-quality deep imaging, perfectly suitable for astrophysical studies where precise photometry is crucial.

The details of KiDS data reduction are provided in the relevant papers \citep{KiDS-DR2,KiDS-DR3}; of importance for this work is that the basic catalogues are produced using the SExtractor \citep{SExtractor} software {in dual-image mode}, which provides several magnitude types for each band{, measured directly on astrometrically and photometrically calibrated, stacked images (``coadds'').} Among them are Kron-like automatic aperture magnitudes \ttt{MAG\_AUTO}, as well as isophotal ones, \ttt{MAG\_ISO}. {Two types of catalogues are produced: single-band, with source extraction and photometry done independently in each band, and multiband, which we use here, where source detection is based on the $r$ band, and aperture-matched photometry is derived for the other filters.}

KiDS data reduction also involves  a post-processing stage in which Gaussian Aperture and Photometry \citep[GAaP,][]{GAaP} magnitudes are derived {\citep{KiDS-GL}. For this, the coadds are first ``Gaussianized'', meaning that the point spread function (PSF) is homogenized across each individual coadd. The photometry is then measured using a Gaussian-weighted aperture (the size and shape of which are set by the $r$-band major and minor axis lengths and orientation) that compensates for the seeing differences between the filters because each part of the source gets the same weight across all filters. We will call this procedure ``PSF homogenization'' from now on.}

{Additional ``photometric homogenization'' is achieved by adjusting the zeropoints across the full survey area. This is done using the coadd overlaps in the $r$ and $u$ bands, homogenizing the photometry in these two filters, and then $g$ and $i$ bands are tied to the $r$ band using stellar locus regression, which homogenizes the $g-r$ and $r-i$ colours, and therefore the $g$ and $i$ band zeropoints. The photometric homogenisation is done using the GAaP photometry, and in the final catalogues the resulting zeropoint offsets (`\ttt{ZPT\_OFFSET\_band}' for each filter) are reported in separate columns, together with Galactic extinction corrections which are based on the \cite{SFD} maps. The zeropoint-calibrated and extinction-corrected magnitudes will be denoted as `\ttt{calib}' from now on:
\begin{multline}
\mtt{MAG\_type\_band\_calib} = \\ = \mtt{MAG\_type\_band} + \mtt{ZPT\_OFFSET\_band} - \mtt{EXT\_SFD\_band}\;,
\end{multline}
where the uncalibrated measurements were taken directly from the KiDS multiband catalogue. However, since the zeropoint offsets were derived from GAaP measurements, they work better for the GAaP photometry than for other types.}

The GAaP magnitudes are the default ones for KiDS, and are used in most of the scientific analyses. They are also applied in the pipeline-\phz\ derivation with BPZ \citep{KiDS-GL}{, as they provide very good galaxy colours}. Our studies presented here will also use GAaP magnitudes as defaults. In \S\ref{Sec:GAMA-depth} we show quantitatively that indeed this type of photometry is the most optimal for \phz\ estimation among the 3 tested types available from KiDS multiband data (the other being ISO and AUTO), even for bright sources. One should bear in mind, though, that the GAaP magnitudes cannot be generally used as proxies for \textit{total} fluxes of galaxies, especially at the bright end where they severely underestimate the total flux (by $\sim1$ mag or more).

Unless indicated otherwise, the KiDS data we use have undergone appropriate cleaning of bad photometry{. First of all, in all the analysis we used only those sources which have GAaP magnitudes measured for each band, to guarantee that \phzs\ are estimated using the full $ugri$ information. These cuts apply mostly to the $u$ and $i$ bands, in which respectively 13\% and 7\% of KiDS sources do not have magnitude measurements in the multiband catalogue because of a combination of intrinsically lower source brightnesses in $u$ and decreased depth in both $u$ and $i$ bands, as compared to $g$ and $r$ (cf.\ table 3 in \citealt{KiDS-DR3}). Once this filtering is applied in all the bands, the DR3 sample is reduced to 39.2 million objects.}

{Such a four-band requirement is obviously a limitation for the current analysis, especially compared to the BPZ approach where the \phzs\ are derived for \textit{all} the KiDS sources, and upper limits, non-detections, and lacking measurements are handled appropriately. However, the \phzs\ using fewer bands will be obviously of worse overall quality than the $ugri$-based ones, which would lead to inhomogeneities in the eventual ML \phz\ catalogue. We postpone a detailed analysis of the influence of missing bands on KiDS \phzs\ to the forthcoming KiDS-VIKING 9-band data release, where this situation will be much more common.}

{Furthermore, we defined a `CLEAN' sample by additionally requiring that magnitude errors are provided in each band, as well as by removing artefacts with any of the following masking flags set: readout spike, saturation core, diffraction spike, secondary halo, or bad pixels\footnote{This was done by applying the bitwise operator $\mtt{IMAFLAGS\_ISO}\_\mathrm{band} \, \& \, 01010111 = 0$ for each band. See appendix A.2 of \cite{KiDS-DR3} for more details of these flags.}, following \cite{Radovich17}. The resulting CLEAN dataset includes 36.9 million KiDS-DR3 objects out of 48.7 million in the full multi-band catalogue.}

For the purpose of \phz\ derivation in DR3 we also define a `FIDUCIAL' dataset, which is based on the CLEAN sample additionally purified of stars (by applying the $\mtt{SG2DPHOT}=0$ flag\footnote{{\ttt{SG2DPHOT} is a KiDS star/galaxy classification flag derived from the $r$-band source morphology \citep{KiDS-DR2,KiDS-DR3}. Extended sources are assigned a value of 0.}}) and trimmed at the faint end to encompass the magnitude ranges of the spectro-photo training set described in \S\ref{Sec:KiDS spectroscopic}. More precisely, we removed from the KiDS DR3 those sources for which any of the $ugri$ magnitudes were beyond the 99.9th percentile of the spectroscopic catalogue distribution. These cuts are $\mtt{MAG\_GAAP\_u\_calib}<25.4$, $\mtt{MAG\_GAAP\_g\_calib}<25.6$, $\mtt{MAG\_GAAP\_r\_calib}<24.7$ \& $\mtt{MAG\_GAAP\_i\_calib}<24.5$. Applying these cuts on the artefact-purified DR3 dataset gives 20.5 million sources in the FIDUCIAL sample. This sample will be used as the reference set for weighting the spectroscopic catalogue, used for training of the global DR3 \phz\ solution, as discussed in \S\ref{Sec:DR3-release}.

We emphasise that in the released full-depth catalogue, the \phzs\ are derived for all the sources that have the 4 $ugri$ GAaP magnitudes measured, although they will be most likely unreliable outside the FIDUCIAL dataset, and of course do not have any meaning for stars. In order not to propagate residual bad photometry to \phz\ calibration, in the training and validation (optimisation) phase we additionally applied $\mtt{MAGERR\_GAAP}\_\mathrm{band}<1$ for each band, but not in the tests nor the final evaluation in the target catalogue. {Such an additional cut affects mostly the $u$ filter, and removes an extra $\sim3\%$ from the training data.}
 
We also used KiDS-like observations outside of the nominal KiDS footprint, namely from VST imaging of deep spectroscopic fields described in \S\ref{Sec:KiDS spectroscopic}: CDFS (from the VOICE survey, \citealt{VOICE}) and two DEEP2 fields (2h and 23h).  Details of observing conditions of these observations are provided in \cite{KiDS-450}, appendix C1. Here it is sufficient to note that they were of comparable quality as the full KiDS.

\subsection{VIKING photometry}
\label{Sec:VIKING}

We also tested how going beyond KiDS photometry can improve the \phzs. The planned KiDS footprint is practically fully covered by the VISTA Kilo-degree Infrared Galaxy survey \citep[VIKING,][]{VIKING} providing five near-IR bands $z Y J H K_s$ at a similar depth to KiDS, and a joint KiDS-VIKING data reduction is ongoing.  
At the time of performing the experiments described in this paper, we did not yet have access to these joint data, and thus limit our tests to GAMA-LAMBDAR \citep{GAMA-LAMBDAR} forced VIKING photometry on the GAMA sources. These tests are therefore currently limited to KiDS-GAMA objects {in the equatorial fields}, and apply only to GAMA depth in KiDS ($r\lesssim20$ mag). The input photometry, and in particular the apertures used for these forced-photometry VIKING measurements, came from SDSS DR7. They are therefore of worse quality than what can be expected from a similar approach using KiDS sources instead. {They also had no homogenisation of a similar form as in KiDS applied.}

The LAMBDAR measurements come in the form of fluxes, and we also used those that were negative or zero\footnote{We did not have to convert the fluxes to any magnitude system, because ML \phz\ methods are agnostic to physical units. What matters is that each particular photometric parameter is measured self-consistently. This is a useful advantage of these methods over the SED-fitting ones.}. We discarded only those sources where at least one of the VIKING bands had no measurement at all ($\mtt{band\_flux} = -999$); at GAMA depth this is however a small number, $\sim3\%$, of all the objects. No extinction corrections nor zero-point offsets were applied in this test phase. In the near future, once joint optical -- near-IR photometry becomes available for KiDS sources, also outside the GAMA regions, these experiments will be extended. In particular, we expect the \phzs\ derived from KiDS+VIKING to improve over what is presented in \S\ref{Sec:GAMA-depth} thanks to incorporating VIKING GAaP magnitudes, zero-point calibrated and extinction-corrected in the same manner as the KiDS $ugri$ measurements.

\subsection{WISE}
\label{Sec:WISE}
In the GAMA-depth experiments, we also used date from the Wide-field Infrared Survey Explorer \citep[WISE,][]{WISE}, which cover the full sky in four mid-IR bands ($W1$ -- $W4$) ranging from 3.4 $\mu$m to 23 $\mu$m. WISE is the most sensitive in its two shorter-wavelength channels, $W1$ (3.4 $\mu$m) and $W2$ (4.6 $\mu$m), reaching respectively 54 $\mu$Jy and 71 $\mu$Jy (5$\sigma$), which in $W1$ is equivalent to $\sim21$ mag in the AB system. The public WISE catalogue\footnote{Available from \url{http://irsa.ipac.caltech.edu/Missions/wise.html}.} is however limited to sources with a $5\sigma$ detection in at least one band. Therefore, rather than using that dataset, which is very incomplete even at GAMA depth \citep{Cluver14,Jarrett17}, we employed the GAMA-LAMBDAR catalogue which includes forced-photometry WISE flux measurements for all the GAMA sources in the equatorial fields.

Because of the much lower sensitivity of the $W4$ (23 $\mu$m) channel than the three others, it has a very high number of non-detections ($\mtt{W4\_flux}=0$) even in the LAMBDAR catalogue and will not be used. Also the $W3$ band (12 $\mu$m) has a considerable number of  measurements lacking {(17\%)}, so part of our experiments employing WISE use either the $W1+W2$ bands or $W1+W2+W3$. At present such WISE forced photometry for KiDS sources is not available, so these tests were limited only to the GAMA depth (\S\ref{Sec:GAMA-depth}) and cannot currently be extended beyond that. We are planning to obtain WISE measurements for a subsample of KiDS sources, but this will be limited to the bright end of the latter survey because of its much larger depth \citep[cf.][]{Lang16}.

\subsection{Spectroscopic: compilation of various datasets}
\label{Sec:KiDS spectroscopic}

As any other ML \phz\ tool, ANNz2 and MLPQNA used in this study require \textit{training sets} of sources from the target photometric sample which have also spectroscopic redshifts measured. Empirical \phz\ methods perform optimally if the training set is representative of the target data. Ideally, the former should be a random subset of the latter to provide the same distributions in magnitudes, colours, and redshift. However, even if this ideal setup cannot be met, ML will perform well as long as the important parameters such as magnitudes span the same range in training and target data, especially if some weighting is applied on the training data to mimic the target set. 

On the other hand, MLMs usually do badly in extrapolating; for instance, training on a bright subset of much deeper target data is likely to give very biased results at the faint end. In addition, it must be remembered that  ML \phzs\ usually perform best at the median redshift (where they should provide practically zero  bias), and by construction they tend to overestimate the redshifts at low $z$ and underestimate them at high $z$ \citep[e.g.][]{2MPZ}. On the other hand, if applied properly, MLMs should give unbiased redshift as a function of $\zph$ in a sense that $\langle \zsp | \zph \rangle = \zph$, which is not necessarily the case for template-fitting approaches.

In modern deep photometric surveys we hardly ever have spectroscopic subsets that are sufficiently representative for \phz\ training at the full depth \citep[e.g.][]{Sanchez14,Masters15,Beck16} and the situation will get worse with planned campaigns such as LSST or Euclid \citep[cf.][]{Newman15}, especially when one takes into account the requirements that \phzs\ must meet in order not to heavily degrade cosmological constraints \citep{Ma06}.

In the case of KiDS, the original footprint was optimized to first cover four GAMA fields as well as the COSMOS area. Of these, only the latter offers spectroscopy at a depth comparable to KiDS photometric data. On the other hand, the whole KiDS footprint is covered by either SDSS or 2dFLenS spectroscopic observations {(see below), and these two samples have very similar properties in terms of their target selection for spectroscopy. Although very useful as a part of the overall training set, } neither of these reach the full KiDS depth, and both offer only sparse sampling of colour-preselected objects (mostly luminous red galaxies, LRGs) beyond the local volume of $z<0.1$. There are however several deep spectroscopic fields in the southern sky, and for the purpose of extending our spectroscopic calibration data, we have either included external measurements or asked for dedicated observations of some of them, as discussed in \cite{KiDS-450}.

Below we provide details of the spectroscopic data integrated into the training/calibration set used in this study. {Their basic properties are summarised in Table \ref{Tab: spectro data} and their redshift distributions are shown in Fig.~\ref{Fig:dNdz_spectro_all}. All the \spz\ samples had appropriate redshift quality cuts applied to preserve only science-grade measurements.} Cross-matches between KiDS photometric sources and the spectroscopic objects were done using a $1\arcsec$ matching radius.

\subsubsection{GAMA}
\label{Sec:GAMA spec-z}
Galaxy And Mass Assembly \citep[GAMA,][]{GAMA} is a spectroscopic survey of five fields, which employed the AAOmega spectrograph on the Anglo-Australian Telescope, with targets selected mostly from the Sloan Digital Sky Survey (SDSS), as well as from other surveys, including KiDS. It spans 3 equatorial fields (G09, G12 and G15) and two southern ones (G02 and G23) of which only G02 is outside the KiDS footprint. GAMA is  $98.5\%$ complete spectroscopically for SDSS galaxies with $r_\mathrm{Petro}<19.8$ mag in the equatorial fields, and $94.2\%$ complete for KiDS galaxies to $i<19.2$ mag in G23 \citep{GAMA-II}. Some of the measured sources are however fainter, and there additionally exists an unpublished catalogue of deeper observations in the G15 field (2,300 sources of good redshift quality, with $\meanz = 0.34$) which we also use here.

These four fields give us in total almost 230,000 KiDS sources with GAMA spectroscopic redshift measurements, and their $\meanz = 0.23$. This, together with the excellent spectroscopic completeness of GAMA and no colour preselection therein other than star and quasar removal, makes GAMA \textit{the} photometric redshift calibration set at the bright end of KiDS. Indeed, we will devote \S\ref{Sec:GAMA-depth} to a GAMA-depth analysis, where GAMA \spzs\ were used to calibrate KiDS ML \phzs\ with excellent accuracy and precision.

\input{table-specz-fullDR3.tex}

\subsubsection{SDSS}
The Sloan Digital Sky Survey \citep[SDSS,][]{SDSS} is a photometric and spectroscopic survey of $\sim\pi$ steradians of the Northern sky, performed from the Apache Point Observatory in New Mexico, USA. SDSS is currently in Stage IV of its operations \citep{SDSS-IV} and we use its spectroscopic sources from the Data Release 13 \citep[DR13,][]{SDSS.DR13} which encompasses and supersedes all the earlier releases.

SDSS overlaps with KiDS in the equatorial fields above \mbox{$\delta=-3^\circ$.} 
{From the SDSS spectroscopic dataset, we only use sources with class `GALAXY', and do not include those which are `QSO', as training on the latter might bias the \phzs. We verified that it is indeed the case: training with SDSS QSOs included gives slightly worse overall results than if they are not used (but see \citealt{Soo17}).} 
There are almost 57,000 SDSS DR13 spectroscopic galaxies with KiDS DR3 photometric measurements, however those with $r<19.8$ are mostly included in GAMA, and eliminating them gives about 43,000 unique KiDS$\times$SDSS galaxies. While the full SDSS-matched sample has a mean redshift of only $\meanz\sim0.35$, those that remain after removal of GAMA are at much higher redshifts, $\meanz\sim0.71$. This is mostly thanks to the completed Baryon Oscillation Spectroscopic Survey \citep[BOSS,][]{BOSS} and first data from the extended BOSS \citep[eBOSS,][]{eBOSS}, both targeting preselected higher-$z$ galaxies. A caveat is that these are mostly LRGs, which are not representative of the whole population and could bias the \phzs\ {if used as the sole calibration sample \citep{redMaGiC}. In our analysis we employ them as part of the overall training set, and the \spz\ sample weighting applied in the \phz\ derivation procedure should mitigate the related effects of an unevenly populated colour space.} 

\begin{figure}
\centering
\includegraphics[width=0.48\textwidth]{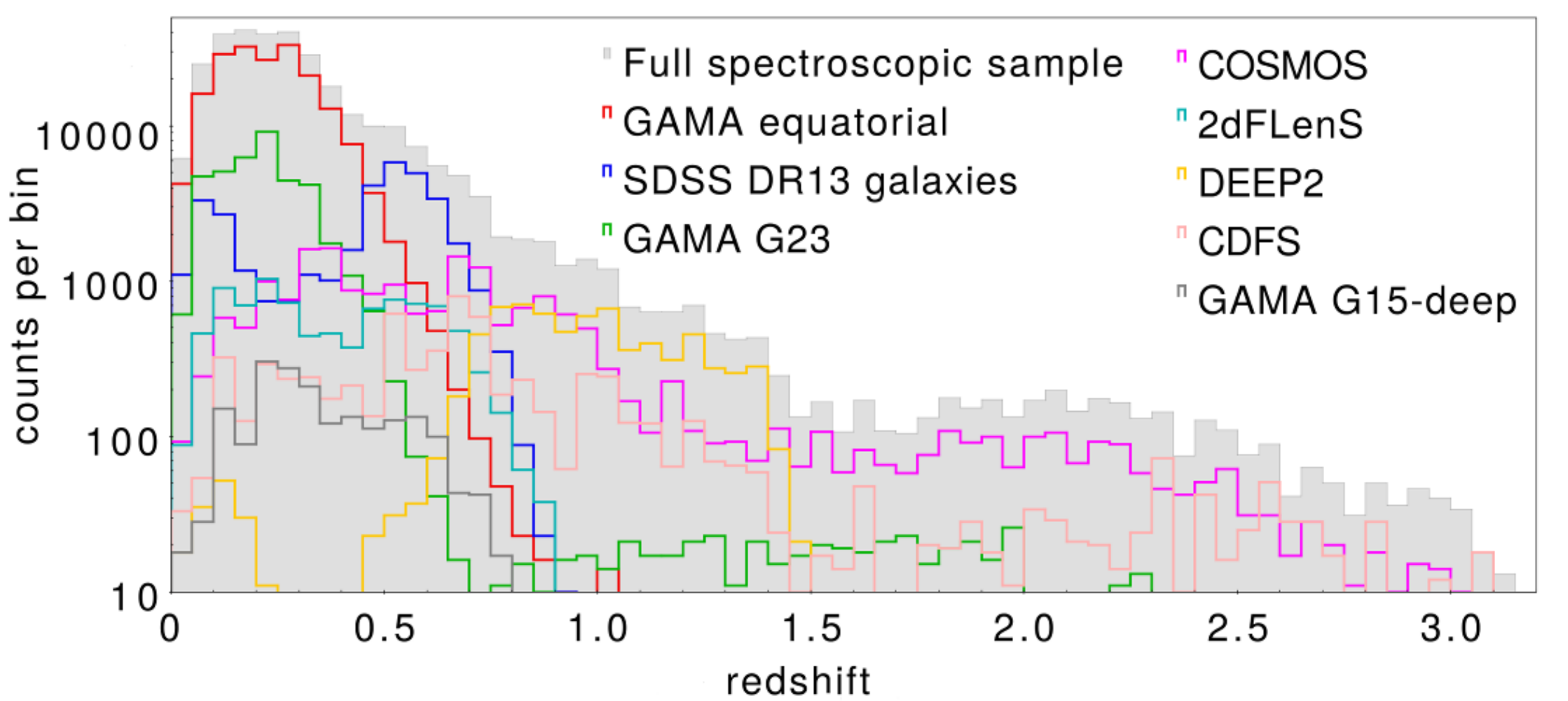}
\caption{Redshift distribution of the full KiDS DR3 spectroscopic training sample and of particular datasets included. {The histograms show sources with 4-band $ugri$ photometry in KiDS or in auxiliary datasets outside the nominal footprint.}}
\label{Fig:dNdz_spectro_all} 
\end{figure}

\subsubsection{2dFLenS}
The 2-degree Field Lensing Survey \citep[2dFLenS,][]{2dFLenS} is a spectroscopic survey conducted at the Australian Astronomical Observatory between September 2014 and January 2016, covering an area of 731 deg$^2$ principally located in the KiDS regions.  By expanding the overlap area between galaxy redshift samples and gravitational lensing imaging surveys, 2dFLenS aims to facilitate the joint analysis of lensing and clustering observables including all cross-correlation statistics \citep[e.g.][]{Joudaki18}, and to assist with \phz\ calibration by direct training methods \citep{2dFLenS-photo-z} and by cross-correlation \citep{Johnson17}.  The 2dFLenS spectroscopic dataset contains two main target classes: $\sim 40,000$ LRGs across a range of redshifts $z < 0.9$, selected by SDSS-inspired cuts, and a magnitude-limited sample of $\sim 30,000$ objects in the range $17 < r < 19.5$.

In KiDS DR3 we have almost 12,000 2dFLenS galaxies, of which 9,000 are unique (after excluding sources in common with SDSS and GAMA). The mean redshift of 2dFLenS, after eliminating the SDSS and GAMA overlap, is $\meanz \sim 0.39$. As in the case of SDSS, a caveat of using the 2dFLenS sources for \phz\ training is that outside the local volume they are mostly LRGs.

\begin{figure*}
\begin{center}
\includegraphics[width=0.3\textwidth]{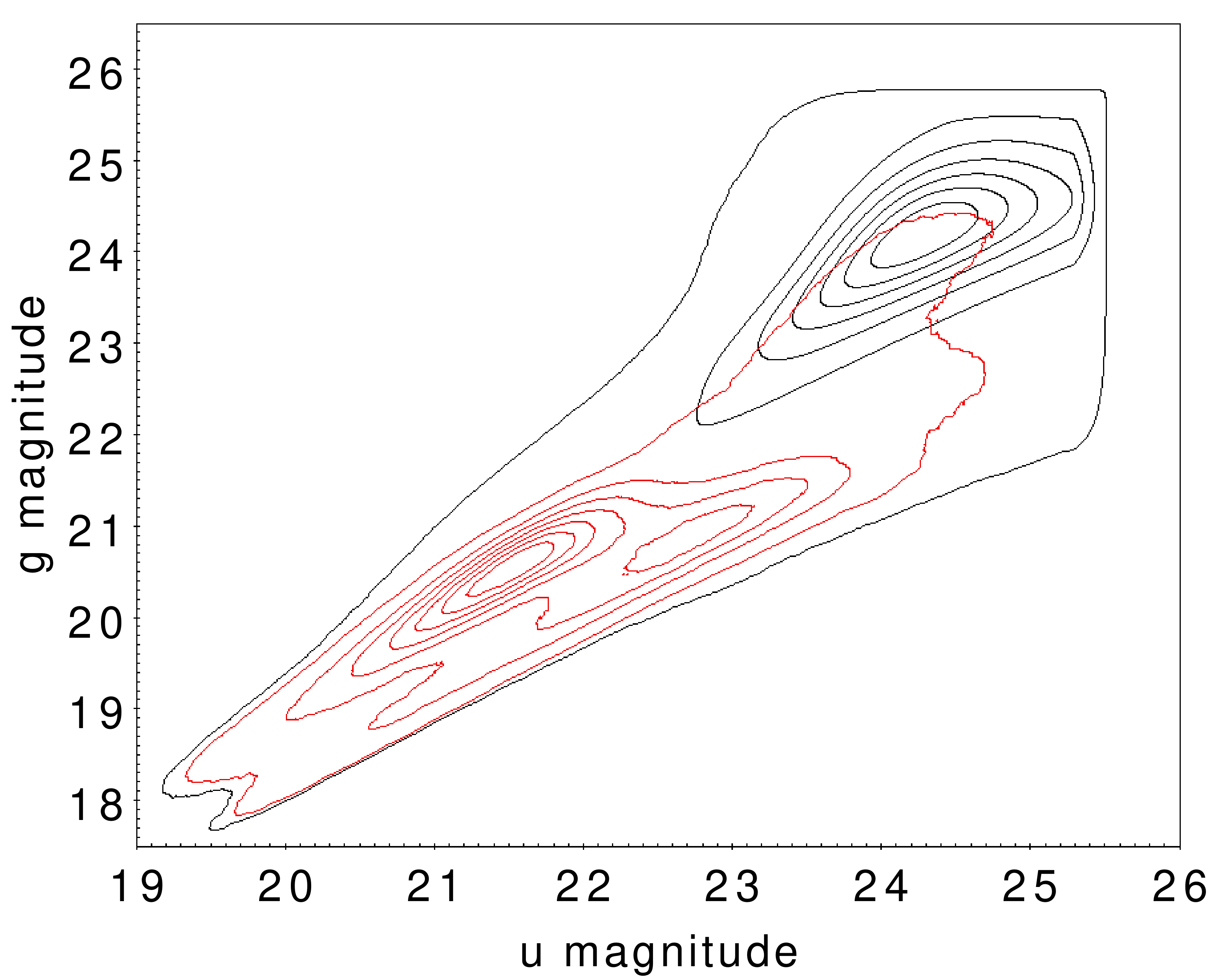}
\hspace{2mm}
\includegraphics[width=0.3\textwidth]{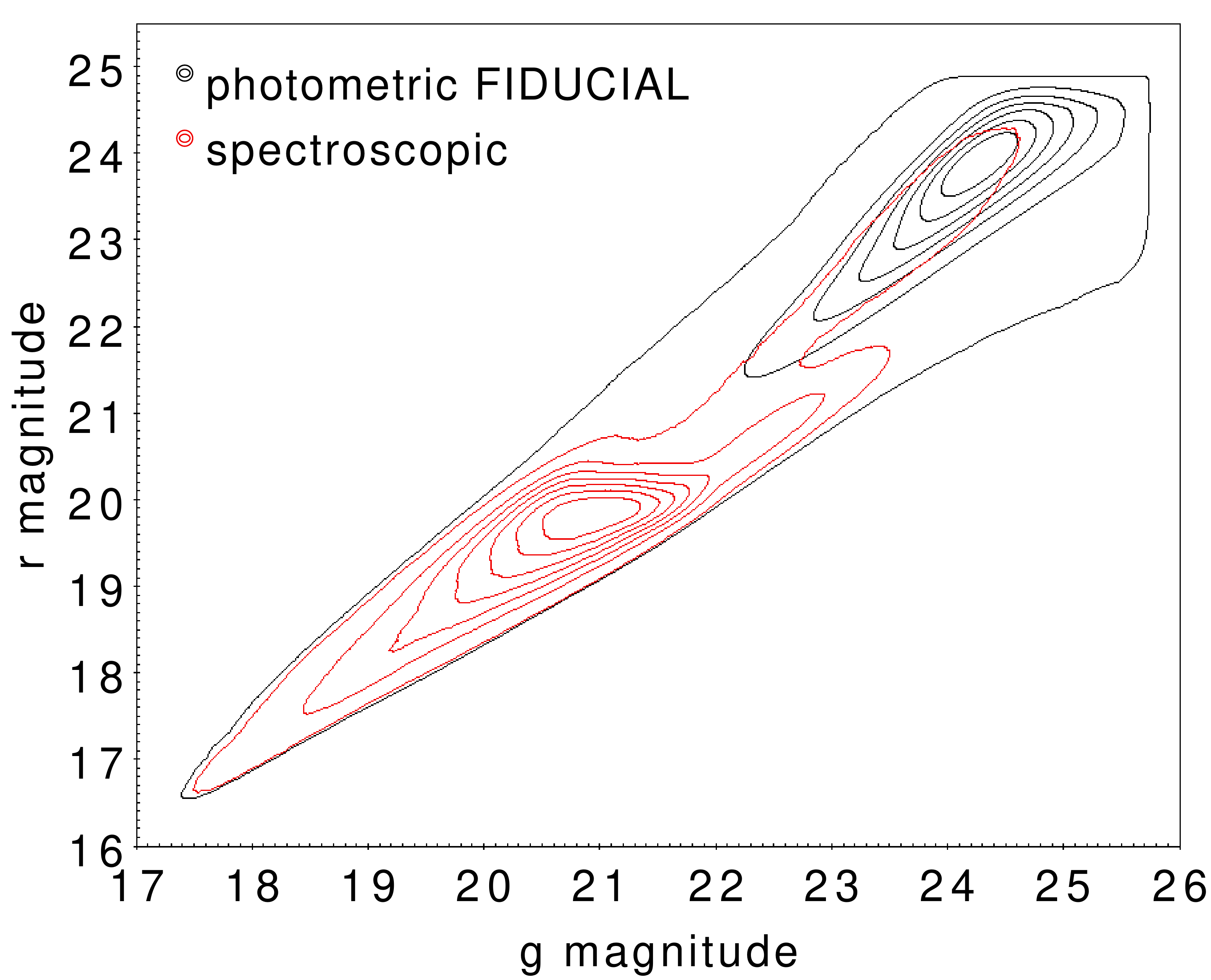}
\hspace{2mm}
\includegraphics[width=0.3\textwidth]{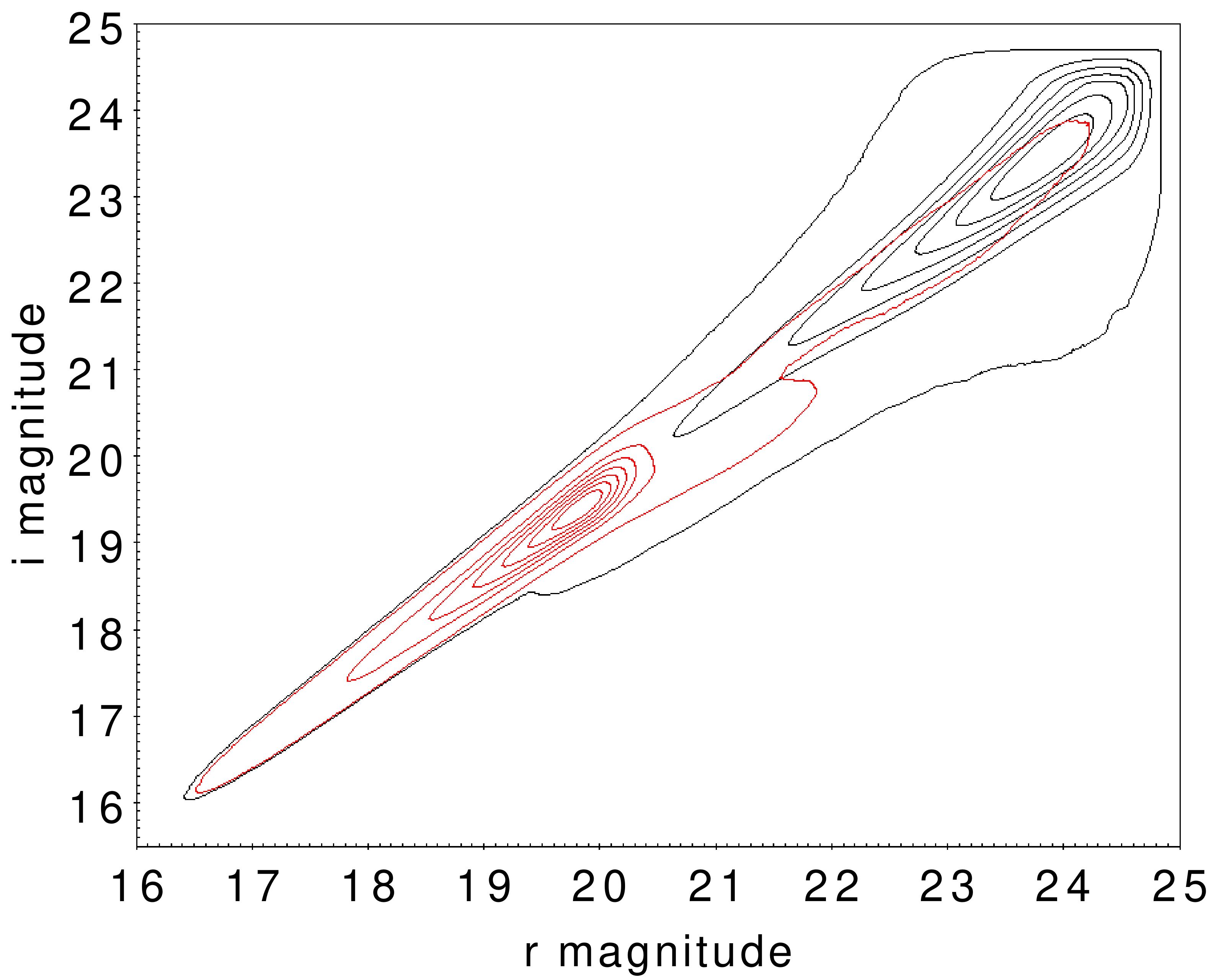}\\

\vspace{2mm}

\includegraphics[width=0.3\textwidth]{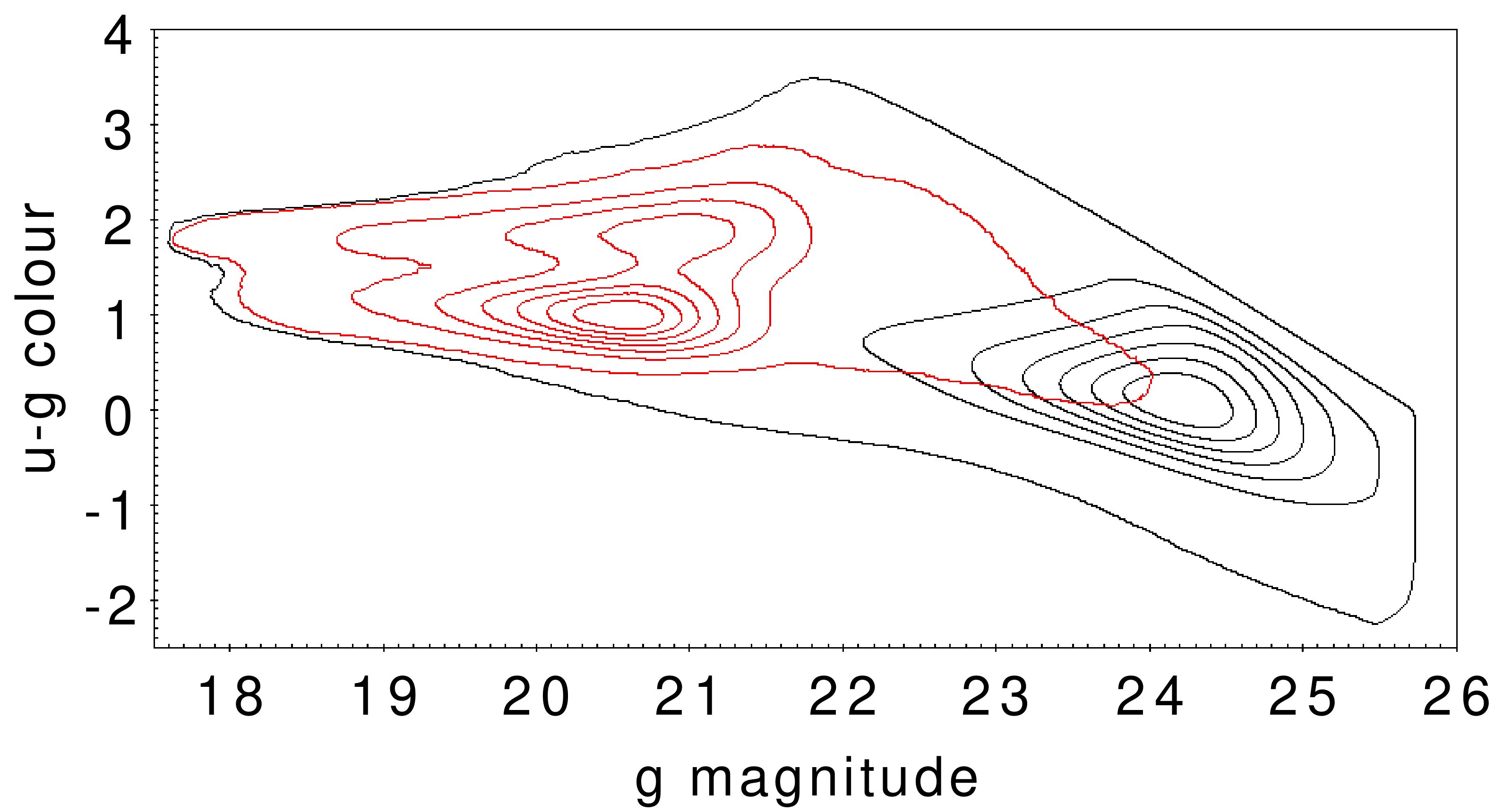}
\hspace{2mm}
\includegraphics[width=0.3\textwidth]{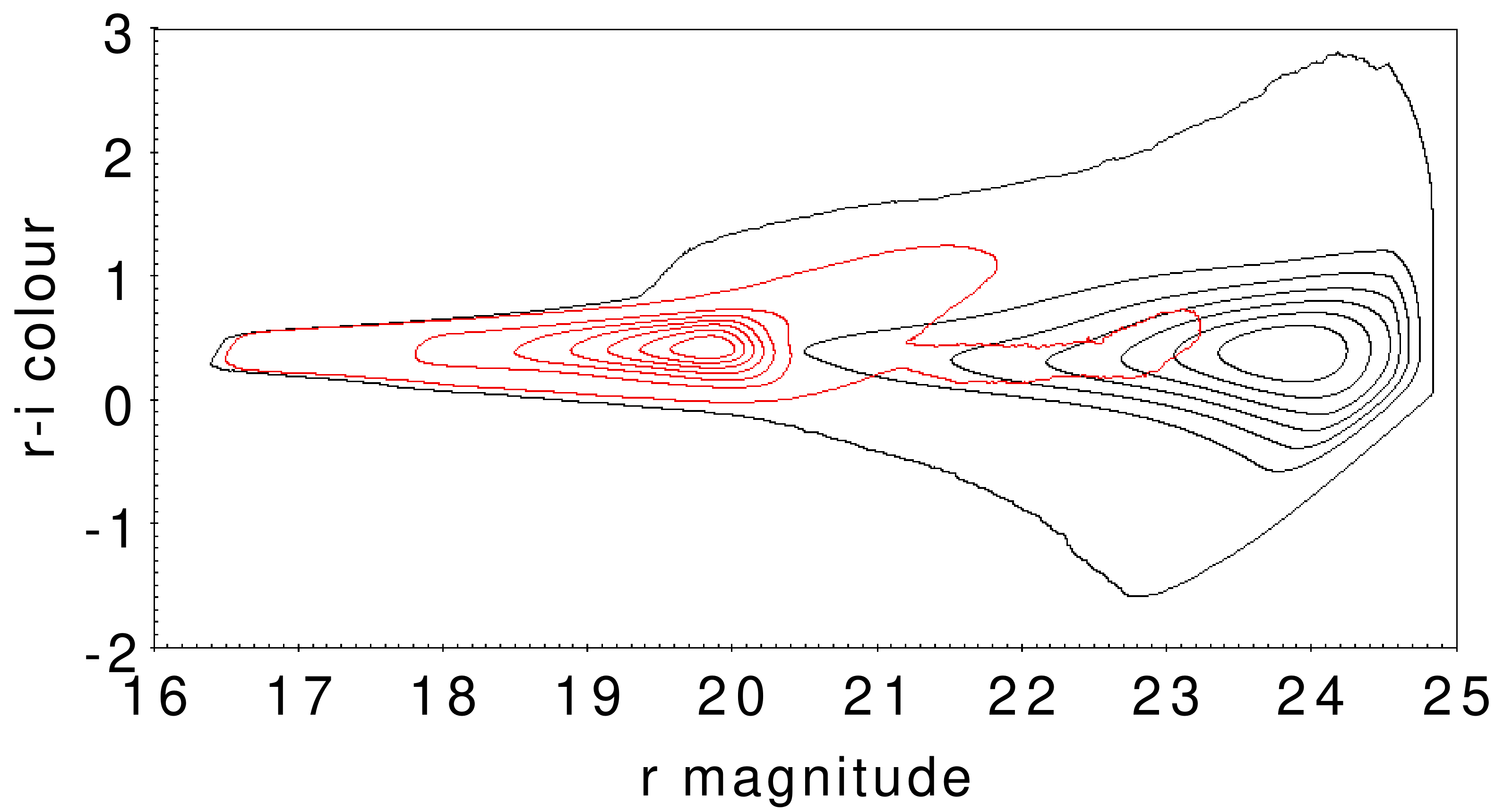}
\hspace{2mm}
\includegraphics[width=0.3\textwidth]{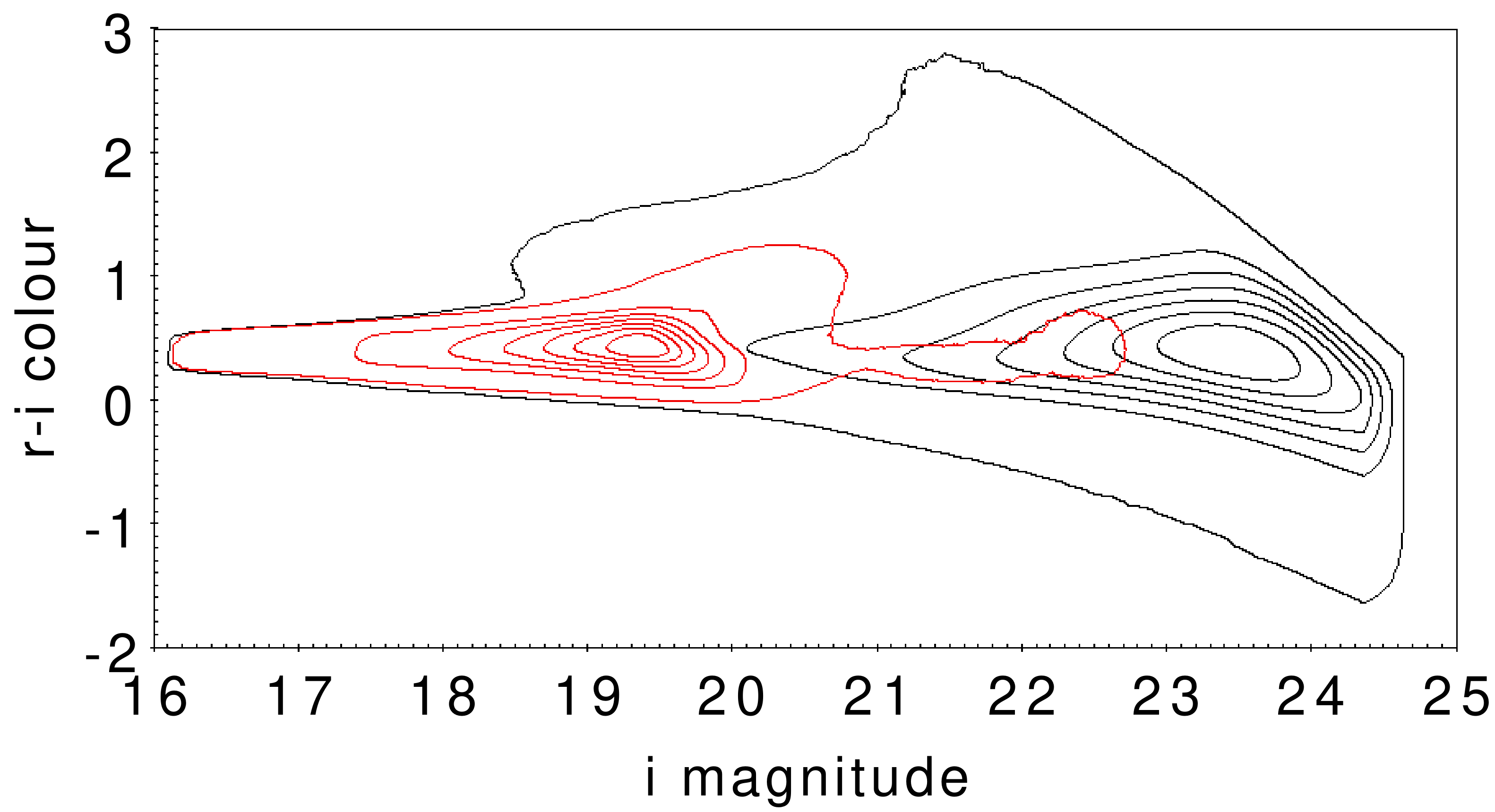}

\end{center}

\caption{{Top row: comparison of magnitude distributions for the KiDS-DR3 photometric FIDUCIAL sample (black) and the spectroscopic redshift calibration dataset (red). Bottom row: similar comparison but for selected magnitude-colour planes. The contours are linearly spaced. The FIDUCIAL sources are used as the reference for weighting the \spz\ training set in the derivation of \phzs\ for the full catalogue (\S\ref{Sec:DR3-release}). See also Fig.~\ref{Fig:col-col_training_weighting} for colour-colour plots where weighting of the training is additionally illustrated.}}
\label{Fig:mag_color_photo_spectro}
\end{figure*}

\subsubsection{zCOSMOS}
The COSMOS field, centred roughly at $\alpha=150^\circ, \delta=2.2^\circ$, is currently one of the most comprehensively sampled areas in terms of deep spectroscopy. The original KiDS footprint was designed to cover 1 deg$^2$ of this field, so the photometric data here come from the main KiDS pipeline. For the \phz\ experiments, we joined two main spectroscopic datasets in this field. The first one is a non-public dataset from the zCOSMOS team, that is deeper than the public release \citep{zCOSMOS}, kindly shared by the zCOSMOS team. It incorporates spectroscopic data from various other observational campaigns in this field. After cleaning of bad-quality redshifts, this catalogue includes almost 28,000 sources, of which over 19,000 have a counterpart in KiDS-DR3 with $\meanz = 0.87$.

We supplement this catalogue with a GAMA-team reanalysis of public COSMOS data, dubbed G10 \citep{Davies15}, which includes almost 24,000 spectroscopic measurements of appropriate quality. As there is large overlap between the zCOSMOS and G10 samples, we removed the duplicates, and eventually were left with about 6,700 unique sources from a G10 cross-match with KiDS, of $\meanz = 0.61$, which were added to zCOSMOS. 

The two samples together give about 25,900 sources with KiDS measurements, of which 21,100 have all four $ugri$ bands measured. These data have $\meanz = 0.71$ but span up to $z = 3$ (Fig.~\ref{Fig:dNdz_spectro_all}) which makes them crucial for \phz\ calibration at the high-$z$ tail of KiDS.

\subsubsection{CDFS}
The Chandra Deep Field South (CDFS), centred at $\alpha\simeq53.1^\circ$, $\delta\simeq-27.8^\circ$, is another area  surveyed by VST that has deep spectroscopy available. Unlike COSMOS, however, it is located outside the KiDS footprint and the photometric data we use here come from a KiDS-like reduction of VST imaging from the VOICE project \citep{VOICE}. As in the zCOSMOS case, here the spectroscopic data were also composed of two datasets: an ESO-released compilation of GOODS/CDFS spectroscopy\footnote{\url{http://www.eso.org/sci/activities/garching/projects/goods/MasterSpectroscopy.html}}, including about 5,600 sources with `secure' or `likely' redshifts (of which 3,500 with KiDS measurements, $\meanz = 1.04$), supplemented with data from the Arizona CDFS Environment Survey \citep[ACES,][]{ACES} (6,400 with quality flag $\geq3$, of which 4,440 in KiDS, $\meanz = 0.59$).

After removing duplicates we have 7,000 \spzs\ in the CDFS area, of which 5,600 with all four bands available. This sample is slightly deeper on average ($\meanz = 0.74$) than KiDS-zCOSMOS, but has much fewer spectroscopic sources; it however also spans to large redshifts of $z\sim3$ (Fig.~\ref{Fig:dNdz_spectro_all}), which makes it equally important for \phz\  calibration and helps mitigate sample variance related to using very small areas for this purpose.

\subsubsection{DEEP2}

The DEEP2 Galaxy Redshift Survey \citep{DEEP2} covers 2.8 deg$^2$ in four patches and is colour-selected in a way to target high redshift ($z\sim1$) galaxies. Although not appropriate for \phz\ calibration on its own, it is very useful when joined with the other samples, adding data in the $0.5<z<1.5$ range.

Two of the DEEP2 fields are within reach of VST and we have KiDS-like observations for them: the 2h field, centred at $\alpha\simeq37.2^\circ, \delta\simeq0.5^\circ$, and the 23h field at $\alpha\simeq352.0^\circ, \delta\simeq0.0^\circ$. There are over 16,000 DEEP2 sources with $\mtt{ZQUALITY}\geq3$ in there, of which some 9,000 have KiDS-like measurements. Among these, 7,100 have measurements in all the four $ugri$ filters, with $\meanz = 0.97$, but limited almost entirely to $0.6 \lesssim z \lesssim 1.4$.

\subsection{Properties of the photo-spectro compilation}
\label{Sec:KiDS_photo-spectro}

In total we have over 310,000 sources with good-quality spectroscopic redshift measurements available for KiDS DR3. However, for these to be applicable as a \phz\ training set, the data had to be cleaned of bad photometry as discussed in \S\ref{Sec:KiDS-photometric}. 
We also required $z>0.001$ to avoid residual stellar contamination and local volume galaxies with a possibly significant contribution of peculiar velocity to measured redshift.
After these cuts, the full DR3 spectroscopic set used in this paper includes almost 280,000 objects. We reiterate though that this sample, having $\meanz = 0.33$, is dominated by GAMA with $z<0.6$, and at higher redshifts it is very limited -- see Fig.~\ref{Fig:dNdz_spectro_all} and Table \ref{Tab: spectro data} for details. 
{We would also like to emphasise that, what is a general problem for \phz\ estimation and calibration, deep spectroscopic surveys preferentially measure redder galaxies. This is also the case for our training compilation beyond the GAMA depth, where it includes mostly red objects, unlike the target data, dominated by blue galaxies at the faint end (the ``faint blue galaxy problem'', \citealt{Ellis97}).}

We illustrate the non-representativeness of our spectroscopic data in Fig.~\ref{Fig:mag_color_photo_spectro}, which {compares selected magnitude-magnitude (top) and magnitude-colour (bottom) distributions for the spectroscopic (red) and photometric (black) data. For the latter we show the FIDUCIAL sample (as defined in \S\ref{Sec:KiDS-photometric}), which is the one used as reference for weighting the training set employed for the full-depth DR3 \phz\ catalogue (\S\ref{Sec:DR3-release}).} Clearly, both in magnitude and colour space of KiDS DR3   there are regions not well sampled by the {current} \spz\ data. This issue cannot be {fully} overcome without adding further deep {and appropriately preselected} spectroscopic data to the calibration sample {\citep{Masters17}}, although we mitigate its importance by {the aforementioned} weighting using the kNN procedure \citep{Lima08} implemented in ANNz2.
{On the other hand, as far as weak lensing analyses using KiDS data are concerned, the objects that are missing in the overlapping \spz\ samples are mostly faint galaxies at high redshift, which are unresolved by KiDS and are thus either not included or are heavily downweighted when measuring lensing shear.}

%%%
\section{KiDS DR3 experiments and associated photometric redshift catalogue}
\label{Sec:KiDS-DR3-photozs}

{In this Section we quantify the performance of ML \phzs\ in KiDS DR3, and compare them to the pipeline solution from BPZ. This is done by running several \phz\ experiments in which we applied ANNz2 and MLPQNA to different training and test subsets of the KiDS DR3 spectro-photo compilation presented above. We also describe the publicly released \phz\ catalogue derived with ANNz2, which includes all the DR3 sources that have the four $ugri$ bands measured (39.2 million objects). An earlier version of this catalogue was made available with the DR3 publication \citep{KiDS-DR3}. Here we update that dataset and provide more details on its properties.}  

The tests below will be obviously limited to the spectroscopic data, so the conclusions based on them may not be easily extrapolated to the full photometric set. This is however a general truth in \phz\ performance checks if incomplete \spz\ samples are used as calibrators, as is the case for most of the modern photometric surveys \citep{PHAT}. {Due to the nature of spectroscopic campaigns, which either explicitly target or are more efficient at measuring spectra of red and intrinsically luminous galaxies, the colour space of \spz\ samples is undersampled in some areas \citep{Masters15} which may lead to biases in direct comparisons of spectroscopic and photometric redshifts.}

In what follows, by a test sample we will always mean data not used in the training and validation phase.  Note that if both are selected randomly, the training and test samples will be statistically equivalent, so such tests will mostly tell how well the MLMs did for representative training data but not necessarily how well they do for the target photometric sample. We thus performed two types of experiments: \textit{i)} where the training and test data were statistically equivalent (\S\ref{Sec:Random_DR3_subsample}--\ref{Sec:Downweight_bright}), as well as \textit{ii)} those where the training and test samples were very different (\S\ref{Sec:COSMOS-CDFS-tests}); in the latter case, {in some of the tests} weighting was applied to the training data. {Such comparisons with available spectroscopic redshifts do not however provide the full picture on \phz\ performance due to biases in the calibration data such as their preference for red galaxies over blue ones, and limited depth. Therefore, in \S\ref{Sec:DR3-release} and Appendix \ref{App:dNdz_red_blue} we also analyse output \phz\ redshift distributions of the target photometric sample.}

The performance of the \phzs\ will be measured using the following statistics:
\begin{itemize}
\item[*]  bias, $\langle \delta z \rangle = \langle \zph - \zsp \rangle$, unclipped;
\item[*]  normalised bias, $\langle \delta z / (1+\zsp) \rangle$, unclipped;
\item[*] standard deviation of normalised error, $\sigma_{\delta z/ (1+\zsp)}$, unclipped;
\item[*] scaled median absolute deviation of normalised error, $\mrm{SMAD}\left(\delta z / (1+\zsp)\right)$, where \mbox{$\mrm{SMAD}(x) = 1.4826\; \mrm{median}\left(|x-\mrm{median}(x)|\right)$};
\item[*] percentage of catastrophic outliers for which $|\delta z / (1+\zsp)| > 0.15${; we use this particular definition of outliers to be consistent with other KiDS \phz\ analyses \citep{KiDS-GL,KiDS-DR2,KiDS-DR3}}.
\end{itemize}
For non-Gaussian distributions which usually characterise \phz\ errors, the unclipped scatter is not always informative, and SMAD, converging to the standard deviation (SD) for Gaussians, is preferred as the measure of the actual scatter. We also provide the SD as its comparison with SMAD helps judge how non-Gaussian the distribution is.

\input{table-phz-fullDR3.tex}

The statistics for MLM results will be computed for the test sets unseen by the algorithm in the training phase. They will also be compared to the results from the fiducial KiDS \phz\ solution, BPZ, which is independent of any training; for consistency, in such comparisons, we will use exactly the same test sets for the MLM and BPZ cases.
The BPZ statistics will be based on the central $Z\_B$ values only. In the case of ANNz2, 
we use the unweighted MLM-average (\ttt{ANNZ\_MLM\_avg\_1}) generally found to perform best among the {five types of} point estimates from this software (\S\ref{Sec:ANNz2}). For MLPQNA, we use the output of the regression network without any further manipulation.

\subsection{Random subsample of the spectroscopic data}
\label{Sec:Random_DR3_subsample}

In the first experiment we chose a random subsample (1/3) of the full spectroscopic data for training and validation and used the remaining 2/3 as a blind test set. {We have checked that the exact proportions of this split do not have a large importance for the results, provided that there are enough sources both in training and test samples to guarantee good statistics.}
The results for this test, compared with BPZ, are provided in the top rows of Table \ref{Tab: photo-z tests full DR3}. Except for the  normalised bias, both ANNz2 and MLPQNA clearly outperform BPZ for this low-$z$ dominated sample, the two ML approaches having statistics very comparable between each other. We have to note that in this case, the test data had the same properties as the training set, which means that this particular experiment shows only the performance of the MLMs in an ideal setup of the training being fully representative for the target data, which is not the case in KiDS. This experiment is thus mostly useful to judge the performance of the methods for the bright end of the sample. See also \cite{Cavuoti15b} and \textcolor{blue}{Amaro et al. (subm.)} for a more detailed discussion of how MLPQNA performs in this regime, as well as \S\ref{Sec:GAMA-depth} of this paper for a dedicated study of ANNz2 performance at the bright end of KiDS.

\subsection{Downweighting the bright end}
\label{Sec:Downweight_bright}

As the training set is dominated by bright galaxies (cf.\ Fig.~\ref{Fig:mag_color_photo_spectro}), in the second step we constructed {a sample in which we artificially down-weighted the bright end. This was done by randomly selecting 10\% of the bright-end ($r<20$) sources from the full KiDS spectro-photo compilation, while keeping all the objects with $r\geq 20$. {The subsampling percentage was chosen to obtain the mean redshift of the joint sample in between that of the fully random one from \S\ref{Sec:Random_DR3_subsample} and those of the COSMOS and CDFS datasets analysed in \S\ref{Sec:COSMOS-CDFS-tests}. This procedure gave us a joint} sample of 118,000 galaxies with $\meanz = 0.49$ and $\langle r \rangle = 21$~mag. This dataset was again divided into training and test sets in proportions 1:2. Photo-$z$ statistics are provided in Table \ref{Tab: photo-z tests full DR3}, second set of rows. In this case all the computed statistics for ANNz2 and MPLQNA are better than for BPZ, and the two empirical methods gave results very comparable to each other.

\subsection{COSMOS and CDFS as independent test samples}
\label{Sec:COSMOS-CDFS-tests}

{The most informative approach to judge the performance of KiDS ML \phzs\ is to use separate deep training end test data. Therefore,} as a next step, we trained ANNz2 and MLPQNA on KiDS spectroscopic sources from outside the COSMOS field and tested the results on KiDS-COSMOS \spz\ data; then we repeated the exercise this time with CDFS (train excluding CDFS, test on KiDS-CDFS). This way the test sets were fully independent from the training ones, and had very different characteristics. On the other hand, these two target samples have similar mean redshifts of $z\sim0.75$, closer to what we expect from the full KiDS than the mean redshift of the current spectroscopic calibration data would suggest. 
Therefore, these experiments provide the most insight into the true performance of the \phz\ methods {at the full depth available from \spz\ samples overlapping with KiDS}.

\begin{figure*}
\centering
\includegraphics[width=0.35\textwidth]{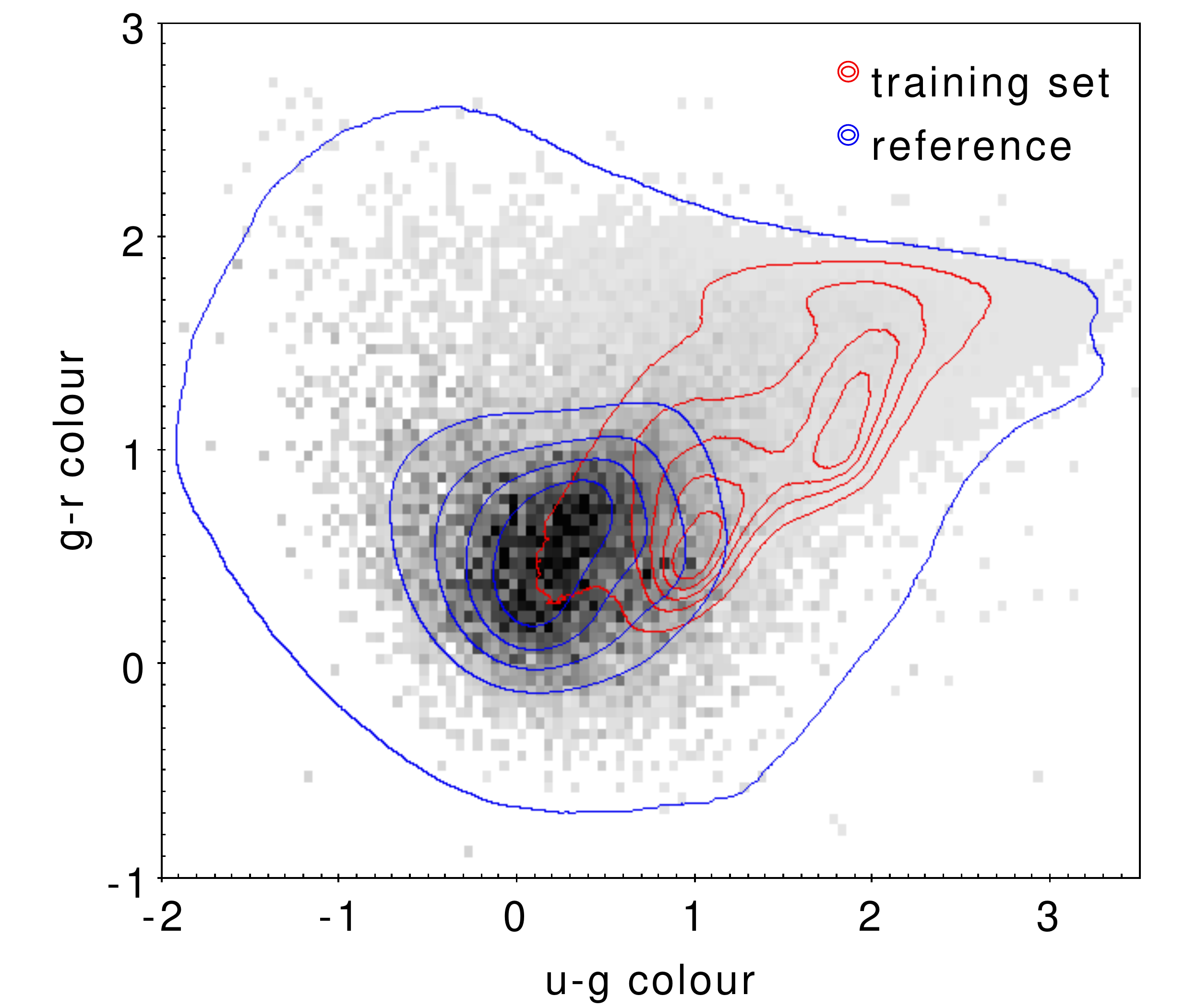}
\hspace{1cm}
\includegraphics[width=0.35\textwidth]{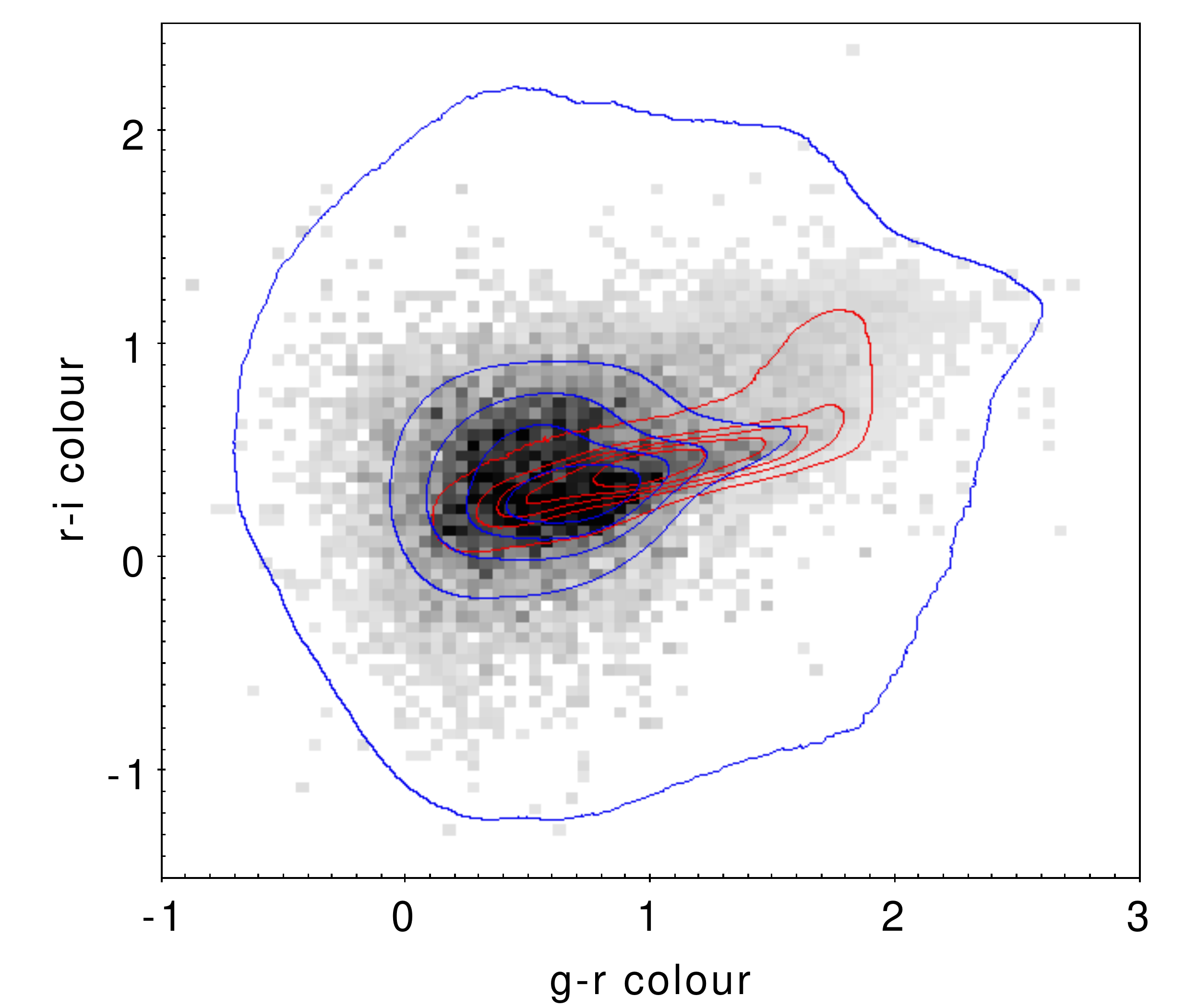}
\caption{Illustration of the kNN weighting procedure applied to the training data for the KiDS DR3 \phzs, as projected to colour-colour planes. Red contours show the unweighted spectroscopic training data, while the blue ones are for the reference photometric sample (FIDUCIAL). The greyscale pixels show the distribution of the weighted spectroscopic sample. The contours are linearly spaced.}
\label{Fig:col-col_training_weighting}
\end{figure*}

In the case of the ANNz2 experiments, two approaches were taken: in the first one we trained on a random subsample of non-COSMOS/non-CDFS data (respectively 10\% and 3\%) without any weighting; in the second one we trained on all the non-COSMOS/non-CDFS data but this time weighting the training sample in GAaP $ugri$ magnitude space with the kNN method (as implemented in the ANNz2 code) to mimic the properties of the target COSMOS/CDFS data, respectively. These weights were then used in the whole \phz\ procedure. The reason for taking just a small random subsample for the no-weighting experiments was that otherwise there would be a huge, unrealistic imbalance between the size of the training and test sets; the subsampling percentages used made the training and test sets comparable in size. On the other hand, in the weighting case, the weights for most of the training objects were much smaller than unity, so the effective weighted number of training sources was also comparable to the target set sizes. For MLPQNA, the experiments had the same setup as ANNz2 without any weighting.

The results of these experiments are compared in the two bottom set of rows of Table \ref{Tab: photo-z tests full DR3}. If no weighting is applied, then both MLPQNA and especially ANNz2 perform worse than BPZ in terms of scatter, but better in terms of  bias. Weighting does improve the ANNz2 results, although not significantly; in the COSMOS case, they provide similar scatter to the BPZ case while still have much smaller  bias. For CDFS, MLPQNA performed generally better than both the unweighted and weighted ANNz2 experiments, but the scatter from both ML approaches remains visibly worse than measured from BPZ. {The large fraction of outliers for these two deep comparison datasets is partly due to how these outliers were defined, namely with respect to a fixed normalised error value of 0.15. For BPZ, these results are consistent with what was shown in \cite{KiDS-GL} where test samples of similar depth as in here were used (CDFS and non-public zCOSMOS). On the other hand, \cite{KiDS-DR3} used a shallower public zCOSMOS sample and consequently found a smaller outlier fraction both for BPZ and ANNz2.}

\subsection{KiDS DR3 ANNz2 \phz\ catalogue release}
\label{Sec:DR3-release}

\begin{figure}
\centering
\includegraphics[width=0.45\textwidth]{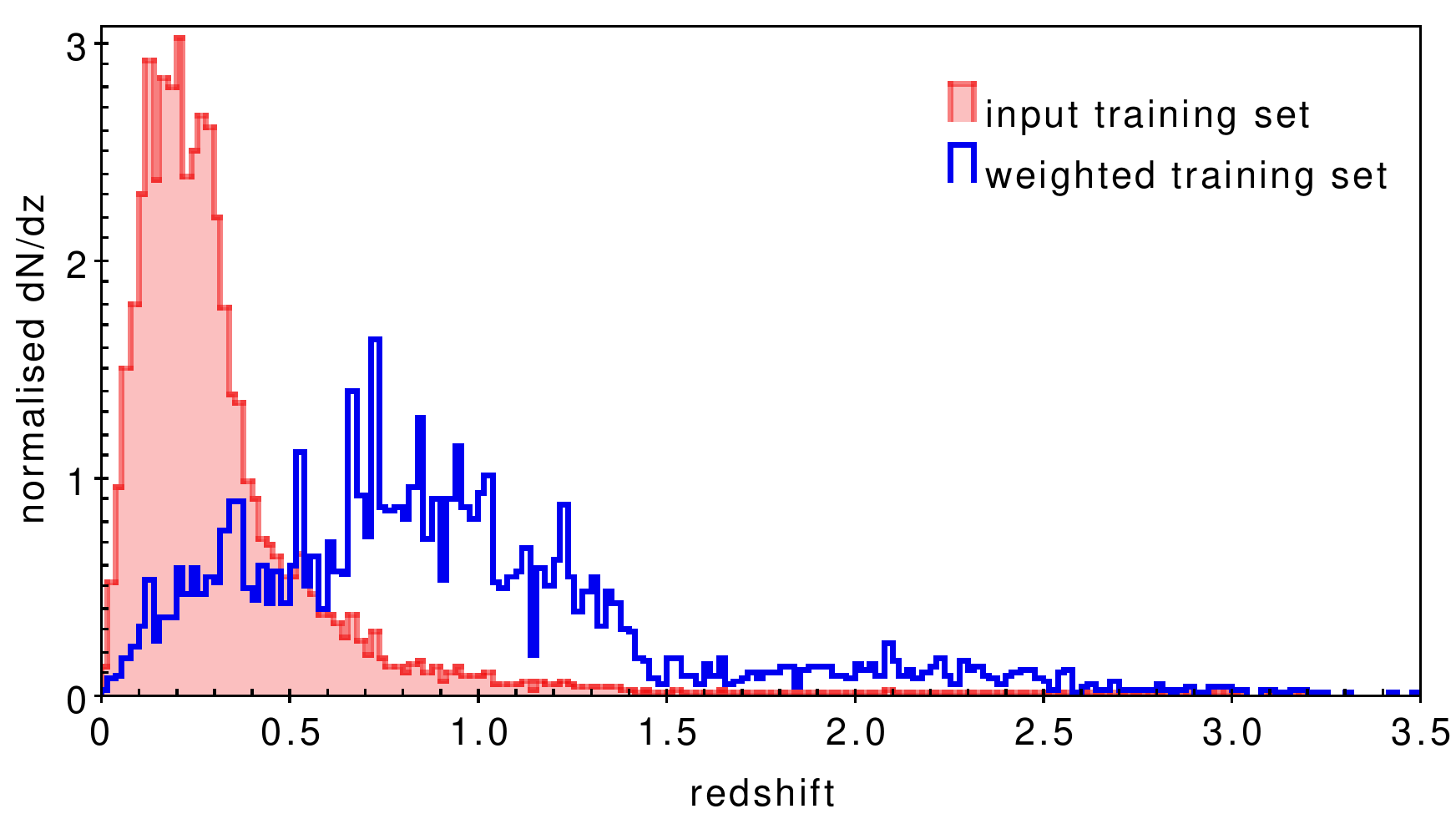}
\caption{Comparison of spectroscopic redshift distributions of the unweighted training set (red) and after applying the kNN weights to it (blue). {The weights were derived with reference to the KiDS DR3 FIDUCIAL dataset, and subsequently used in ANNz2 training and evaluation for the public KiDS DR3 \phz\ data release. Histograms are normalised to unit area.}}
\label{Fig:dNdz_spectro_weighted}
\end{figure}

\begin{figure*}
\centering
\includegraphics[width=0.28\textwidth]{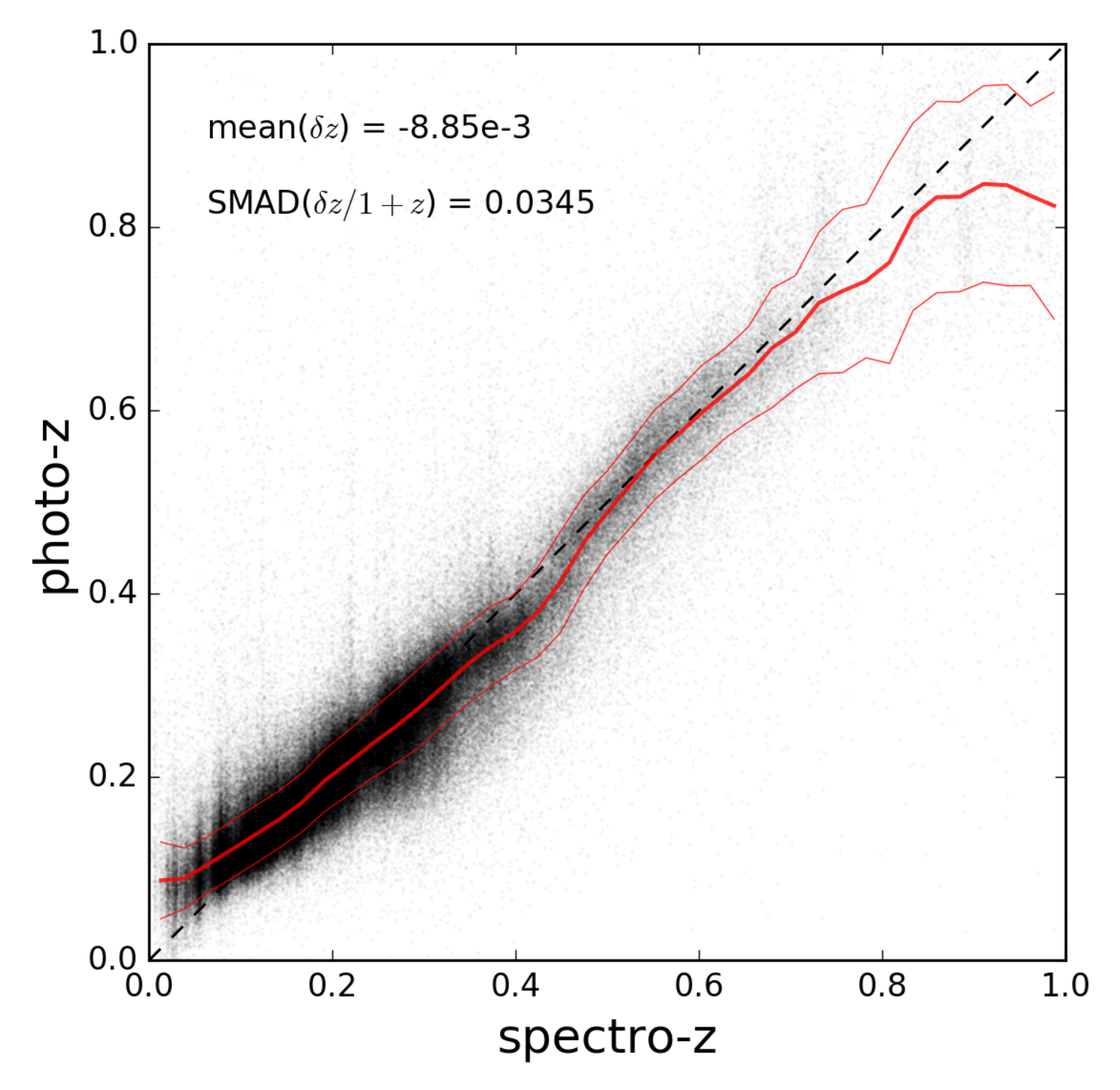}
\includegraphics[width=0.35\textwidth]{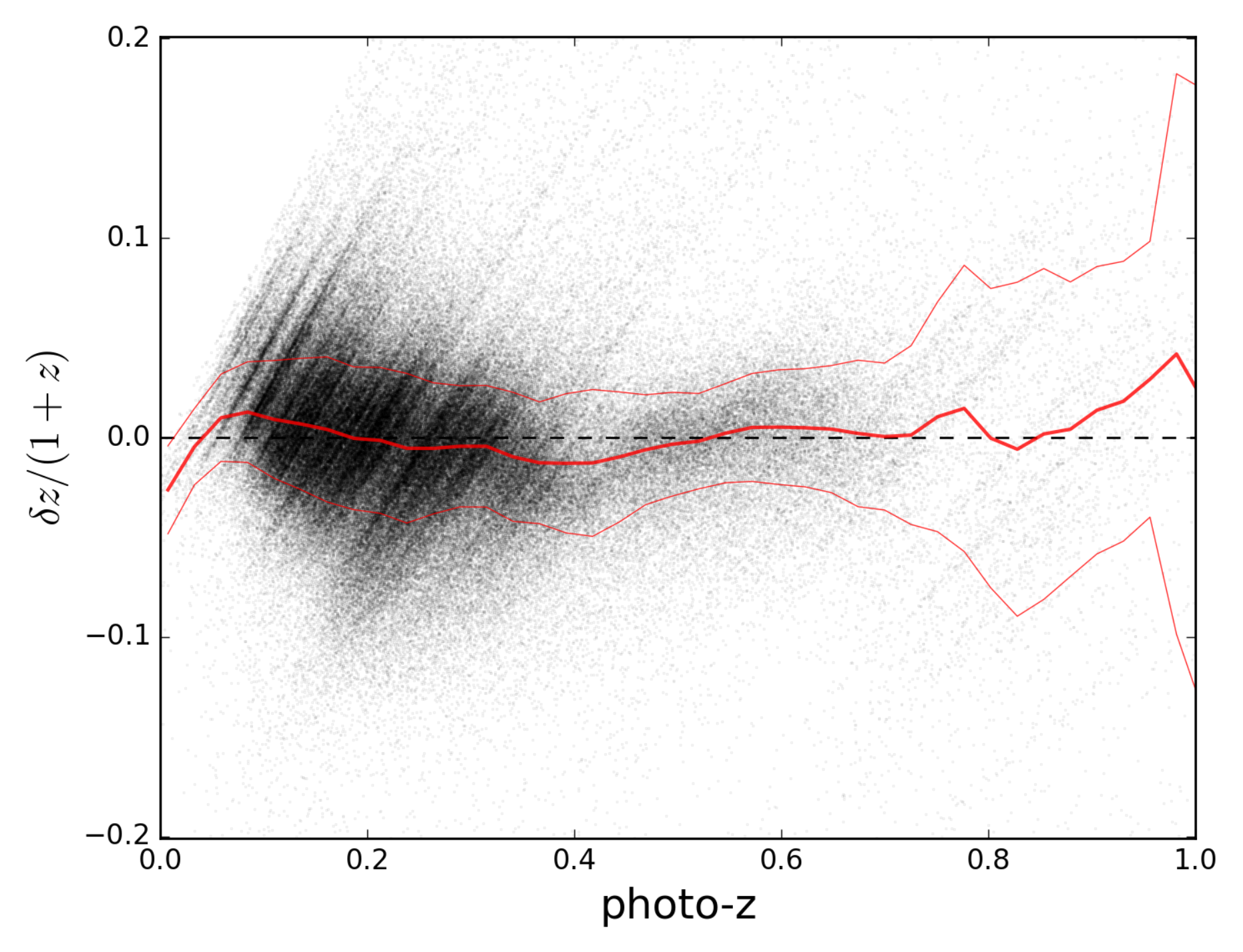}
\includegraphics[width=0.35\textwidth]{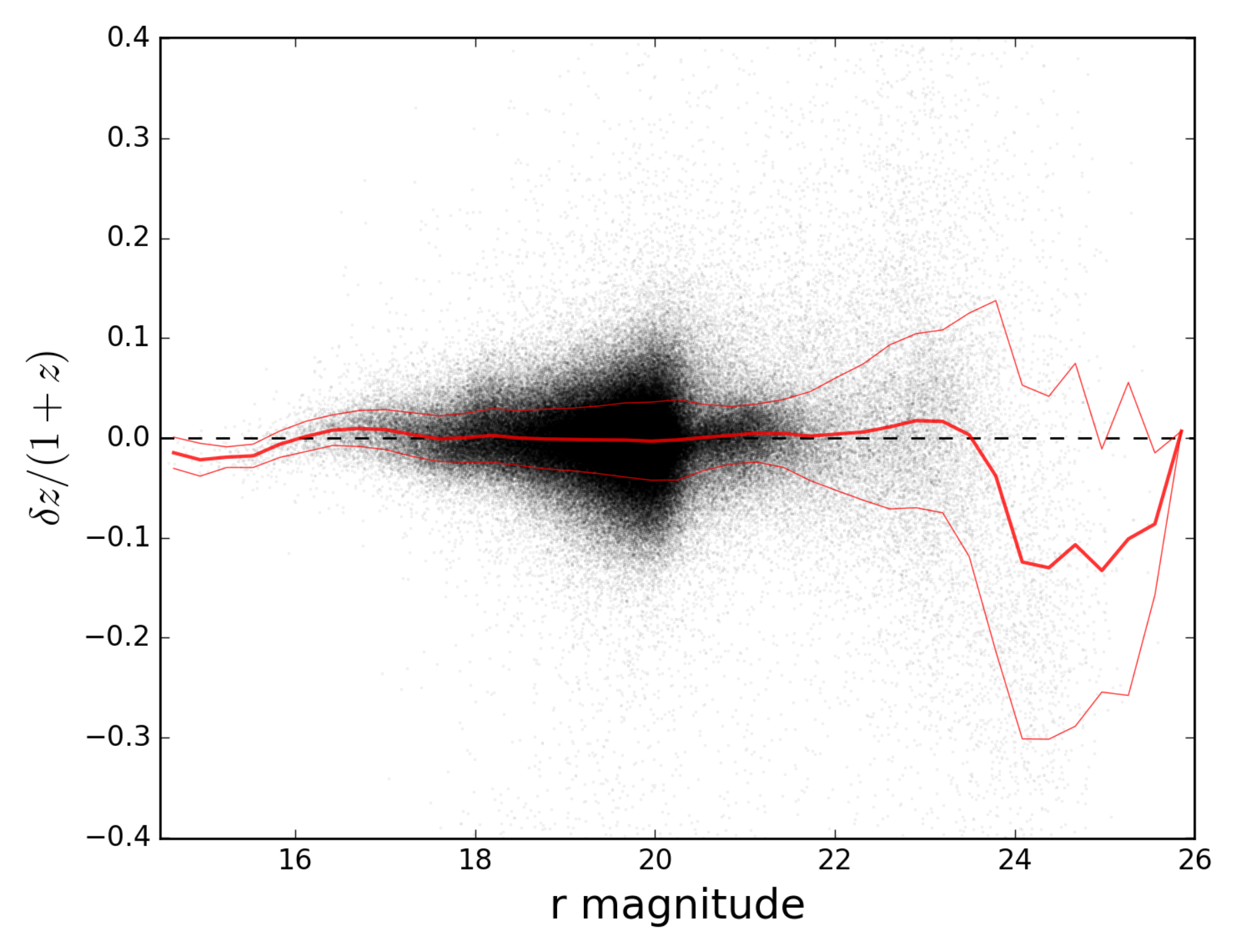}
\caption{Performance of the KiDS DR3 ANNz2 \phzs\ {from the released catalogue} as compared to the overlapping spectroscopic samples. Left-hand panel: direct \spz--\phz\ comparison; central panel: \phz\ error as a function of \phz; right-hand panel: \phz\ error as a function of the $r$-band magnitude. The thick solid line shows the running median while the thin lines encompass the scatter (SMAD). Note different scalings of the $\delta z/(1+z)$ axes. {Based on these comparisons, we judge the published \phzs\ to be reliable within $\zph<0.9$ and $r<23.5$.}}
\label{Fig:DR3_photoz_performance}
\end{figure*}

Having performed the above tests, we used the full KiDS-matched spectroscopic sample as the training+validation set to train ANNz2{, and produced the full-depth DR3 \phz\ catalogue, originally released with the DR3 paper \citep{KiDS-DR3}, and now updated\footnote{Data available from \url{http://kids.strw.leidenuniv.nl/DR3/ml-photoz.php\#annz2}.}. This catalogue includes all the 39.2 million DR3 sources that have the full set of $ugri$ bands measured, but only part of them will have \phzs\ of sufficient quality to be considered reliable. Below we quantify the performance of these ML \phzs.}

In the whole \phz\ procedure we used the kNN weighting of the training data, as implemented in ANNz2 (\S\ref{Sec:ANNz2}), applied in the $ugri$ magnitude space. The reference dataset was the FIDUCIAL sample described in \S\ref{Sec:KiDS-photometric}, constructed in such a way to include only likely galaxies and encompass magnitude ranges of the training data. Fig.~\ref{Fig:col-col_training_weighting} compares the 2D contours of the training sample (red) in colour space to those of the reference FIDUCIAL dataset (blue), {and to the weighted distribution of the \spz\ sources illustrated as background greyscale pixels.} Fig.~\ref{Fig:dNdz_spectro_weighted} shows the unweighted (red) and weighted (blue) input spectroscopic redshift distributions of the training set. The latter, of weighted $\langle z^{(w)} \rangle = 0.93$, can be regarded as a proxy for what should be expected from the true redshift distribution of the target sample \citep{KiDS-450,Soo17}, although at high redshifts $z>1.5$ this might be just a crude approximation due to sample variance in the very limited calibration data (Fig.~\ref{Fig:dNdz_spectro_all}).

In \cite{KiDS-DR3} we described an earlier version of the KiDS DR3 ANNz2 \phz\ catalogue, for which 100 ANNs were used in the training phase. Here we update that catalogue, having found a small issue in selecting the FIDUCIAL sources for weighting the spectroscopic sets. The changes are very small and all the conclusions from \cite{KiDS-DR3} regarding the performance of ANNz2 \phzs\ remain valid; the catalogue is updated for consistency.

Figure \ref{Fig:DR3_photoz_performance} summarises the properties of the KiDS DR3 ANNz2 \phzs\ as compared to \spzs\ from the datasets overlapping with the DR3 footprint (i.e.\ a subset of the full training sample, excluding CDFS and DEEP2). They show that the \phzs\ are stable for $\zsp\lesssim 0.9$ and $\zph\lesssim 0.9$, as well as for $r\lesssim23.5$, above which their quality quickly deteriorates. These could be then considered the limits up to which the presented here ANNz2 \phzs\ are relatively reliable.
{In \cite{KiDS-DR3} the performance of \phzs\ was illustrated as a function of \spz\ and the $r$-band magnitude, but for shallower calibration samples than here (public GAMA DR2 and public zCOSMOS). Here we prefer to focus on the error behaviour as a function of magnitude and \phz, as these latter are the quantities available to the end user of the catalogue. Table \ref{Tab: photo-z stats full DR3} quantifies this performance in bins of \phz, for both ANNz2 and BPZ (binning is done in the respective \phz\ type). The statistics were derived using the same overlapping \spz\ samples as employed for Fig.~\ref{Fig:DR3_photoz_performance}, which become very incomplete at $z\sim1$. At present there are no sufficiently deep and complete spectroscopic datasets available in the KiDS footprint that would allow for a reliable quantification of \phz\ performance at the full depth of the survey.}
\input{table-stats-DR3.tex}

{We have also verified that both ANNz2 and BPZ perform better for red galaxies than for blue ones. For instance, if we split the overlapping spectroscopic sample according to the colour-colour line $g-r = 0.8 - 0.8(r-i)$, then sources redwards of this division have $\delta z_\mrm{BPZ}=0.015$, $\mrm{SMAD}\left(\delta z_\mrm{BPZ} / (1+z)\right) = 0.030$ for BPZ and $\delta z_\mrm{ANNz2}=-8.0\times10^{-3}$, $\mrm{SMAD}\left(\delta z_\mrm{ANNz2} / (1+z)\right) = 0.032$ for ANNz2, while those on the blue side have $\delta z_\mrm{BPZ}=-0.098$, $\mrm{SMAD}\left(\delta z_\mrm{BPZ} / (1+z)\right) = 0.049$ and $\delta z_\mrm{ANNz2}=-0.014$, $\mrm{SMAD}\left(\delta z_\mrm{ANNz2} / (1+z)\right) = 0.050$. Similar worsening when going from red to blue is also observed for other statistics as well as for differently defined red/blue separations. This general behaviour should not be surprising: the observed optical colours of red galaxies are a strong function of redshift while those of the blue ones much less depend on $z$. Regardless of the approach used, this means that \phzs\ for blue galaxies are expected to be worse than for red ones. Let us however reiterate that red galaxies dominate at the faint end of our spectroscopic calibration sample. This means that our possibility to reliably quantify \phz\ performance for faint blue galaxies is limited.}

{The limitations of the spectroscopic calibration data mentioned above mean that classic \spz--\phz\ comparisons do not give the complete picture on the performance of the latter. Therefore, a useful test, even if rather qualitative, is provided by the verification of \phz\ distributions of target photometric samples. In Fig.~\ref{Fig:dNdz_DR3_photometric} we first compare $dN/dz_\mrm{phot}$ of the FIDUCIAL sample, for BPZ (grey bars) and ANNz2 (blue line). We see that they are very different with the ANNz2 one being smooth and extending to high redshifts, while the BPZ $dN/dz$ shows several significant peaks, likely resulting from aliasing in colour-redshift space (i.e., emission lines moving between the filters). The SED-fitting solution has here practically no redshifts beyond $z_\mrm{BPZ}>1.5$; many sources are instead assigned low $z_\mrm{BPZ}$. This latter behaviour of BPZ is related to the prior which has been used in its implementation for KiDS purposes, directly propagated from an earlier CFHTLenS analysis \citep{Hildebrandt12}, where the original prior from \cite{BPZ} was modified to behave better for that survey. It was not optimized further for KiDS as the default redshift calibration in KiDS cosmological analyses is not based on individual redshift estimates but on external spectroscopic samples (the `DIR' method of \citealt{KiDS-450}). However, this prior is now being revised for new KiDS releases to provide also more reliable individual \phzs\ from BPZ. As far as the abundance of low-$z_\mrm{BPZ}$ sources is concerned, these are mostly galaxies with observed blue colours. More discussion on this can be found in Appendix~\ref{App:dNdz_red_blue}.} 

{The ANNz2 redshift distribution in Fig.~\ref{Fig:dNdz_DR3_photometric} is much more regular than that of BPZ}, although probably not trustworthy beyond $z>1$ or so {(as discussed a couple paragraphs above)}, where practically all the sources have $r>23$. We also observe a flattening of $dN/dz_\mrm{ANNz2}$ at $\zph\sim0.5$. which may reflect worse performance of ANNz2 in this regime (cf.\ Fig.~\ref{Fig:DR3_photoz_performance}) {and is probably related to the properties of the training set. Namely, at low redshifts the calibration data are dominated by the complete flux-limited ($r<19.8$) GAMA sample. Its $dN/dz$ quickly drops off at $z\sim0.5$, beyond which the training is composed of various deeper but not as complete datasets (\S\ref{Sec:KiDS spectroscopic}). Despite the weighting applied to the training data, this imbalance is apparently propagated into the \phz\ solution. Part of the reason might be also under-performance of the weighting procedure, which in 4-dimensional parameter space could be prone to biases from the large-scale structure and noise, as evidenced by various peaks and dips in Fig.~\ref{Fig:dNdz_spectro_weighted}. We will be testing if these issues could be mitigated in future KiDS releases, which will include more spectroscopic calibration data and will be extended with VIKING near-IR measurements, providing thus 9-dimensional magnitude space.}

In the same Fig.~\ref{Fig:dNdz_DR3_photometric} we also show $dN/dz_\mrm{phot}$ for the FIDUCIAL sample but trimmed at $r<23.5$ (green bars for BPZ and red line for ANNz2), which we have judged above to be the limit up to which the published ANNz2 redshifts are reliable. Indeed, we see that the main peak observed in the FIDUCIAL sample at $z_\mrm{ANNz2}\sim0.95$ as well as most of the tail at $z>1$ come from sources fainter than this magnitude cut, which is probably a sign of extrapolation. Interestingly, such a flux limit removes also several high peaks in the distribution of BPZ \phzs, although the $dN/dz$s for the $r<23.5$ sources remain very different between the BPZ and ANNz2 solutions. {Noting here that we do not expect this particular flux-limited selection to provide improved $z_\mrm{BPZ}$ over other cuts, as we show in Appendix~\ref{App:dNdz_red_blue} the main source of this persisting discrepancy seems to be very different treatment of blue galaxies by the two \phz\ approaches. Namely, ANNz2 assigns them a flat and extended $dN/dz$ while BPZ limits the output redshifts to a couple of rather narrow ranges. In particular, a significant fraction of blue galaxies are allocated to relatively low \phzs\ ($z_\mrm{BPZ}<0.4$) by the KiDS DR3 BPZ implementation. On the other hand, applying a colour cut on the sample to separate out redder galaxies allows us to largely mitigate the \phz\ differences.}

We note that until now, galaxies with \phzs\ beyond {$z_\mrm{BPZ}>0.9$} have not been used for KiDS scientific analyses mostly due to the inability of their proper calibration at this high-$z$ end. Forthcoming developments from using additional VIKING data as well as extending the spectroscopic training samples should provide the possibility of deriving better \phzs\ at this range {both using BPZ and MLMs}, which is certainly of great interest for lensing studies.

\begin{figure}
\centering
\includegraphics[width=0.48\textwidth]{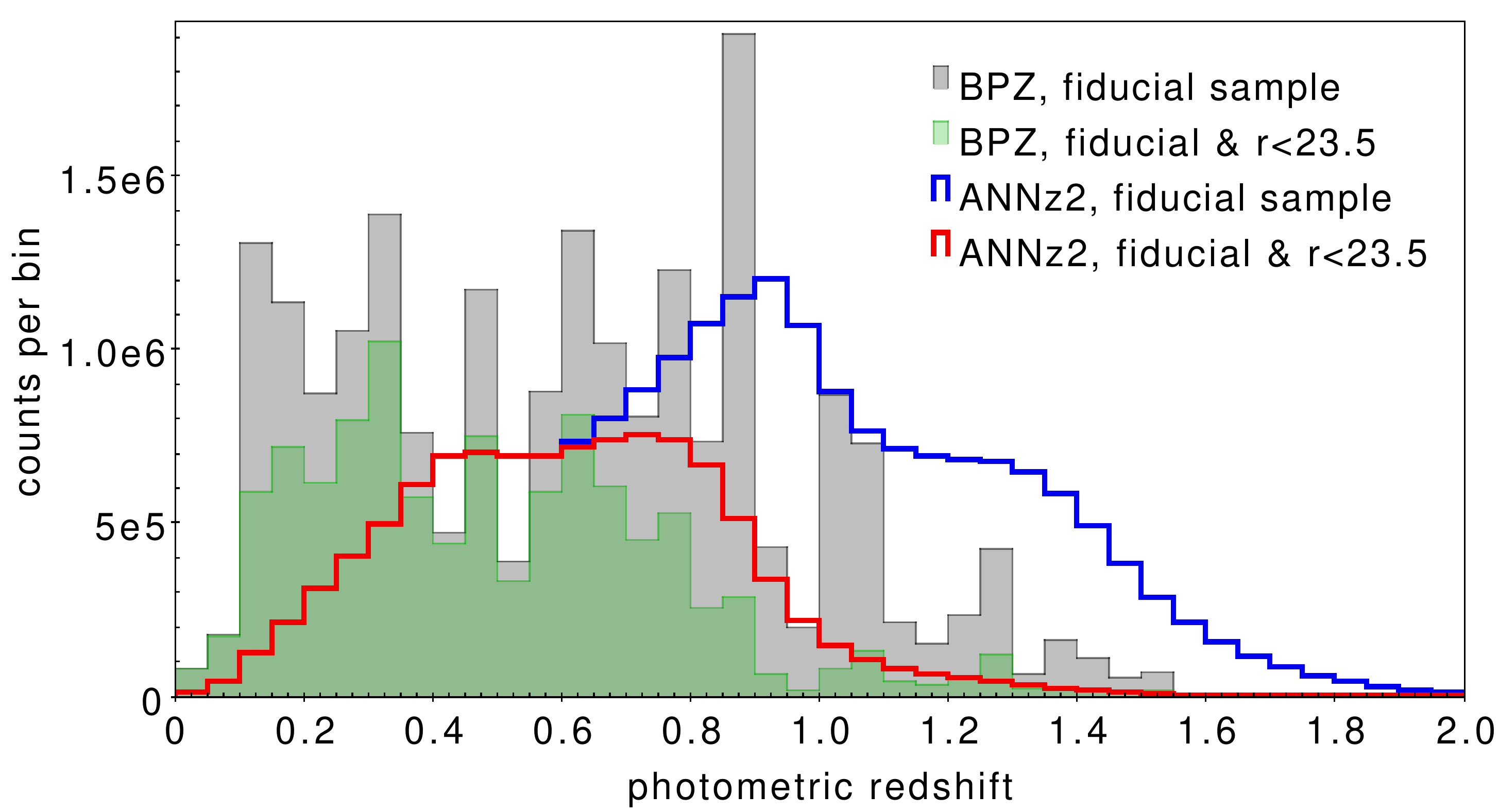}
\caption{Comparison of photometric redshift distributions for two KiDS DR3 samples: FIDUCIAL (see \S\ref{Sec:KiDS-photometric} for details), and FIDUCIAL with an additional cut of $r<23.5$, for BPZ (bars) and ANNz2 (lines). The comparison suggests that for $r>23.5$ the ANNz2 \phzs\ in DR3 may not be trustworthy as they are based on extrapolation. {The shape of the BPZ $dN/dz$ is driven by the prior that was adopted (see text for details).}}
\label{Fig:dNdz_DR3_photometric}
\end{figure}

As we showed in this Section, the currently derived ANNz2 \phzs\ are{, at least to $\zph<0.9$ and $r<23.5$,} of quality comparable  to the default KiDS BPZ ones in terms of the overall statistics, and fare considerably better in terms of  bias in most of the regimes, and also in terms of scatter at the bright, low-redshift end. In the near future we expect both the ML and template-fitting KiDS solutions to improve. For the ML case, extending the training sample is important, and will be made possible thanks to currently processed or ongoing KiDS-like observations of some of the VVDS \citep{VVDS} and VIPERS \citep{VIPERS} fields. These will give additional calibration samples spanning redshifts $0 < z < 1.6$, which will help mitigate sample variance plaguing the derivation of high-$z$ \phzs.

Both the ML and SED-fitting methods will benefit from the major extension of photometry, that is -- the addition of five VIKING NIR bands.
In the following Section we present the improvement possible thanks to adding the VIKING data to ANNz2 derivation at low redshifts, but our early tests show that similar gains should be also expected at larger depths. In future KiDS releases, starting from DR4, we plan to derive the ANNz2 \phzs\ from 9-band KiDS+VIKING photometry employing the extended training data.

{To summarise this comparison of the two \phz\ solutions, it is clear that both have their limitations which the user should be aware of. Photo-$z$ accuracy is expected to be a function of apparent magnitude, colour, and true redshift. This is inevitable given errors in photometry and the SEDs of galaxies  -- the optical colours of blue galaxies have a relatively weak dependence on redshift. We can mitigate these differences to some extent, for instance by weighting the training data in ML \phz\ derivation, but not entirely remove their impact. As far as the ANNz2 \phzs\ are concerned, they should be preferred in the range where sufficient spectroscopic training data are available, which is brighter and redder sources. Outside of this range, where MLMs suffer from biases incurred by extrapolation beyond the training coverage, the recommended solution is an SED-fitting one such as BPZ. However, the lack of sufficiently deep and complete, especially in terms of the blue population, calibration data overlapping with KiDS DR3 does not allow us to reliably quantify the performance of both these approaches at $z>1$.}

%%%
\section{GAMA-depth experiments and associated photometric redshift catalogue}
\label{Sec:GAMA-depth}

{In this Section we analyse ANNz2 \phz\ performance at the bright end of KiDS, and describe the associated catalogue release which includes 800,000 sources with $r<20.3$.} 
As already mentioned, ML \phzs\ usually work best at the median redshift of the training set and tend to over-/underestimate redshifts at the low-/high-$z$ regime, i.e. at the bright/faint end of the sample. This is also the case for KiDS DR3 where, partly due to the training set weighting to mimic the target data, the ANNz2 \phz\ solution is not optimized for the bright end of the sample. Also the BPZ \phzs\ calculated in the KiDS DR3 pipeline do not perform very well at the bright end, especially in terms of  bias. 

However, there is considerable interest in obtaining a KiDS dataset with well-constrained \phzs\ in the relatively nearby Universe, as such a sample could be then used for such measurements as galaxy-galaxy lensing \citep[e.g.][]{Velander14}, general galaxy evolution studies {\citep[e.g.][]{Tortora16,CostaDuarte17}} or for studying the effects of the cosmic web \citep[e.g.][]{Gruen16}. Indeed, this type of study have already been undertaken by the KiDS team, using \textit{spectrosopic} data for the foreground sample, thanks to the (by-design) full overlap of KiDS with GAMA equatorial fields \citep{Viola15,Sifon15,vanUitert16,Brouwer16,vanUitert17a,Dvornik17}.

The GAMA survey has however already finished, its overlap with KiDS ($\sim200$ deg$^2$) will thus not increase.
Having a GAMA-like catalogue within the full {planned} KiDS coverage {of $\sim1500$ deg$^2$} would make it possible to reduce the statistical errors of the aforementioned KiDS analyses by a factor of $\sim2.5$. We therefore examined what accuracy and precision can be obtained by training KiDS \phzs\ on GAMA \spzs, and how these could be improved by extending the parameter space with {redshift-dependent} measurements other than magnitudes, as well as by adding IR photometry. This is also of interest for ongoing and  future photometric surveys such as the HSC or LSST, which will overlap with GAMA, but will provide even deeper photometry. 
An alternative route towards a KiDS foreground sample with well-constrained \phzs\ could be via identifying LRGs, for instance using the `redMaGiC' algorithm \citep{redMaGiC}. {This type of analysis is currently ongoing (\textcolor{blue}{Vakili et al. in prep.}).}

All the tests described hereafter will be using the ANNz2 software only, and will be restricted to the GAMA equatorial fields to guarantee very high completeness of the training data (except for the catalogue release of \S\ref{Sec:GAMA catalogue release}). Unlike in \S\ref{Sec:KiDS-DR3-photozs}, here we kept the same training and test sets for all the experiments; what was varied were the parameters used in the \phz\ derivation. We tested practically all the KiDS multiband measurements from the DR3 public release that correlate with redshift, such as (observed) magnitudes, colours, angular sizes, and other related photometric parameters. In addition, when analysing various extensions to the basic KiDS $ugri$ magnitudes, we also took advantage of the availability of GAMA LAMBDAR catalogues \citep{GAMA-LAMBDAR}, which include VIKING and WISE forced photometry measurements on GAMA targets. {These extra features were first added `individually' to the basic $ugri$ setup (in relevant groups, e.g.\ magnitudes in a single fixed aperture but from all the bands) and once those providing the most amelioration had been determined, they were combined into multi-parameter setup used at a further stage.}
Below we present the main results of these tests, focusing on those photometric measurements which brought the most improvement to the \phzs\ over using only the default (GAaP) $ugri$ magnitudes.

{We note that a possibly more optimal way of extending the parameter space would be to first apply a dimensionality reduction algorithm such as Principal Component Analysis (PCA) or related \citep[e.g.][]{Singal11}. This would remove redundancy from the feature space and speed up the training process. At this stage of data exploration, we preferred however to work directly with the parameters provided in the catalogues, to verify which among them are the most useful for \phz\ estimation; PCA might however blur such information. In future releases of KiDS data, where also the 5-band VIKING information will be added by default, and perhaps also WISE for the bright sources, the parameter space will grow considerably, so such preprocessing may indeed become necessary.}

The usefulness of parameters other than magnitudes, especially the morphological ones, on \phz\ estimation, has been studied by several authors {\citep[e.g.][]{Wray08,Way11,Singal11,Hoyle15,Jones17,Gomes18}}, with mixed conclusions. We refer the reader to the recent `Morpho-z' analysis by \cite{Soo17} and references therein for an overview of these earlier efforts. As far as we are aware, {perhaps with the exception of the more recent analysis by \cite{Gomes18},} none of these previous studies of that kind operated at the same regime as ours here, namely for relatively bright galaxies with excellent photometry (very high signal-to-noise) and using a complete spectroscopic sample for \phz\ calibration.

{As far as the the recent results from \cite{Soo17} are concerned,} they found improvement in \phzs\ only if morphology was used with a very limited set of passbands (as small as one in some cases), while for a more complete setup such as their fiducial $ugriz$, even some deterioration of \phzs\ was obtained after adding non-magnitude parameters. Our experiments discussed here apply to a very different regime of magnitudes and redshifts than those by \cite{Soo17}, and unlike in that paper, we also incorporate colours in addition to purely morphological parameters such as sizes. These two independent analyses, and their conclusions, should then be regarded as largely complementary.

There are over 190,000 GAMA galaxies with good-quality redshifts ($\mrm{NQ}\geq3$) that have a KiDS counterpart in the equatorial fields. We split this sample randomly in proportion 1:2 into a training and test set; after cleanup of bad photometry and masking we are left with $\sim56,000$ galaxies for training and validation, and $\sim112,000$ for most of the tests. Due to occasionally missing IR fluxes, these numbers decrease by typically $\sim 1000$ sources for some of the experiments where for instance VIKING or WISE measurements were additionally required; this has no influence on our conclusions.

\input{table-phz-GAMA-short.tex}

\subsection{Choice of the magnitude type}
\label{Sec:GAMA magnitude type}
We start by looking at the properties of GAMA-depth \phzs\ based on $ugri$ magnitudes only. As the benchmark we will use the default (pipeline) KiDS results from BPZ, as well as those from the ANNz2 training of the full KiDS-DR3 described in \S\ref{Sec:KiDS-DR3-photozs}, but evaluated on the GAMA test set only. The detailed statistics for these two cases are provided in the top rows of Table \ref{Tab: photo-z tests GAMA}. At this bright end, BPZ performs $\sim 10\%$ better than the overall ANNz2 DR3 solution in terms of scatter, but {over one order} of magnitude worse in terms of  bias. The third row of Table \ref{Tab: photo-z tests GAMA} shows that training the ANNs on GAMA sources only allows us to significantly reduce the scatter (by over 35\%) while the bias remains minimal. These \phz\ results for KiDS-GAMA sources are considerably better than those of {WISE~$\times$~SuperCOSMOS} \citep{WISC} where practically the same spectroscopic training data were used, also employing 4 bands at similar depth as here, although of worse photometric quality {(in the optical based on digitised photographic plates)}. This highlights the importance of photometry for \phzs\ even if the training sets are the same and the target data have similar depths.

As discussed in \S\ref{Sec:KiDS-photometric}, the default KiDS GAaP magnitudes \citep{GAaP}, used for instance in the BPZ pipeline and for the ANNz2 solution of \S\ref{Sec:DR3-release}, underestimate the fluxes of bright galaxies. We have therefore tested if this influences the photometric redshifts and whether other choices of magnitudes could be used to improve the results. This was done by training and testing on the same data but with $ugri$ GAaP magnitudes replaced by first ISO and then AUTO measurements {provided in the DR3 catalogue, zero-point calibrated and extinction corrected (`\ttt{calib}')}. The results {for both these options} are shown in the second set of rows of {Table \ref{Tab: photo-z tests GAMA full} in Appendix \ref{App:Full_GAMA_table} (the shortened Table \ref{Tab: photo-z tests GAMA} only lists the ISO case)}. It is obvious that both cases gave significantly worse results in terms of scatter (respectively by almost 20\% and over 25\%) than the GAaP solution. {We would like to reiterate however that these ISO and AUTO magnitudes have not undergone the PSF homogenisation discussed in \S\ref{Sec:KiDS-photometric}, unlike the GAaP ones. As was shown by \cite{Hildebrandt12}, performing PSF homogenisation on the image level guarantees very accurate colours also for the ISO magnitudes, and therefore should improve the \phzs\ based on such measurements. However, as such processing has not been done in KiDS except for the GAaP magnitudes, we will be thus using only the latter} as the basic set of parameters, and other quantities will be supplemented to them to look for \phz\ improvement. We did test the option of using the GAaP and ISO magnitudes together (i.e. 8 training parameters), but the improvement was minimal, and several other combinations worked much better, as described hereafter.

\subsection{GAaP magnitudes with one additional set of parameters}
\label{Sec:GAMA-depth_mags+one_set}

Having determined that the GAaP magnitudes are optimal for \phz\ estimation, we checked how adding other parameters, correlated with redshift, can improve the results. We tested the following measurements from the KiDS multi-band catalogue (see table A.2 in \citealt{KiDS-DR3} for details on these quantities):

\begin{itemize}

\item[*] \ttt{A} \& \ttt{B}, i.e. linear semi major and minor axes of the galaxy{, derived from $r$-band imaging}; at the GAMA redshift range, these are expected to trace the monotonically increasing angular diameter distance;

\item[*] \ttt{FLUX\_APER\_\textit{size}\_\textit{band}}, i.e. flux in \textit{size} aperture, where $size \in \lbrace 4, 6, 10, 14, 25, 40, 100 \rbrace$ pixels. The tests were performed for each of the $ugri$ bands in two configurations:

	\begin{itemize}

	\item fixed \textit{size}, all possible \textit{bands}, i.e. four parameters used together with the GAaP $ugri$ magnitudes, for instance: \ttt{FLUX\_APER\_4\_u}, \ttt{FLUX\_APER\_4\_g}, \ttt{FLUX\_APER\_4\_r} \& \ttt{FLUX\_APER\_4\_i}, and so on for each \textit{size};

	\item fixed \textit{band}, all possible \textit{sizes}, i.e.\ seven parameters used together with the GAaP $ugri$ magnitudes, for instance: \ttt{FLUX\_APER\_4\_u}, \ttt{FLUX\_APER\_6\_u}, ..., \ttt{FLUX\_APER\_40\_u} \& \ttt{FLUX\_APER\_100\_u}, and so on for each \textit{band}; this configuration can be regarded as a proxy for surface brightness profile, expected to provide very good constraints on \phzs\ \citep{Kurtz07};

	\end{itemize}

\item[*] \ttt{ISOAREA\_IMAGE\_\textit{band}}, i.e.\ isophotal aperture in pixel$^2$, for each \textit{band} (four parameters used together with the $ugri$ GAaP magnitudes);

\item[*] \ttt{colour\_GAAPHOM\_\textit{band1}\_\textit{band2}}, i.e.\ homogenised and extinction corrected GAaP colours, for all combinations of \textit{band1} and \textit{band2} (six parameters used together with the $ugri$ GAaP magnitudes).

\end{itemize}

{Except for the GAaP-based magnitudes and colours, the remaining parameters used in these tests were derived by SExtractor. This software may not be optimal for source extraction in crowded fields and new, presumably better, tools are being developed for such a purpose, e.g.\ The Tractor \citep{TheTractor} or ProFound \citep{ProFound}. Testing their performance would be however beyond the scope of this work as source extraction is embedded deeply in KiDS processing pipeline. In other words, we cannot verify at present which of the SExtractor-based measurements are unreliable. Moreover, these parameters were not corrected for PSF variations so the \phz\ improvement they provide should be considered as lower limits.}

{We note however that the `GAMA-like' galaxies we are concerned with in this Section are typically much larger than the PSF, so the variations of the latter are not expected to significantly bias the measurements that we use here. This is especially true in the most stable $r$ band, in which galaxy sizes, used in the \phz\ derivation for the released catalogue (\S\ref{Sec:GAMA catalogue release}), are measured. The PSF in this band is well-behaved both within individual coadds \citep{KiDS-GL}, as well as between the tiles \citep{KiDS-DR3}. A possibly more important PSF-related effect could be for those features which use multi-band information, such as \ttt{FLUX\_APER\_\textit{size}\_\textit{band}}. Here indeed differences in the PSF between the bands could affect the \phz\ measurements. Being unable to quantify this effect at present, we note that overall, even if small, improvement in \phzs\ after adding these various parameters suggests that the related noise does not dominate.} 

The third set of rows of Table \ref{Tab: photo-z tests GAMA} presents {selected} results from the combinations of the $ugri$ magnitudes with one set of those of the above additional parameters which performed best in terms of scatter{; a more extended set of results is given in Table \ref{Tab: photo-z tests GAMA full} in Appendix \ref{App:Full_GAMA_table}}. Some of the parameters do not appear in the Tables because they brought less improvement than those listed, but all of them were tested. The best results were obtained by adding GAaP colours, and a bit worse by using the {sizes (semi-axes)}, as well as for several \ttt{FLUX\_APER} combinations. The improvement is not huge (no more than by 5\% in scatter with respect to the basic $ugri$ setup) but clearly visible.

The fact that using colours together with magnitudes improves the magnitude-only results may seem puzzling at first, because the two sets of parameters are redundant. {We verified that the same effect exists not only for ANNs but also for BDTs, so it seems to be a general property of this type of MLMs. Moreover, similar improvement was observed by \cite{Hoyle15} where SDSS magnitudes were also combined with colours, and ANNs used for \phz\ experiments.}
We interpret this improvement as due to the fact that the parameter space of magnitudes only is not constraining enough for the MLMs to converge on a solution sufficiently close to the truth. {The MLMs do not "know" a priori that colours are simple combinations of magnitudes.  
Using both together} adds physical information on how galaxy observed properties are related to redshift and forces the MLMs to work in a much better constrained region of the parameter space, which improves the mapping between photometry and redshift.

We also tested how adding photometry external to KiDS (namely from VIKING and WISE) can improve the \phzs, and the results are very promising in view of the forthcoming KiDS-VIKING data, as well as of the forced photometry of WISE on KiDS sources that we are planning to obtain. As is clear from Table \ref{Tab: photo-z tests GAMA}, already adding WISE $W1$ and $W2$ to KiDS $ugri$ gives results better than the best KiDS-only solution discussed in this Subsection. Extending the parameter space by adding $W3$ brings in further improvement, with a caveat that part of the objects ($\sim16\%$) have no flux measurements in this band (i.e. $\mtt{W3\_flux}=0$), which means that for them the $W3$ information was effectively ignored. {As was already discussed in \S\ref{Sec:ANNz2}, the default setup of our \phz\ experiments does not use the parameter errors provided in the catalogues but relies on an internal error model built by ANNz2. In case of WISE we tested however the option of directly using the provided errors, which are here considerable especially in the $W3$ channel. We found negligible (sub-percent) differences in the \phz\ statistics between the two setups, as is shown in Table \ref{Tab: photo-z tests GAMA full} in Appendix \ref{App:Full_GAMA_table}.}

The KiDS+WISE statistics can be compared for instance with the 2dFLenS study by \cite{2dFLenS-photo-z} where it was found that adding the $W1,W2$ bands to the $ugriz$ optical setup (based on VST-ATLAS photometry) improved the results by 5-10\%, which was however a smaller difference than between some of the \phz\ methods tested there. Last but not least, using the five additional bands from VIKING (without employing WISE) gives the best results of all considered so far, although the improvement is not dramatic ($\sim -8\%$ in scatter over the fiducial $ugri$ setup). This is somewhat similar to some of the results from a DES+VHS analysis by \cite{Banerji15}, although those experiments used a very different setup from that presented here, so the results are not directly comparable.

It is also worth noting that  the  normalised bias of \phzs\ is improved by adding IR, by an order of magnitude over most of the so-far discussed experiments using only optical data. Furthermore, these results should be treated as lower limits to the improvement in \phzs\ possible by adding the IR data{, for several reasons}. Firstly, the LAMBDAR forced photometry was based on SDSS DR7 apertures as input, which are more noisy than KiDS measurements. Secondly, we may expect -- similarly as found in \S\ref{Sec:GAMA magnitude type} -- that using GAaP magnitudes for the IR bands (being currently derived by the KiDS team for VIKING) will improve the derived \phzs. Last but not least, of some importance should be proper zero-point calibration and extinction corrections, not applied to these bands in the present tests.

We would like to emphasise however that the NIR data are expected to help with \phz\ estimation mostly when the Balmer break is redshifted into the appropriate filters, leaving the KiDS $ugri$ coverage, which happens for $z>1.35$. Therefore, data much deeper than used in this Section are needed for the NIR to bring the most benefit for \phzs. 

\subsection{GAaP magnitudes with multiple parameter set combinations}
\label{Sec:GAMA-depth_mags+more_sets}

Having determined the single set of additional parameters bringing the most improvement to \phzs\ when used with the basic $ugri$ setup, we proceeded to joining them. Here we explored only the combinations of these sets from the previous tests that gave the best results. From the third block of rows of Tables \ref{Tab: photo-z tests GAMA} and \ref{Tab: photo-z tests GAMA full} it is clear that the best option of two additional parameter sets should be by adding VIKING and WISE; however, due to the current unavailability of proper photometry for these surveys outside of the GAMA equatorial patches, it is of interest to study also KiDS-only parameter combinations. In particular, we examined the unions of the \ttt{A}, \ttt{B} sizes, GAaP colours, \ttt{FLUX\_APER\_10\_\textit{band}} and \ttt{FLUX\_APER\_\textit{size}\_r} measurements, first pairwise (i.e.\ by adding two sets of parameters to GAaP magnitudes), and then in multiple combinations based on the results of the GAaP+pairwise experiments.

\begin{figure*}
\centering
\includegraphics[width=0.25\textwidth]{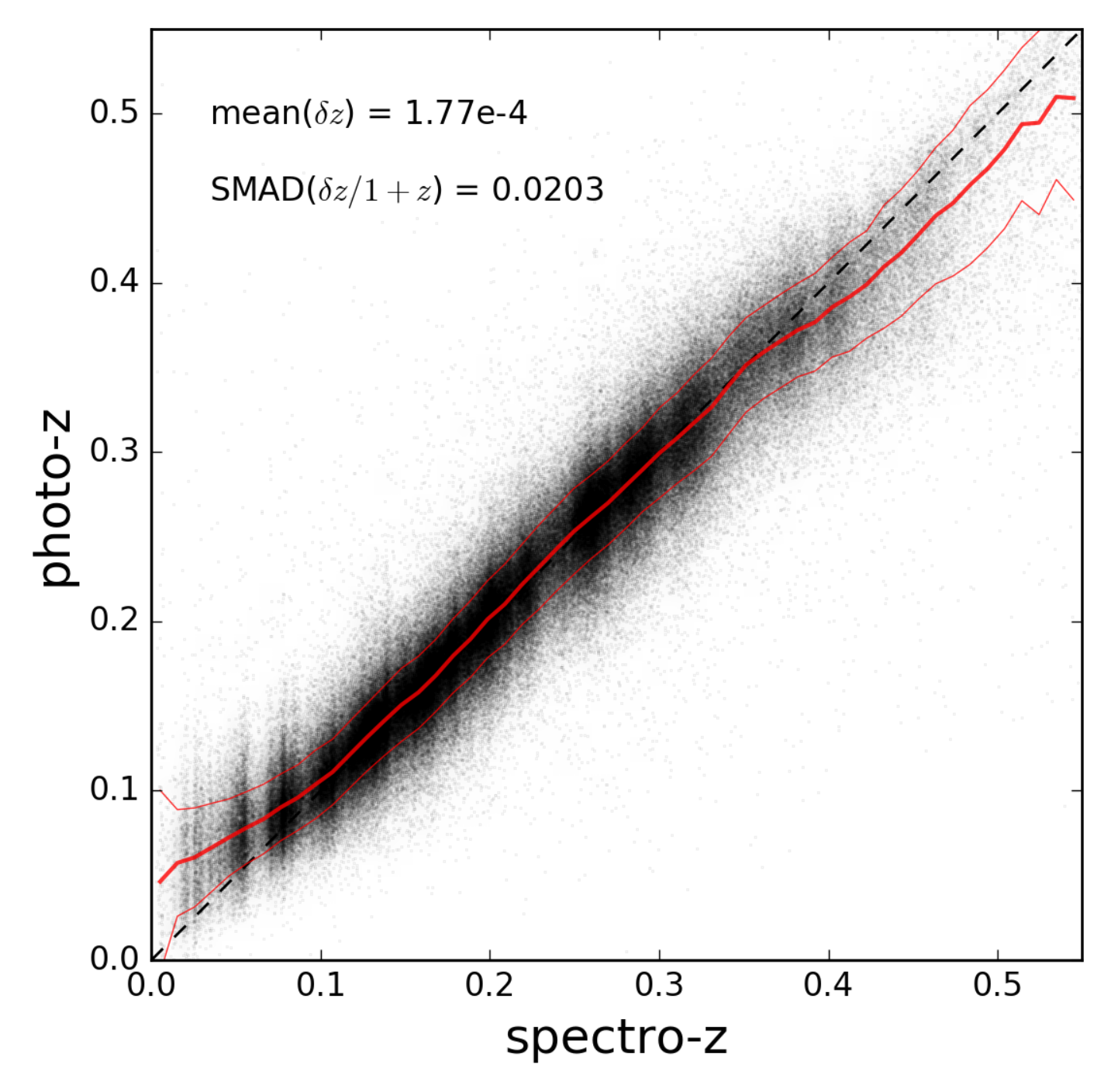}
\includegraphics[width=0.37\textwidth]{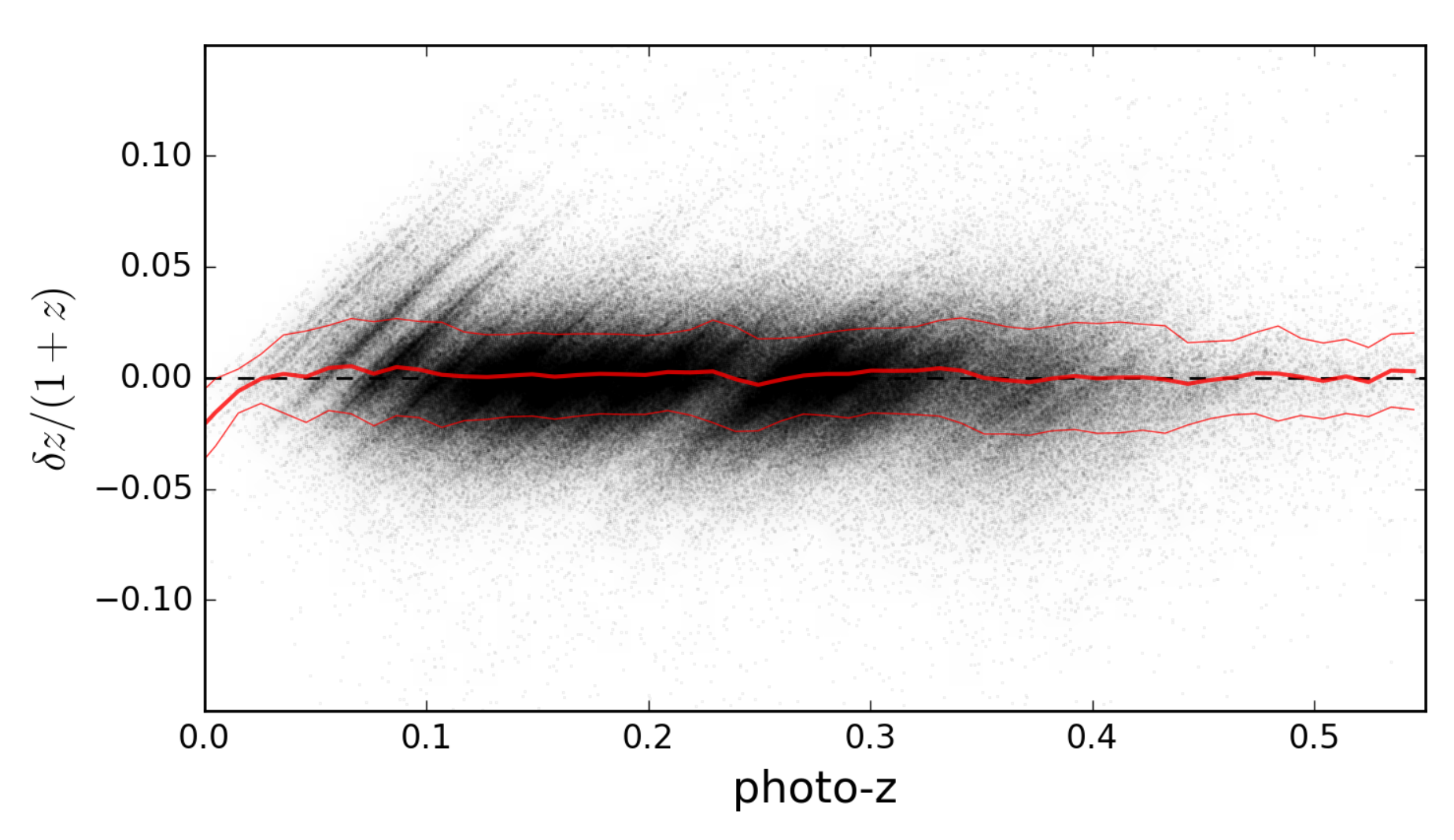}
\includegraphics[width=0.35\textwidth]{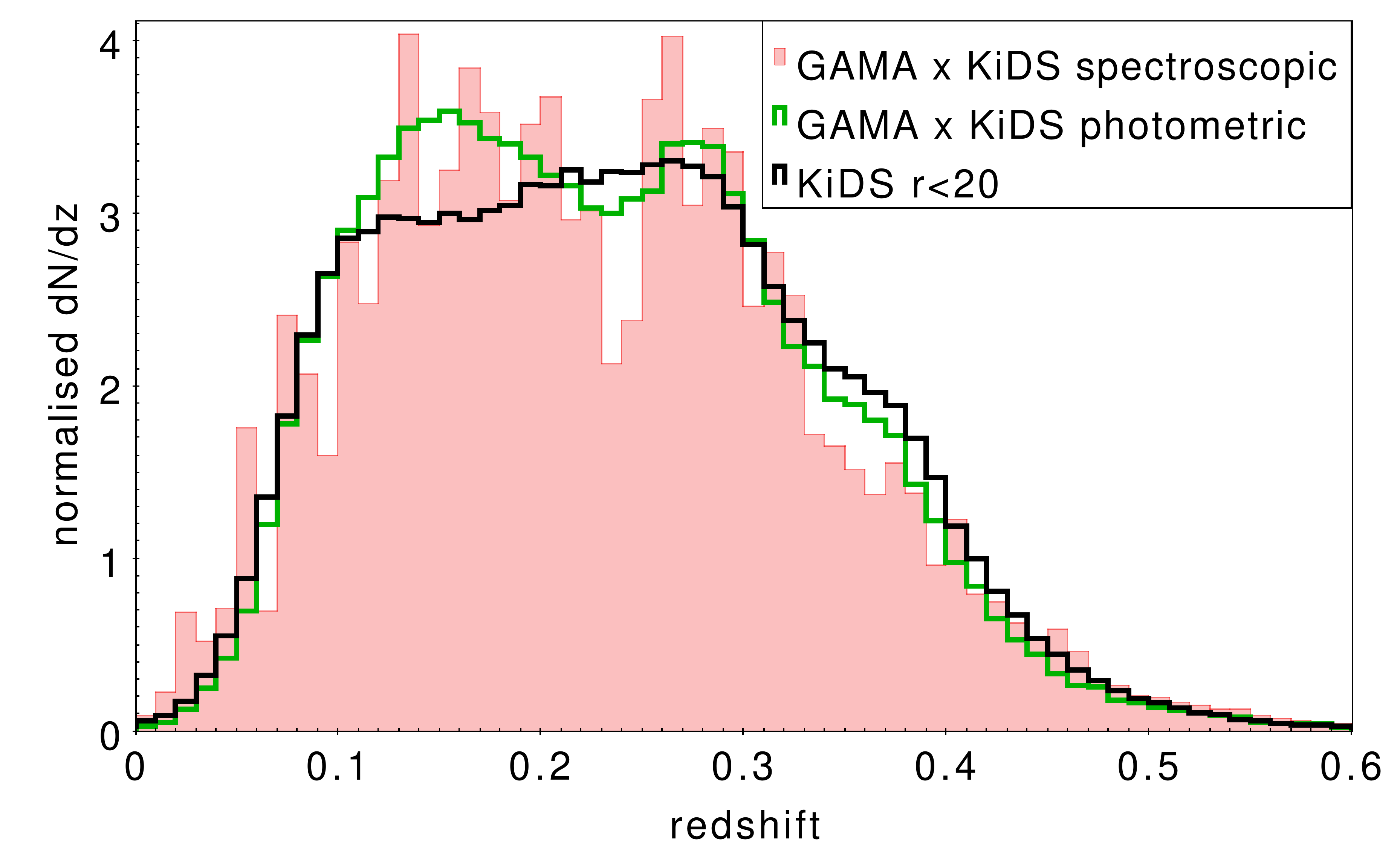}
\caption{Performance of the KiDS-GAMA ANNz2 \phzs\ as compared to the GAMA spectroscopic redshifts in the equatorial fields.
Left-hand panel: direct \spz--\phz\ comparison; central panel: \phz\ error as a function of \phz; right-hand panel: comparison of redshift distributions for the same set of KiDS$\times$GAMA sources {(red bars for \spzs, green line for \phzs), with also $dN/dz_\mrm{phot}$ of the full bright-end KiDS sample ($r<20$) overplotted (black line), all normalised to unit area under the histograms.}}
\label{Fig:KiDS-GAMA_photoz_performance}
\end{figure*}

In the fourth set of rows of Tables \ref{Tab: photo-z tests GAMA} and \ref{Tab: photo-z tests GAMA full} we provide results for the best cases of the GAaP+pairwise options, first for KiDS-only and then for KiDS+VIKING+WISE ones. In the latter case we were able to break the barrier of $0.02$ in scatter, which is more than 10\% improvement over using only $ugri$ magnitudes. But even without IR photometry, combining GAaP magnitudes, colours, and linear sizes gives very well constrained \phzs\ with  bias $\sim 10^{-3}$ and scatter $\sim0.02$. {This will be taken advantage of in the publicly released \phz\ catalogue described in \S\ref{Sec:GAMA catalogue release}.} 
Interestingly, whenever $ugri$ magnitudes are supplemented with optical colours and/or morphological parameters, adding WISE $W1$--$W3$ seems more beneficial than adding VIKING $z$--$K_s$ bands.

Using more extended parameter setups further improves the results, although it is at the expense of the computation time, which for ANNs scales non-linearly with the number of training parameters. As far as KiDS-only quantities are concerned, we stopped our experiments on four-set combinations, the best results of which came from using GAaP magnitudes, colours, linear sizes, and 10-pixel aperture magnitudes in the four bands. The improvement over using $ugri$ magnitudes+colours+sizes is not large. On the other hand, combining the optical magnitudes, colours, and sizes with VIKING and WISE measurements allowed us to obtain $\mrm{SMAD}\left(\delta z / (1+z)\right) < 0.019$ with  normalised bias $\sim 10^{-5}$. In the case of further adding the 10-pixel aperture $ugri$ magnitudes, some further improvement is seen (Table \ref{Tab: photo-z tests GAMA full}), but we have reasons to believe that this setup of even 24-dimensional parameter space might be too large for the ANNs to work efficiently (see e.g.\ \citealt{Soumagnac15} for a related discussion).

\subsection{KiDS-GAMA \phz\ catalogue release}
\label{Sec:GAMA catalogue release}

Based on the results of the above analysis, we computed very accurate and precise \phzs\ for a sample of {800,000 bright ($r \lesssim 20$)} KiDS galaxies {over the whole DR3 area of 450 deg$^2$}. This time the training set included both the GAMA equatorial data, as well as those from the G23 field. Although the latter are less complete than the former, the G23-GAMA galaxies were also selected for spectroscopic measurements based on their apparent magnitudes and not colours. This guarantees that we do not introduce biases into the training that could be related to using for instance LRGs or other sources for which \phzs\ are usually better constrained. 

The target GAMA-like KiDS catalogue will have somewhat different selections than the spectroscopic GAMA sample. The main reason are differences in photometry: GAMA input sources were selected from SDSS DR7 using the $r$-band Petrosian magnitude measurements as reference (except for G02 and G23 where additional photometry from CFHTLenS, KiDS and VIKING was used; \citealt{GAMA-II}). In KiDS, we do not have Petrosian magnitudes, and the closest to them among the 3 possibilities discussed in \S\ref{Sec:KiDS-photometric} are the \ttt{AUTO} ones. Using the GAMA equatorial fields with an $r_\mrm{Petro}\leq 19.8$ cut as the reference, we obtained $\langle r_\mrm{AUTO}^\mrm{KiDS} - r_\mrm{Petro}^\mrm{SDSS} \rangle = 0.018 \pm 0.201$ mag (mean with standard deviation). These differences are related not only to different ways of measuring the magnitudes, but also partly to the lower photometric quality of GAMA-input SDSS photometry than that of the KiDS measurements.

{Nevertheless}, in order to maximise the \textit{completeness} of KiDS `GAMA-like' galaxies with respect to the actual GAMA sources, we need to take a fainter cut in KiDS than $r<19.8$. For instance, $r_\mrm{AUTO} \leq 20.3$ retains $99.5\%$ of GAMA galaxies, while using $r_\mrm{AUTO} \leq 19.8$ would decrease the completeness to only 95.1\%. On the other hand, going on average 0.5 mag fainter than the fiducial GAMA limits will introduce many sources into the KiDS GAMA-like sample that are not well covered by the spectroscopic training sample (although it is partly alleviated due to the fact that GAMA \textit{does} include some fainter objects with redshift measurements). We emphasise however that all these considerations do not influence the ANNz2 \phz\ training process itself, which uses only confirmed KiDS$\times$GAMA galaxies. One should only bear in mind that the evaluated \phzs\ of sources fainter than the GAMA limits might be not reliable, and cuts on the presented here GAMA-like KiDS catalogue might be necessary to mitigate this.

To summarise, the released data product\footnote{Data available from \url{http://kids.strw.leidenuniv.nl/DR3/ml-photoz.php\#annz2}.} includes KiDS DR3 sources cut at $r_\mrm{AUTO} \leq 20.3$, extinction-corrected and zero-point calibrated, i.e. using `\ttt{MAG\_AUTO\_r\_calib}' for selection. Together with this, star removal was also applied by using {KiDS star-galaxy separator} $\mtt{SG2DPHOT}=0$. At the bright end of KiDS, this parameter is reliable enough to guarantee practically 100\% purity of the galaxy sample, while it only minimally influences the completeness with respect to GAMA ($\sim99.2\%$ once combined with the $r_\mrm{AUTO} \leq 20.3$ magnitude cut). Such a selection from the full KiDS-DR3, together with the requirements of $|\mtt{MAG\_GAAP\_band}|<99$ and $\mtt{MAGERR\_GAAP\_band}>0$ for each band, gives 801,000 sources over the KiDS DR3 footprint. Applying additionally the masking flag of $\mtt{IMAFLAGS\_ISO\_band} \, \& \, 01010111 = 0$ for each band leaves 695,000 galaxies in the GAMA-like KiDS-clean sample. We do not apply this latter flagging to the published data, leaving this to the end-users.

As mentioned, the GAMA-depth \phzs\ released with this paper are based on the training set composed of KiDS$\times$GAMA galaxies from the equatorial (G09, G12, G15) and southern (G23) fields. This sample includes almost 227,500 galaxies with $\meanz = 0.23$ and $\langle r_\mrm{GAaP} \rangle = 19.4$. The parameters supplied to the ANNs were GAaP $ugri$ magnitudes, related colours, and the \ttt{A} and \ttt{B} linear sizes; this is the setup with the best performance among the KiDS-only combinations of the magnitudes with two additional sets of parameters (fourth set of rows in Table \ref{Tab: photo-z tests GAMA}). As in all other experiments, this training sample was split randomly in two halves, one for the actual training, and the other for validation/optimisation. A total  of 250 ANNs were trained, of architectures generated randomly each time.

Figure \ref{Fig:KiDS-GAMA_photoz_performance} illustrates the performance of the GAMA-depth \phzs\ in the released catalogue, as judged from a comparison with the spectroscopic GAMA data in the equatorial fields. We see that except for the very local volume of $z\sim0$, the \phzs\ are extremely stable and well-constrained up to the limits of GAMA at $z\sim0.6$, and their overall performance for this sample is $\langle \delta z \rangle = 1.77\times 10^{-4}$ and $\mrm{SMAD}(\delta z/(1+z)) = 0.0203$. For comparison, the KiDS pipeline \phzs\ from BPZ give for the same data $\langle \delta z \rangle = 0.0153$ and $\mrm{SMAD}(\delta z/(1+z)) = 0.0317$.
The redshift distributions in the right-hand panel of Fig.~\ref{Fig:KiDS-GAMA_photoz_performance} show that the ML \phzs\ are so good to trace even a `dip' in $dN/dz$ of the GAMA {equatorial} catalogue which is caused by the large-scale structure crossing the GAMA fields \citep{Eardley15}. This dip is of course not observed in the full KiDS-GAMA \phz\ dataset{, as shown with the black line illustrating all the KiDS $r<20$ sources for which \phzs\ were derived as described in the present Section.}

This `GAMA-like' KiDS catalogue has been already used in scientific analyses, and its first published application is presented in \cite{Brouwer18}, where it is employed as the foreground for a weak lensing analysis of galaxy troughs and ridges.

\section{Conclusions and future prospects}
\label{Sec:conclusions_and_prospects}

In this paper we presented an analysis of machine learning photometric redshifts in the Kilo-Degree Survey Data Release 3{, and quantified the properties of two accompanying \phz\ catalogue releases, one at the full depth of the survey and the other limited to its bright end. In the latter case, we additionally studied possible extensions of the fiducial $ugri$ parameter space, both by adding extra imaging information (galaxy colours, sizes, fluxes in fixed apertures), as well as by using infrared photometry from VIKING and WISE.}

{At the full depth available from overlapping spectroscopy we} made a comparison of two MLMs used in KiDS -- ANNz2 and MLPQNA -- between each other as well as against the KiDS pipeline \phz\ solution from BPZ. This was done for various samples extracted from the current overlap between KIDS DR3 (plus some auxiliary photometric data) and external spectroscopic catalogues. We showed that at the bright, low-redshift end ($z<0.5$) of KiDS, the two ML \phz\ methods perform better than BPZ in most statistics, which is expected, as this is where the spectroscopic calibration data (mostly from GAMA) is the most abundant. But also for dimmer and higher-redshift sources (up to $z\sim1$) the MLMs provide well-constrained \phzs\, of comparable quality to the BPZ solution, despite much worse training data coverage there. 

These {general} conclusions apply also to the publicly-released KiDS DR3 \phz\ catalogue derived using ANNz2. This dataset includes all the KiDS DR3 sources having 4-band $ugri$ measurements {(over 39 million objects), although for part of these the \phzs\ are based on extrapolation over the limits of the training set and must be used with caution.} For scientific applications we {therefore defined} a FIDUCIAL {subsample of 20.5 million extended sources}, which is limited to the photometric coverage of the training sets used by ANNz2, and provides more secure \phzs\ (additionally improved thanks to weighting the training set). We judge that the \phzs\ in this catalogue are trustworthy to at least {$\zph \lesssim 0.9$} and $r \lesssim 23.5$.

In the second part of the paper we focused on the bright end of the KiDS catalogue ($r<20$, $\meanz = 0.23$) and made a comprehensive analysis of ML \phzs\ in this regime, taking advantage of the excellent KiDS photometry and of the high spectroscopic completeness of the largely overlapping GAMA survey. Having obtained very accurate, $\langle \delta z / (1+z) \rangle \sim 10^{-4}$, and precise, $\sigma_{\delta z} < 0.022 (1+z)$, \phzs\ for these KiDS sources when training the ANNz2 algorithm on 4-band $ugri$ magnitudes, we further studied how extending this basic parameter space can improve the redshifts. We looked at both adding KiDS-only quantities, such as colours, galaxy sizes, and fixed-aperture magnitudes, but also -- in view of the forthcoming or planned forced-photometry reduction of VIKING and WISE data in KiDS fields -- at additionally using near- and mid-infrared photometry.

The general conclusion {from the bright-end study} is that extending the parameter space used for \phz\ derivation with galaxy colours as well as morphological quantities does improve the results, both in terms of  bias as well as scatter. These improvements, although noticeable, are  not huge, and in the best case of using $ugri$ magnitudes + colours + linear sizes + 10-pixel-aperture magnitudes, the scatter in $\delta z / (1+z)$ is reduced by $\sim 9 \%$ over the fiducial case of magnitudes only. Adding IR measurements, on the other hand, is more promising, especially if they are also combined with the optical colours and morphological parameters. Photo-$z$s derived from 12-band photometry (from $u$ up to WISE 12 $\mu$m) have scatter smaller by $>10\%$ than in the optical-only case. Combined further with colours and sizes, \phzs\ of $\sigma_{\delta z} < 0.019 (1+z)$ will be possible. 

{It is worth emphasising that these} improvements should be considered as lower limits {for KiDS-based estimates}, as the IR photometry we used was based on SDSS apertures, and we lacked the GAaP magnitudes, shown in this paper to provide much more robust \phzs\ than other magnitude types available in KiDS. 

Future prospects for KiDS \phzs, in particular the ML ones, look bright. Both the photometry coverage as well as the training sets are being significantly extended. The availability of nine KiDS+VIKING bands at practically the full depth of KiDS will allow us to improve the accuracy and precision of \phzs\ both at the bright end of the sample (as shown in this paper), but also -- and perhaps more importantly -- at the faint end where the NIR bands start to play a significant role in constraining \phzs\ \citep{Banerji15}. For this amelioration to be achieved, more high-redshift spectroscopic training data is however needed to mitigate the sample variance. And indeed, observations of relevant fields both in the optical with VST and in the NIR with VISTA have been already made (VVDS fields) or are ongoing, or solicited for (VIPERS areas). We also plan to add the overlapping WiggleZ \citep{WiggleZ} to the spectroscopic sample, not used until now in KiDS analyses.

It is worth emphasising that the forthcoming 9-band KiDS+VIKING \phzs\ will place these surveys in a unique position as far as datasets of such angular extent and depth are concerned. Until now, \phzs\ with (at least) that many bands have been available either for wide-angle but very shallow samples (e.g. SDSS-based, \citealt{Way09}; 2MPZ, \citealt{2MPZ}) or for deep but small-area ones (e.g. EGS, \citealt{EGS}; COSMOS, \citealt{COSMOS2015}). The joint KiDS+VIKING \phz\ analysis and related data products will fill this important gap.

\begin{acknowledgements}
We thank Iftach Sadeh for the assistance in using the ANNz2 software, and for its continuous development, as well as Jonn Soo for very useful suggestions regarding the practicalities of the code. Ben Hoyle is acknowledged for enlightening discussions on machine-learning photo-z estimation. We also thank an anonymous referee for very useful comments and suggestions that allowed us to improve this paper.\\

Based on data products from observations made with ESO Telescopes at the La Silla Paranal Observatory under programme IDs 177.A-3016, 177.A-3017 and 177.A-3018, and on data products produced by Target/OmegaCEN, INAF-OACN, INAF-OAPD and the KiDS production team, on behalf of the KiDS consortium. OmegaCEN and the KiDS production team acknowledge support by NOVA and NWO-M grants. Members of INAF-OAPD and INAF-OACN also acknowledge the support from the Department of Physics \& Astronomy of the University of Padova, and of the Department of Physics of Univ. Federico II (Naples).\\

GAMA is a joint European-Australasian project based around a spectroscopic campaign using the Anglo-Australian Telescope. The GAMA input catalog is based on data taken from the Sloan Digital Sky Survey and the UKIRT Infrared Deep Sky Survey. Complementary imaging of the GAMA regions is being obtained by a number of independent survey programs including GALEX MIS, VST KiDS, VISTA VIKING, WISE, Herschel-ATLAS, GMRT, and ASKAP providing UV to radio coverage. GAMA is funded by the STFC (UK), the ARC (Australia), the AAO, and the participating institutions. The GAMA website is \url{http://www.gama-survey.org/}.\\

2dFLenS is based on data acquired through the Australian Astronomical Observatory, under program A/2014B/008.  It would not have been possible without the dedicated work of the staff of the AAO in the development and support of the 2dF-AAOmega system, and the running of the AAT.\\

Funding for SDSS-III was provided by the Alfred P. Sloan Foundation, the Participating Institutions, the National Science Foundation, and the U.S. Department of Energy Office of Science. The SDSS-III website is \url{http://www.sdss3.org/}. SDSS-III is managed by the Astrophysical Research Consortium for the Participating Institutions of the SDSS-III Collaboration including the University of Arizona, the Brazilian Participation Group, Brookhaven National Laboratory, Carnegie Mellon University, University of Florida, the French Participation Group, the German Participation Group, Harvard University, the Instituto de Astrofisica de Canarias, the Michigan State/Notre Dame/JINA Participation Group, Johns Hopkins University, Lawrence Berkeley National Laboratory, Max Planck Institute for Astrophysics, Max Planck Institute for Extraterrestrial Physics, New Mexico State University, New York University, Ohio State University, Pennsylvania State University, University of Portsmouth, Princeton University, the Spanish Participation Group, University of Tokyo, University of Utah, Vanderbilt University, University of Virginia, University of Washington, and Yale University.\\

MBi is supported by the Netherlands Organization for Scientific Research, NWO, through grant number 614.001.451, and by the Polish National Science Center under contract UMO-2012/07/D/ST9/02785.

MBi and HHo acknowledge support from the European Research Council FP7 grant number 279396.

HHo acknowledges support from Vici grant 639.043.512, financed by the NWO.

CB acknowledges the support of the Australian Research Council through the award of a Future Fellowship.

JTAdJ is supported by the Netherlands Organisation for Scientific Research (NWO) through grant 621.016.402.

HHi is supported by an Emmy Noether grant (No. Hi 1495/2-1) of the Deutsche Forschungsgemeinschaft.

MBr acknowledges financial contribution from the agreement ASI/INAF I/023/12/1.

CH acknowledges support from the ERC under grant number 64711.

KK acknowledges support by the Alexander von Humboldt Foundation.

DP acknowledges the support of the Australian Research Council through the award of a Future Fellowship.

GVK acknowledges financial support from the Netherlands Research School for Astronomy (NOVA) and Target. Target is supported by Samenwerkingsverband Noord Nederland, European fund for regional development, Dutch Ministry of economic affairs, Pieken in de Delta, Provinces of Groningen and Drenthe.\\

This work has made use of \textsc{TOPCAT} \citep{TOPCAT} and \textsc{STILTS} \citep{STILTS} software, as well as of \textsc{python} (\url{www.python.org}), including the packages \textsc{NumPy} \citep{NumPy}, \textsc{SciPy} \citep{SciPy}, and \textsc{Matplotlib} \citep{Matplotlib}.

\end{acknowledgements}

%%%%%%%%%%%%%%%%%%%%%%%%%%%%%%%%%%%%%%%%%%%%%%%%%%

%%%%%%%%%%%%%%%%%%%% REFERENCES %%%%%%%%%%%%%%%%%%

\bibliographystyle{mnras}
\bibliography{KiDS-photo-z}

%%%%%%%%%%%%%%%%%%%%%%%%%%%%%%%%%%%%%%%%%%%%%%%%%%

%%%%%%%%%%%%%%%%% APPENDICES %%%%%%%%%%%%%%%%%%%%%
\onecolumn
\begin{appendix}

{
\section{Data release details}
Here we provide details of the two photometric redshift catalogues released with this paper, available via \url{http://kids.strw.leidenuniv.nl/DR3/ml-photoz.php\#annz2}. We include only basic parameters in them, in order to enable end users to apply selections as described in the paper. For more sophisticated filtering, these datasets can be cross-matched with the overall DR3 data using the unique source identifier \ttt{ID}.
}

{
\subsection{Full-depth photometric redshift catalogue}
Table \ref{Tab:full-depth columns} lists the columns included in the publicly-released full-depth KiDS DR3 photometric redshift catalogue, in which \phzs\ were derived using ANNz2 trained on the full spectro-photo compilation (\S\ref{Sec:KiDS_photo-spectro}), as detailed in \S\ref{Sec:DR3-release}. The dataset includes 39.2 million sources extracted from DR3 by requiring that all the four $ugri$ GAaP magnitudes are measured. The catalogue is meant for general-purpose uses, but additional filtering as described in \S\ref{Sec:KiDS-photometric} and \S\ref{Sec:DR3-release} is needed to remove artefacts and to guarantee reliable \phzs; we provide the `fiducial' flag to be applied as the minimum requirement. For details of the listed columns, please see appendix A.2 of \cite{KiDS-DR3}.
}

\begin{table}[!h]
\begin{center}
\caption{\label{Tab:full-depth columns} Columns provided in the full-depth ANNz2 photometric redshift catalogue.}
\begin{tabular}{llll}
\hline\hline
Label & Format & Unit & Description \\
\hline
ID & 23A & & Source identifier \\
RAJ2000 & D & deg & Right ascension (J2000) \\
DECJ2000 & D & deg & Declination (J2000) \\
SG2DPHOT & K &  & Source classification  \\
IMAFLAGS\_ISO\_band & J & & Mask flag \\
MAG\_GAAP\_band\_calib & E & mag & Calibrated GAaP magnitude\tablefootmark{a} \\
MAGERR\_GAAP\_band & E & mag & Error in GAaP magnitude \\
zphot\_ANNz2 & D &  & Photometric redshift derived with ANNz2 \\
fiducial & I & & Flag defining fiducial selection \\
\hline
\end{tabular}
\tablefoot{
\tablefoottext{a}{Extinction-corrected and zero-point calibrated GAaP magnitudes: \\
$\mtt{MAG\_GAAP\_band\_calib} = \mtt{MAG\_GAAP\_band} + \mtt{ZPT\_OFFSET\_band} - \mtt{EXT\_SFD\_band}$.}
}
\end{center}
\end{table}

{
\subsection{Bright-end photometric redshift catalogue}
Table \ref{Tab:bright-end columns} lists the columns included in the publicly-released bright-end KiDS DR3 photometric redshift catalogue, in which \phzs\ were derived using ANNz2 trained on the GAMA spectroscopic sources (\S\ref{Sec:GAMA spec-z}), as detailed in \S\ref{Sec:GAMA catalogue release}. The dataset includes $800,830$ sources extracted from DR3 by applying the magnitude cut of $\mtt{MAG\_AUTO\_R\_calib}\leq 20.3$, star-galaxy separation parameter $\mtt{SG2DPHOT}=0$, and by requiring that all the four $ugri$ GAaP magnitudes, and their errors, are measured. Additional filtering as described in \S\ref{Sec:GAMA catalogue release} is needed to remove artefacts. For details of the listed columns, please see appendix A.2 of \cite{KiDS-DR3}.
}

\begin{table}[!h]
\begin{center}
\caption{\label{Tab:bright-end columns} Columns provided in the bright-end ANNz2 photometric redshift catalogue.}
\begin{tabular}{llll}
\hline\hline
Label & Format & Unit & Description \\
\hline
ID & 23A & & Source identifier \\
RAJ2000 & D & deg & Right ascension (J2000) \\
DECJ2000 & D & deg & Declination (J2000) \\
IMAFLAGS\_ISO\_band & J & & Mask flag \\
MAG\_AUTO\_band\_calib & E & mag & Calibrated Kron-like elliptical aperture magnitude\tablefootmark{a} \\
MAGERR\_AUTO\_band & E & mag & RMS error for MAG\_AUTO  \\
MAG\_ISO\_band\_calib & E & mag & Calibrated isophotal magnitude\tablefootmark{a}   \\
MAGERR\_ISO\_band & E & mag & RMS error for MAG\_ISO  \\
MAG\_GAAP\_band\_calib & E & mag & GAaP `calibrated' magnitude\tablefootmark{a} \\
MAGERR\_GAAP\_band & E & mag & Error in GAaP magnitude \\
zphot\_ANNz2 & D &  & Photometric redshift derived with ANNz2 \\
\hline
\end{tabular}
\tablefoot{
\tablefoottext{a}{Extinction-corrected and zero-point calibrated magnitudes:\\
$\mtt{MAG\_type\_band\_calib} = \mtt{MAG\_type\_band} + \mtt{ZPT\_OFFSET\_band} - \mtt{EXT\_SFD\_band}$.}
}
\end{center}
\end{table}

{
\section{Photometric redshift distributions for red and blue sources}
\label{App:dNdz_red_blue}
In \S\ref{Sec:DR3-release} we compared \phz\ distributions for the BPZ and ANNz2 solutions of the FIDUCIAL DR3 photometric sample and found visible discrepancies at various redshift ranges. To probe further the source of these disagreements, we examined the differences between individual \phzs\ from the two available full-depth solutions, i.e. $z_\mrm{ANNz2} - z_\mrm{BPZ}$, as a function of observed colours. By examining various colour-space projections we found that the two solutions are more consistent for redder sources. In particular by projecting on the $r-i$ vs. $g-r$ colour-colour plane, we determined that a good distinction between more consistent and less consistent \phzs\ (in terms of $|z_\mrm{ANNz2} - z_\mrm{BPZ}|$) is given by $g-r = 0.8 - 0.8(r-i)$ and $r-i=0.5$ lines. Sources lying above these two lines (i.e. redder) have overall much smaller \phz\ differences than the bluer ones. We note that this is just a general division irrespectively of morphology, and some of the ``red'' galaxies can in fact be spirals, or even irregulars of Magellanic type \citep{Fukugita95}. In Fig.~\ref{Fig:dNdz_red_blue} we compare the resulting $dN/dz_\mrm{phot}$ distributions for thus selected ``red'' and ``blue'' galaxies, for the two \phz\ solutions. This comparison is limited to the FIDUCIAL photometric sample, additionally cut at $r<23.5$ to avoid ANNz2 extrapolation discussed in \S\ref{Sec:DR3-release}. This confirms that redder galaxies have much more consistent \phz\ distributions than the full dataset, while the blue ones are assigned very different $dN/dz$s by the two methods. In particular, for BPZ a large fraction of blue sources have $\zph<0.4$ and these constitute most of the low-$z_\mrm{BPZ}$ peak observed in the full sample without the colour division. On the other hand, ANNz2 produces a very flat $dN/dz_\mrm{phot}$ distribution for blue galaxies and there are almost no red sources allocated to $z_\mrm{ANNz2}>1$. The latter is equally true for $z_\mrm{BPZ}>1$.
}
\begin{figure}
\centering
\includegraphics[width=0.5\textwidth]{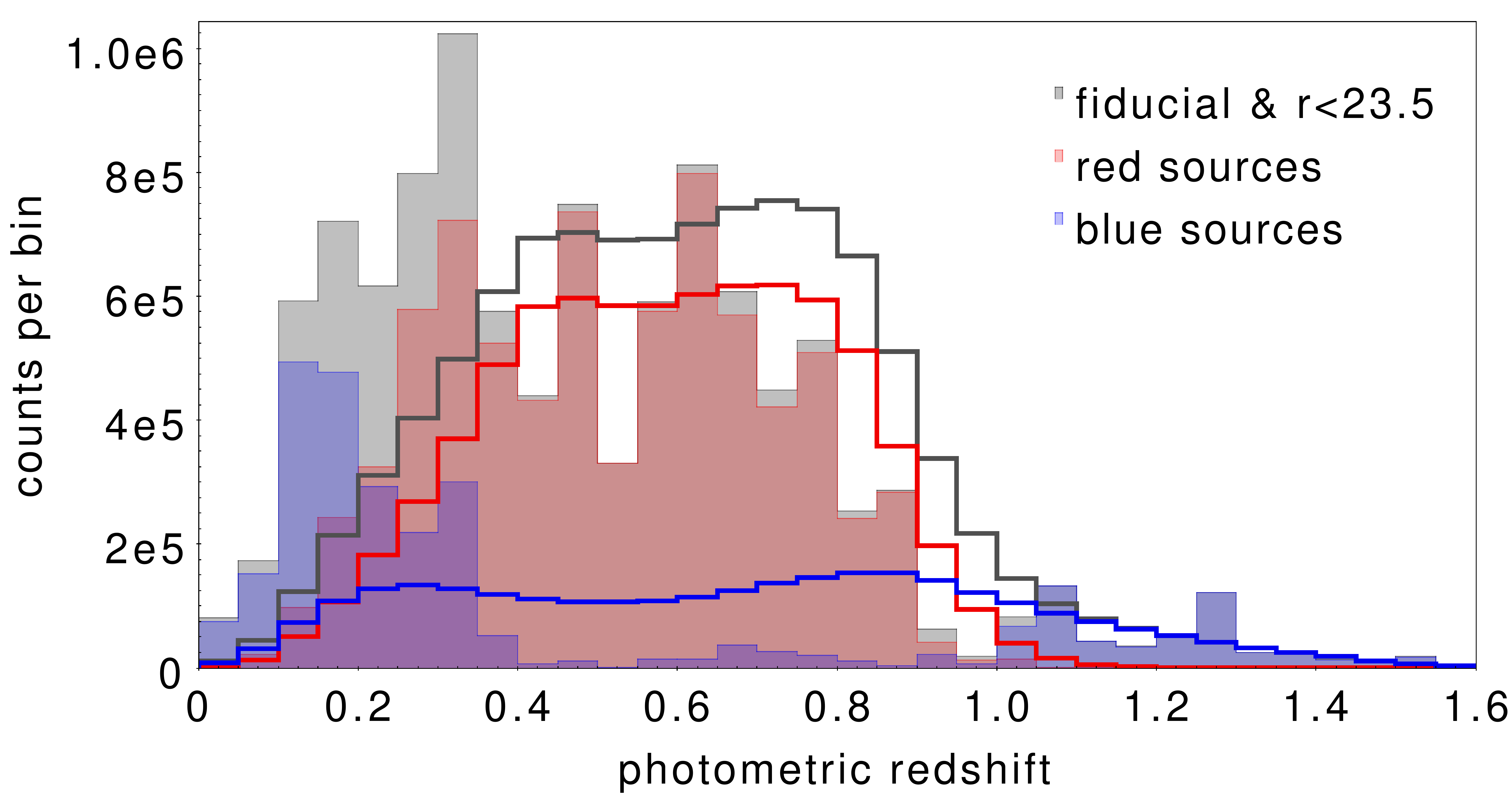}
\caption{{Comparison of photometric redshift distributions for KiDS DR3 galaxies preselected as red and blue, the division line being a join of the $g-r = 0.8 - 0.8(r-i)$ and $r-i=0.5$ conditions. The sample is limited to the FIDUCIAL selection (\S\ref{Sec:KiDS-photometric}) with an additional $r<23.5$ cut. Filled histograms are for the BPZ solution and thick lines are for ANNz2. Grey is for the full sample, while red and blue are for respective colour selections.}}
\label{Fig:dNdz_red_blue}
\end{figure}

{
\section{Extended table for GAMA-depth experiments}
\label{App:Full_GAMA_table}
Table \ref{Tab: photo-z tests GAMA full} is an extended version of Table \ref{Tab: photo-z tests GAMA}.
\input{table-phz-GAMA.tex}
}

\end{appendix}

%%%%%%%%%%%%%%%%%%%%%%%%%%%%%%%%%%%%%%%%%%%%%%%%%%

\end{document}

%% file: table-specz-fullDR3.tex
%% Table with spectroscopic samples within DR3 footprint
\begin{table*}
\caption{Spectroscopic samples constituting the KiDS DR3 \phz\ training set.}
\begin{tabular}{ l  r  c  c  l }
\hline\hline
{\textbf{sample}} & \multicolumn{1}{c}{\textbf{Number of sources}\tablefootmark{a}} & { \textbf{mean $ z $}\tablefootmark{a} } & { \textbf{mean $r$ mag}\tablefootmark{a} } & \textbf{reference(s)} \\
\hline 
GAMA-II equatorial		& 190,741 &   0.234 & 19.5 & \cite{GAMA-II} \\
SDSS DR13 galaxies	&   56,911 &   0.349 & 19.6 & \cite{SDSS.DR13}  \\
GAMA G23					&  38,854  &   0.238 & 19.3 & \textit{proprietary}\\
zCOSMOS \tablefootmark{b}					&  25,888  &   0.813 & 22.2 & \textit{private comm.} \& \cite{Davies15}   \\
2dFLenS						&  11,873  &   0.362 & 19.6 & \cite{2dFLenS} \\
DEEP2 DR4	 (two fields)				&    8,924  &   0.962 &  23.5 & \cite{DEEP2} \\
CDFS \tablefootmark{c}						&    7,044  &   0.846 &  22.9 & \textit{online} \& \cite{ACES} \\
GAMA G15-deep        &     2,286 &   0.340 &  21.1 & \textit{proprietary}\\
\hline 
Total \tablefootmark{d}                   & 312,501 &   0.335 &  19.9 \\  
\hline    
Total cleaned \tablefootmark{e}          &  278,946 &   0.332 &  19.9 \\
\hline
\end{tabular} 
\tablefoot{\\
\tablefoottext{a}{After cross-match with KiDS DR3, without masking nor quality cuts in KiDS.\\}
\tablefoottext{b}{Data from zCOSMOS public and non-public catalogues, as well as from the GAMA-G10 catalogue.\\}
\tablefoottext{c}{Data from GOODS/CDF-S compilation and from ACES.\\}
\tablefoottext{d}{Duplicate entries removed.\\}
{\tablefoottext{e}{After cleaning of bad photometry as described in \S\ref{Sec:KiDS_photo-spectro}.}}
}
\label{Tab: spectro data} 
\end{table*}

%% file: table-phz-fullDR3.tex
%% Table with photo-z statistics - experiments for full DR3
\begin{table*}
\begin{center}
\caption{Statistics of photometric redshift performance obtained for KiDS DR3 experiments with ANNz2 and MLPQNA vs.\ BPZ. Results for the particular tests are provided in blocks of rows. See text for details.}
\begin{small}
\begin{tabular}{ l  l  r  r  r  r  r }
\hline\hline
\centering {\textbf{sample}} & {\textbf{method}} & \multicolumn{1}{c}{ \textbf{mean of} } & { \textbf{mean of} } & \multicolumn{1}{c}{\textbf{st.dev.\ of}} & \multicolumn{1}{c}{\textbf{SMAD \tablefootmark{a} of}} & \multicolumn{1}{c}{ \textbf{\% of outliers} } \\
{} & {} & \multicolumn{1}{c}{$\delta z = z_\mathrm{ph}-z_\mathrm{sp}$ } &  \multicolumn{1}{c}{$\delta z/(1+z_\mathrm{sp})$ } &  \multicolumn{1}{c}{$\delta z/(1+z_\mathrm{sp})$ } &  \multicolumn{1}{c}{$\delta z/(1+z_\mathrm{sp})$} & \multicolumn{1}{c}{ $|\delta z|/(1+z_\mathrm{sp})>0.15$ }\\ 
\hline 
random subsample & ANNz2 & $-3.3\times 10^{-3}$ & $3.3\times 10^{-3}$ & 0.073 & 0.026 & 3.5\% \\
%\textbf{[for comparison]} & mags+sizes & $-3.0\times 10^{-3}$ & $3.2\times 10^{-3}$ & 0.073 & 0.026 & 3.4\% \\
$\langle z \rangle =0.332$  & MLPQNA & $-2.0\times 10^{-3}$ & $3.9\times 10^{-3}$ & 0.079 & 0.026 & 3.4\% \\
 & BPZ \tablefootmark{b} & $-1.9\times 10^{-2}$ & $-1.5\times 10^{-3}$ & 0.089 & 0.035 & 4.1\% \\
\hline 
random 10\% of $r<20$, & ANNz2 & $-2.4\times 10^{-3}$ & $8.3\times 10^{-3}$ & 0.102 & 0.034 & 7.1\% \\        
all from $r\geq20$,  & MLPQNA & $-3.2\times 10^{-3}$ & $7.2\times 10^{-3}$  & 0.116 &  0.034 &  7.4\% \\
$\langle z \rangle =0.489$ & BPZ \tablefootmark{b} & $-5.8 \times 10^{-2}$ & $-1.9 \times 10^{-2}$ & 0.120 & 0.042 & 8.4\% \\
\hline
trained w/o COSMOS,  & ANNz2 & $-4.4 \times 10^{-2}$ & $1.2 \times 10^{-2}$ & 0.183 & 0.091 & 25.0\% \\
tested on COSMOS & ANNz2w \tablefootmark{c} &  $-6.7 \times 10^{-2} $ & $-4.6\times 10^{-4}$ & 0.184 & 0.086 & 22.7\% \\
$\langle z \rangle =0.784$   & MLPQNA & $-8.0 \times 10^{-2}$ & $-2.7\times 10^{-3}$ & 0.204 & 0.086 & 23.6\%\\
& BPZ \tablefootmark{b} & $-2.4 \times 10^{-1}$ & $-8.5 \times 10^{-2}$ & 0.195 & 0.085 & 24.5\% \\
\hline
trained w/o CDFS,      & ANNz2 & $3.0 \times 10^{-2}$ & $5.2 \times 10^{-2}$ & 0.232 & 0.108 & 25.7\% \\
tested on CDFS & ANNz2w \tablefootmark{c} &  $3.9 \times 10^{-2}$ & $5.4 \times 10^{-2}$ & 0.206 & 0.101 & 26.0\% \\
 $\langle z \rangle =0.742$ & MLPQNA & $1.0 \times 10^{-2}$ & $3.8 \times 10^{-2}$ & 0.222 & 0.100 & 25.8\% \\
 & BPZ \tablefootmark{b} & $-1.9 \times 10^{-2}$ & $-7.2 \times 10^{-2}$ & 0.183 & 0.083 & 23.7\% \\
\hline 
\end{tabular} 
\tablefoot{\\
\tablefoottext{a}{SMAD is the scaled median absolute deviation, converging to standard deviation for Gaussian distributions.\\}
\tablefoottext{b}{BPZ is independent of the training sets -- the numbers are given for comparison (for the same test samples). These statistics are based on the KiDS pipeline solution.\\}
\tablefoottext{c}{Training data weighted with the kNN method, weights propagated throughout the training and evaluation procedure.}
}
\end{small}
\label{Tab: photo-z tests full DR3}
\end{center}
\end{table*}

%% file: table-stats-DR3.tex
%% Table with photo-z statistics for BPZ and ANNz2 - full DR3 final
\begin{table*}
\begin{center}
\caption{Statistics of photometric redshift performance for the released KiDS DR3 catalogue, as obtained from a comparison with overlapping spectroscopic redshifts. Results are presented per bin of ANNz2 and BPZ \phzs, respectively.}
\begin{small}
\begin{tabular}{ c  l  r  r  r  r  }
\hline\hline
\centering { \textbf{\phz\ bin} } & {\textbf{method}} & \multicolumn{1}{c}{ \textbf{mean of} } & { \textbf{mean of} } & \multicolumn{1}{c}{\textbf{st.dev.\ of}} & \multicolumn{1}{c}{\textbf{SMAD of}} \\
{} & {} & \multicolumn{1}{c}{$\delta z = z_\mathrm{ph}-z_\mathrm{sp}$ } &  \multicolumn{1}{c}{$\delta z/(1+z_\mathrm{sp})$ } &  \multicolumn{1}{c}{$\delta z/(1+z_\mathrm{sp})$ } &  \multicolumn{1}{c}{$\delta z/(1+z_\mathrm{sp})$} \\ 
\hline 
$0.0 < \zph \leq 0.1$ & ANNz2 & $3.0\times 10^{-3}$ & $6.2\times 10^{-3}$ & 0.045 & 0.025  \\
 & BPZ & $-7.8\times 10^{-2}$ & $-4.4\times 10^{-2}$ & 0.109 & 0.040  \\
\hline 
$0.1 < \zph \leq 0.2$ & ANNz2 & $1.7\times 10^{-3}$ & $3.9\times 10^{-3}$ & 0.044 & 0.034  \\
 & BPZ & $-2.9\times 10^{-3}$ & $6.7\times 10^{-3}$ & 0.069 & 0.032  \\
\hline 
$0.2 < \zph \leq 0.3$ & ANNz2 & $-1.0\times 10^{-2}$ & $-5.1\times 10^{-3}$ & 0.048 & 0.035  \\
 & BPZ & $3.5\times 10^{-3}$ & $5.6\times 10^{-3}$ & 0.044 & 0.028  \\
\hline 
$0.3 < \zph \leq 0.4$ & ANNz2 & $-1.7\times 10^{-2}$ & $-8.3\times 10^{-3}$ & 0.055 & 0.032  \\
 & BPZ & $1.0\times 10^{-2}$ & $1.0\times 10^{-2}$ & 0.043 & 0.028  \\
\hline 
$0.4 < \zph \leq 0.5$ & ANNz2 & $-2.0\times 10^{-2}$ & $-5.8\times 10^{-3}$ & 0.079 & 0.032  \\
 & BPZ & $4.5\times 10^{-2}$ & $3.4\times 10^{-2}$ & 0.044 & 0.035  \\
\hline 
$0.5 < \zph \leq 0.6$ & ANNz2 & $-5.0\times 10^{-3}$ & $3.0\times 10^{-3}$ & 0.074 & 0.026  \\
 & BPZ & $2.6\times 10^{-2}$ & $2.0\times 10^{-2}$ & 0.054 & 0.028  \\
\hline 
$0.6 < \zph \leq 0.7$ & ANNz2 & $2.2\times 10^{-3}$ & $9.3\times 10^{-3}$ & 0.088 & 0.032  \\
 & BPZ & $6.4\times 10^{-3}$ & $1.1\times 10^{-2}$ & 0.081 & 0.040  \\
\hline 
$0.7 < \zph \leq 0.8$ & ANNz2 & $-9.2\times 10^{-3}$ & $9.0\times 10^{-3}$ & 0.123 & 0.057  \\
 & BPZ & $-2.9\times 10^{-2}$ & $-1.1\times 10^{-3}$ & 0.118 & 0.057  \\
\hline 
$0.8 < \zph \leq 0.9$ & ANNz2 & $-1.1\times 10^{-2}$ & $1.4\times 10^{-2}$ & 0.146 & 0.079  \\
 & BPZ & $-1.5\times 10^{-1}$  & $-3.4\times 10^{-2}$ & 0.183 & 0.084  \\
\hline 
$0.9 < \zph \leq 1.0$ & ANNz2 & $2.7\times 10^{-3}$ & $3.8\times 10^{-2}$ & 0.197 & 0.082  \\
 & BPZ & $-4.8\times 10^{-1}$ & $-1.4\times 10^{-1}$ & 0.239 & 0.256  \\
\hline 
\end{tabular} 
\end{small}
\label{Tab: photo-z stats full DR3}
\end{center}
\end{table*}

%% file: table-phz-GAMA-short.tex
%% Table with photo-z statistics - GAMA depth
\begin{table*}
\begin{center}

\caption{Statistics of photometric redshift performance obtained for the KiDS-GAMA spectroscopic sample {(extract)}. See Appendix \ref{App:Full_GAMA_table} for the full table.}

\begin{small}
\begin{tabular}{ l  r  r  r  r  }
\hline\hline
{\textbf{parameter setup}} & \multicolumn{1}{c}{ \textbf{mean of} } & \multicolumn{1}{c}{ \textbf{mean of} } & \multicolumn{1}{c}{\textbf{st.dev.\ of}} & \multicolumn{1}{c}{\textbf{SMAD\tablefootmark{a} of}} \\ 
{} & \multicolumn{1}{c}{$\delta z = z_\mathrm{ph}-z_\mathrm{sp}$ } &  \multicolumn{1}{c}{$\delta z/(1+z_\mathrm{sp})$ } &  \multicolumn{1}{c}{$\delta z/(1+z_\mathrm{sp})$ } &  \multicolumn{1}{c}{$\delta z/(1+z_\mathrm{sp})$} \\ 
\hline 
$ugri$ GAaP magnitudes, BPZ \tablefootmark{b}
& $1.34 \times 10^{-2}$ & $1.17 \times 10^{-2}$ & 0.0481 & 0.0317     \\ 
$ugri$ GAaP magnitudes, DR3 ANNz2 overall solution \tablefootmark{c}
& $-2.03 \times 10^{-3}$ & $8.61 \times 10^{-4}$ & 0.0508 &  0.0345  \\ 
$ugri$ GAaP magnitudes, GAMA-depth ANNz2 solution
& $-1.94 \times 10^{-3}$ & $-1.91 \times 10^{-4}$ & 0.0328  & 0.0219 \\ 
$ugri$ ISO magnitudes
& $-2.45 \times 10^{-3}$ & $-3.26 \times 10^{-4}$ & 0.0360 & 0.0259 \\ 
\hline 
$ugri$ GAaP magnitudes +\\
\hspace{1mm} KiDS \ttt{FLUX\_APER\_10}\_$ugri$
& $-1.98 \times 10^{-3}$ & $-2.71 \times 10^{-4}$ & 0.0321 & 0.0214 \\
\hspace{1mm} KiDS \ttt{A}, \ttt{B} linear semi-axes
& $-1.96 \times 10^{-3}$ & $-2.54 \times 10^{-4}$ & 0.0318 & 0.0213 \\
\hspace{1mm} KiDS six GAaP $ugri$ colours %for each unique $ugri$ combination
& $-2.13 \times 10^{-3}$ & $-3.99 \times 10^{-4}$ & 0.0321 & 0.0210 \\
\hdashline
\hspace{1mm} WISE $W1,W2,W3$ fluxes
& $-1.61 \times 10^{-3}$ & $-1.47 \times 10^{-5}$ & 0.0312 & 0.0206\\
\hspace{1mm} VIKING $z Y J H K_s$ fluxes
& $-1.58 \times 10^{-3}$ & $-2.22 \times 10^{-5}$ & 0.0312 & 0.0202 \\
\hline
%$ugri$ GAaP magnitudes +\\
\hspace{1mm} KiDS \ttt{A}, \ttt{B} + \ttt{FLUX\_APER\_10}\_$ugri$
& $-1.71 \times 10^{-3}$ & $-1.16 \times 10^{-4}$ & 0.0311 & 0.0209 \\
\hspace{1mm} KiDS \ttt{FLUX\_APER\_10}\_$ugri$ + GAaP colours
& $-1.71 \times 10^{-3}$ & $-1.23 \times 10^{-4}$ & 0.0313 & 0.0205 \\
\hspace{1mm} KiDS \ttt{A}, \ttt{B} + GAaP colours
& $-1.59 \times 10^{-3}$ & $-7.61 \times 10^{-5}$ & 0.0305 & 0.0204 \\
\hdashline
\hspace{1mm} KiDS GAaP colours + VIKING fluxes
& $-1.72 \times 10^{-3}$ & $-1.53 \times 10^{-4}$ & 0.0311 & 0.0199 \\
\hspace{1mm} KiDS GAaP colours + WISE $W1,W2,W3$ fluxes
& $-1.50 \times 10^{-3}$ & $-9.24 \times 10^{-6}$ & 0.0303 & 0.0198 \\
\hspace{1mm} VIKING fluxes + WISE $W1,W2,W3$ fluxes
& $-1.18 \times 10^{-3}$ & $2.47 \times 10^{-4}$ & 0.0302 & 0.0197 \\
\hline
%$ugri$ GAaP magnitudes +\\
\hspace{1mm} KiDS \ttt{A}, \ttt{B} + \ttt{FLUX\_APER\_10}\_$ugri$ + GAaP colours 
& $-1.69 \times 10^{-3}$ & $-1.67 \times 10^{-4}$ & 0.0304 & 0.0200 \\
\hdashline
\hspace{1mm} KiDS GAaP colours + VIKING fluxes + WISE $W1,W2,W3$ fluxes
& $-1.32 \times 10^{-3}$ & $7.55 \times 10^{-5}$ & 0.0296 & 0.0191 \\
\hspace{1mm} KiDS \ttt{A}, \ttt{B} + VIKING fluxes + WISE $W1,W2,W3$ fluxes
& $-1.29 \times 10^{-3}$ & $7.42 \times 10^{-5}$ & 0.0290 & 0.0188 \\
\hline
%\multicolumn{5}{l}
{\hspace{1mm} KiDS \ttt{A}, \ttt{B} + GAaP colours + VIKING fluxes + WISE $W1,W2,W3$ fluxes}
& $-1.33 \times 10^{-3}$ & $3.57 \times 10^{-5}$ & 0.0288 & 0.0186 \\
\hline
\end{tabular} 
\tablefoot{
All the experiments used the same training and test sets of $\meanz = 0.23$; what was varied were the photometric parameters used for the \phz\ derivation, as given in the first column. See text for details of the photometric parameters.\\
\tablefoottext{a}{SMAD is the scaled median absolute deviation, converging to standard deviation for Gaussian distributions.\\}
\tablefoottext{b}{BPZ is independent of the training sets -- the numbers are given for comparison (for the same GAMA test sample). These statistics are based on the KiDS pipeline solution.\\}
\tablefoottext{c}{ANNz2 overall KiDS-DR3 solution from \S\ref{Sec:DR3-release} calculated for the same GAMA test set as in all the experiments of this table.}
}
\end{small}
\label{Tab: photo-z tests GAMA} 
\end{center}
\end{table*}

%% file: table-phz-GAMA.tex
%% Table with photo-z statistics - GAMA depth
\begin{table*}[!h]
\begin{center}
\caption{\label{Tab: photo-z tests GAMA full} Statistics of photometric redshift performance obtained for the KiDS-GAMA spectroscopic sample.}
\begin{small}
\begin{tabular}{ l  r  r  r  r  }
\hline\hline
{\textbf{parameter setup}} & \multicolumn{1}{c}{ \textbf{mean of} } & \multicolumn{1}{c}{ \textbf{mean of} } & \multicolumn{1}{c}{\textbf{st.dev.\ of}} & \multicolumn{1}{c}{\textbf{SMAD\tablefootmark{a} of}} \\ 
{} & \multicolumn{1}{c}{$\delta z = z_\mathrm{ph}-z_\mathrm{sp}$ } &  \multicolumn{1}{c}{$\delta z/(1+z_\mathrm{sp})$ } &  \multicolumn{1}{c}{$\delta z/(1+z_\mathrm{sp})$ } &  \multicolumn{1}{c}{$\delta z/(1+z_\mathrm{sp})$} \\ 
\hline 
$ugri$ GAaP mags, BPZ \tablefootmark{b}
& $1.34 \times 10^{-2}$ & $1.17 \times 10^{-2}$ & 0.0481 & 0.0317     \\ 
$ugri$ GAaP mags, DR3 ANNz2 overall solution \tablefootmark{c}
& $-2.03\times 10^{-3}$ & $8.61\times 10^{-4}$ & 0.0508 &  0.0345  \\ 
$ugri$ GAaP mags, GAMA-depth ANNz2 solution
& $-1.94\times 10^{-3}$ & $-1.91\times 10^{-4}$ & 0.0328  & 0.0219 \\ 
$ugri$ ISO magnitudes
& $-2.45\times 10^{-3}$ & $-3.26\times 10^{-4}$ & 0.0360 & 0.0259 \\ 
$ugri$ AUTO magnitudes
& $-2.72\times 10^{-3}$ & $-3.83\times 10^{-4}$ & 0.0384 & 0.0278 \\ 
\hline 
$ugri$ GAaP magnitudes +\\
\hspace{1mm} KiDS \ttt{FLUX\_APER\_10}\_$ugri$
& $-1.98\times 10^{-3}$ & $-2.71\times 10^{-4}$ & 0.0321 & 0.0214 \\
\hspace{1mm} KiDS \ttt{FLUX\_APER\_14}\_$ugri$
& $-1.97\times 10^{-3}$ & $-2.49\times 10^{-4}$ & 0.0323 & 0.0214 \\
\hspace{1mm} KiDS \ttt{FLUX\_APER\_25}\_$ugri$
& $-2.00\times 10^{-3}$ & $-2.84\times 10^{-4}$ & 0.0327 & 0.0214 \\
\hspace{1mm} KiDS \ttt{FLUX\_APER}\_$\lbrace 4, ..., 100 \rbrace$\_$r$
& $-1.85\times 10^{-3}$ & $-1.91\times 10^{-4}$ & 0.0315 & 0.0214 \\
\hspace{1mm} KiDS \ttt{A}, \ttt{B} linear semi-axes
& $-1.96\times 10^{-3}$ & $-2.54\times 10^{-4}$ & 0.0318 & 0.0213 \\
\hspace{1mm} KiDS six GAaP $ugri$ colours 
& $-2.13\times 10^{-3}$ & $-3.99\times 10^{-4}$ & 0.0321 & 0.0210 \\
\hdashline
\hspace{1mm} WISE $W1,W2$ fluxes
& $-1.44\times 10^{-3}$ & $1.28\times 10^{-4}$ & 0.0314 & 0.0209 \\
\hspace{1mm} {WISE $W1,W2$ fluxes, observational errors used\tablefootmark{d}}
& $-1.55\times 10^{-3}$ & $5.28\times 10^{-5}$ & 0.0313 & 0.0209 \\
\hspace{1mm} WISE $W1,W2,W3$ fluxes
& $-1.61\times 10^{-3}$ & $-1.47\times 10^{-5}$ & 0.0312 & 0.0206\\
\hspace{1mm} {WISE $W1,W2,W3$ fluxes, observational errors used\tablefootmark{d}}
& $-1.54\times 10^{-3}$ & $3.45\times 10^{-5}$ & 0.0311 & 0.0205\\
\hspace{1mm} VIKING $z Y J H K_s$ fluxes
& $-1.58\times 10^{-3}$ & $-2.22\times 10^{-5}$ & 0.0312 & 0.0202 \\
\hline
$ugri$ GAaP magnitudes +\\
\hspace{1mm} KiDS \ttt{A}, \ttt{B} + \ttt{FLUX\_APER\_10}\_$ugri$
& $-1.71\times 10^{-3}$ & $-1.16\times 10^{-4}$ & 0.0311 & 0.0209 \\
\hspace{1mm} KiDS \ttt{FLUX\_APER}\_$\lbrace 4, ..., 100 \rbrace$\_$r$ + GAaP colours
& $-1.74\times 10^{-3}$ & $-1.75\times 10^{-4}$ & 0.0309 & 0.0206 \\
\hspace{1mm} KiDS \ttt{FLUX\_APER\_10}\_$ugri$ + GAaP colours
& $-1.71\times 10^{-3}$ & $-1.23\times 10^{-4}$ & 0.0313 & 0.0205 \\
\hspace{1mm} KiDS \ttt{A}, \ttt{B} + GAaP colours
& $-1.59\times 10^{-3}$ & $-7.61\times 10^{-5}$ & 0.0305 & 0.0204 \\
\hdashline
\hspace{1mm} KiDS \ttt{A}, \ttt{B} + WISE $W1,W2$ fluxes
& $-1.45\times 10^{-3}$ & $5.68\times 10^{-5}$ & 0.0303 & 0.0205 \\
\hspace{1mm} KiDS GAaP colours + WISE $W1,W2$ fluxes
& $-1.60\times 10^{-3}$ & $-6.87\times 10^{-5}$ & 0.0307 & 0.0202 \\
\hspace{1mm} KiDS \ttt{A}, \ttt{B} + VIKING fluxes
& $-1.70\times 10^{-3}$ & $-1.25\times 10^{-4}$ & 0.0307 & 0.0201 \\
\hspace{1mm} KiDS \ttt{A}, \ttt{B} + WISE $W1,W2,W3$ fluxes
& $-1.43\times 10^{-3}$ & $4.33\times 10^{-5}$ & 0.0300 & 0.0200 \\
\hspace{1mm} KiDS GAaP colours + VIKING fluxes
& $-1.72\times 10^{-3}$ & $-1.53\times 10^{-4}$ & 0.0311 & 0.0199 \\
\hspace{1mm} KiDS GAaP colours + WISE $W1,W2,W3$ fluxes
& $-1.50\times 10^{-3}$ & $-9.24\times 10^{-6}$ & 0.0303 & 0.0198 \\
\hspace{1mm} VIKING fluxes + WISE $W1,W2,W3$ fluxes
& $-1.18\times 10^{-3}$ & $2.47\times 10^{-4}$ & 0.0302 & 0.0197 \\
\hline
$ugri$ GAaP magnitudes +\\
\hspace{1mm}  KiDS \ttt{A}, \ttt{B} + \ttt{FLUX\_APER}\_$\lbrace 4, ..., 100 \rbrace$\_$r$ + GAaP colours 
& $-1.73\times 10^{-3}$ & $-1.93\times 10^{-4}$ & 0.0302 & 0.0201 \\
\hspace{1mm} KiDS \ttt{A}, \ttt{B} + \ttt{FLUX\_APER\_4}\_$ugri$ + GAaP colours 
& $-1.77\times 10^{-3}$ & $-2.24\times 10^{-4}$ & 0.0303 & 0.0201 \\
\hspace{1mm} KiDS \ttt{A}, \ttt{B} + \ttt{FLUX\_APER\_10}\_$ugri$ + GAaP colours 
& $-1.69\times 10^{-3}$ & $-1.67\times 10^{-4}$ & 0.0304 & 0.0200 \\
\hdashline
\hspace{1mm} KiDS \ttt{A}, \ttt{B} + GAaP colours + VIKING fluxes
& $-1.59\times 10^{-3}$ & $-1.07\times 10^{-4}$ & 0.0299 & 0.0194 \\
\hspace{1mm} KiDS \ttt{A}, \ttt{B} + GAaP colours + WISE $W1,W2,W3$ fluxes
& $-1.59\times 10^{-3}$ & $-1.24\times 10^{-4}$ & 0.0293 & 0.0192 \\
\hspace{1mm} KiDS GAaP colours + VIKING fluxes + WISE $W1,W2,W3$ fluxes
& $-1.32\times 10^{-3}$ & $7.55\times 10^{-5}$ & 0.0296 & 0.0191 \\
\hspace{1mm} KiDS \ttt{A}, \ttt{B} + VIKING fluxes + WISE $W1,W2,W3$ fluxes
& $-1.29\times 10^{-3}$ & $7.42\times 10^{-5}$ & 0.0290 & 0.0188 \\
\hline
$ugri$ GAaP magnitudes +\\
{\hspace{1mm} KiDS \ttt{A}, \ttt{B} + GAaP colours + VIKING fluxes + WISE $W1,W2,W3$ fluxes} & $-1.33\times 10^{-3}$ & $3.57\times 10^{-5}$ & 0.0288 & 0.0186 \\
{\hspace{1mm} KiDS \ttt{A}, \ttt{B}  + \ttt{FLUX\_APER\_10}\_$ugri$ + GAaP colours + VIKING fluxes}
& $-1.75\times 10^{-3}$ & $-2.46\times 10^{-4}$ & 0.0299 & 0.0195 \\
\multicolumn{5}{l}{\hspace{1mm} KiDS \ttt{A}, \ttt{B}  + \ttt{FLUX\_APER\_10}\_$ugri$ + GAaP colours + WISE $W1,W2,W3$ fluxes}\\
& $-1.61\times 10^{-3}$ & $-1.68\times 10^{-4}$ & 0.0291 & 0.0190 \\
\hline
$ugri$ GAaP magnitudes +\\
\multicolumn{5}{l}{\hspace{1mm} KiDS \ttt{A}, \ttt{B}  + \ttt{FLUX\_APER\_10}\_$ugri$ + GAaP colours + VIKING fluxes + WISE $W1,W2,W3$ fluxes}\\
& $-1.56\times 10^{-3}$ & $-1.48\times 10^{-4}$ & 0.0287 & 0.0186 \\
\hline
\end{tabular} 
\tablefoot{All the experiments used the same training and test sets of $\meanz = 0.23$; what was varied were the photometric parameters used for the \phz\ derivation, as given in the first column. See text for details of the photometric parameters.\\
\tablefoottext{a}{SMAD is the scaled median absolute deviation, converging to standard deviation for Gaussian distributions.\\}
\tablefoottext{b}{BPZ is independent of the training sets -- the numbers are given for comparison (for the same GAMA test sample). These statistics are based on the KiDS pipeline solution.\\}
\tablefoottext{c}{ANNz2 overall KiDS-DR3 solution from \S\ref{Sec:DR3-release} calculated for the same GAMA test set as in all the experiments of this table.\\}
\tablefoottext{d}{{Errors on all the input parameters were directly used in \phz\ derivation instead of relying on the internal error model of ANNz2. See \S\ref{Sec:ANNz2} and \S\ref{Sec:GAMA-depth_mags+one_set} for details.}}
}
\end{small}
\end{center}
\end{table*}